\documentclass[12pt]{article}
\pdfoutput=1
\usepackage{jheppub}

\pdfoutput=1
\usepackage{slashed}
\usepackage{color, verbatim}
\usepackage{latexsym}
\usepackage{epsfig}
\usepackage{amsmath,accents}

\usepackage{amssymb}
\usepackage{graphicx}
\usepackage{bm}

\usepackage[font={small}]{caption}

\usepackage{hyperref}
\usepackage{epstopdf}
\definecolor{myred}{rgb}{0.7, 0, 0}
\definecolor{myblue}{rgb}{0, 0, 0.7}
\definecolor{mygreen}{rgb}{0.04, 0.7, 0.5}
\hypersetup{colorlinks,citecolor=myred,linkcolor=myblue,urlcolor=myblue,linktocpage=true}

\makeatletter

\@addtoreset{equation}{section}
\makeatother

\newcommand{\be}{\begin{equation}}
\newcommand{\ee}{\end{equation}}
\newcommand{\bea}{\begin{eqnarray}}
\newcommand{\eea}{\end{eqnarray}}

\newcommand{\tr}{\operatorname{tr}}
\newcommand{\diag}{\operatorname{diag}}

\begin{document}

\thispagestyle{empty}

\begin{center}


\begin{center}

\vspace{.5cm}

{\Large\bf Cosmological Phase Transitions in Warped Space: \\
\vspace{0.3cm}Gravitational Waves and Collider Signatures}\\

\end{center}

\vspace{1.cm}

\textbf{
Eugenio Meg\'ias$^{\,a, b}$, Germano Nardini$^{\,c}$, Mariano Quir\'os$^{\,d}$
}\\

\vspace{.1cm}
${}^a\!\!$ {\em {Departamento de F\'{\i}sica At\'omica, Molecular y Nuclear and \\ Instituto Carlos I de F\'{\i}sica Te\'orica y Computacional, Universidad de Granada,\\ Avenida de Fuente Nueva s/n,  18071 Granada, Spain}}

${}^b\!\!$ {\em {Departamento de F\'{\i}sica Te\'orica, Universidad del Pa\'{\i}s Vasco UPV/EHU, \\ Apartado 644,  48080 Bilbao, Spain}}

${}^c\!\!$ {\em {AEC,  Institute  for  Theoretical  Physics,  University  of  Bern,\\
Sidlerstrasse  5,  CH-3012  Bern,  Switzerland
}}

\vspace{.1cm}
${}^d\!\!$ {\em {Institut de F\'{\i}sica d'Altes Energies (IFAE),\\ The Barcelona Institute of  Science and Technology (BIST),\\ Campus UAB, 08193 Bellaterra (Barcelona) Spain
}}



\end{center}

\vspace{0.8cm}

\centerline{\bf Abstract}
\vspace{2 mm}

\begin{quote}\small
  We study the electroweak phase transition within a 5D warped model including a scalar potential with an exponential behavior, and strong back-reaction over the metric, in the infrared. By means of a novel treatment of the superpotential formalism, we explore parameter regions that were previously inaccessible. We find that for large enough values of the t'Hooft parameter (e.g.~$N\simeq 25$) the holographic phase transition occurs, and it can force the Higgs to undergo a first order electroweak phase transition, suitable for electroweak baryogenesis. The model exhibits gravitational waves and colliders signatures. It typically predicts a stochastic gravitational wave background observable both at the Laser Interferometer Space Antenna and at the Einstein Telescope. Moreover the radion tends to be heavy enough such that it evades current constraints, but may show up in future LHC runs.
\end{quote}

\vfill

\newpage

\tableofcontents

\newpage
\section{Introduction}
\label{sec:introduction}
The Standard Model (SM) of ElectroWeak (EW) and strong interactions has been put on solid grounds by the past and current experimental data collected at e.g.~the Large Hadron Collider (LHC) or the Large Electron Positron collider~\cite{ALEPH:2005ab,Olive:2016xmw}. Still the model is unable to cope with some cosmological observables and suffers from theoretical drawbacks. For instance, it fails to explain a number of observational and consistency issues such as the baryon asymmetry of the universe, the strong CP problem, the origin of the flavor structure, the origin of inflation and the strong sensitivity to high scale physics. In particular the latter problem, a.k.a.~the hierarchy problem, has motivated the introduction of Beyond the SM (BSM) physics which makes nowadays the subject of active experimental searches at the LHC. 

One of the best motivated BSM frameworks was introduced years ago by Randall and Sundrum~\cite{Randall:1999ee}. In this scenario the hierarchy between the Planck and EW scales is generated by the Anti de Sitter (AdS) warp factor involved in the extra dimension. An appealing feature of this framework is that the five-dimensional (5D) model is holographically dual to a non-perturbative four-dimensional (4D) Conformal Field Theory (CFT) and the dynamics of the strongly-coupled states of the 4D theory can be investigated perturbatively by means of the 5D theory.

Once the extra dimension is integrated out, the Randall-Sundrum theory contains towers of heavy states, the Kaluza-Klein (KK) modes of all SM particles, propagating in the bulk. It also contains a light state, the radion, dual to the dilaton, a Goldstone boson of the conformal invariance of the dual 4D theory. In the absence of a potential stabilizing the brane distance (see e.g.~Ref.~\cite{Goldberger:1999uk}), the radion (and equivalently the dilaton) is massless but, as soon as the extra dimension is stabilized, it acquires a mass.  Still the radion typically remains the lightest BSM state and it can play a relevant role in the collider and early-universe phenomenology. In particular, it undergoes a phase transition during which it acquires a Vacuum Expectation Value (VEV) and which, in the dual language, corresponds to a (holographic) phase transition from the deconfined to the confined phase. In other words, the dilaton condenses.

The holographic phase transition has been studied by a number of authors and it has been concluded to be of first-order~\cite{Creminelli:2001th,Randall:2006py,Kaplan:2006yi,Nardini:2007me,Hassanain:2007js,Konstandin:2010cd,Bunk:2017fic,Dillon:2017ctw,vonHarling:2017yew}. However, in models with small back-reaction on the gravitational metric, in order to avoid the graceful exit problem, one has to consider scenarios where the number of degrees of freedom in the CFT phase (i.e.~the number of ``colors'' $N$ of the $SU(N)$ symmetry) is small, thus jeopardizing the perturbativity of the 5D gravitational theory. It is hence worth investigating models where the conformal symmetry is strongly broken in the infrared (IR), but the corresponding large back-reaction can be conveniently treated. 	In this way one expects to avoid the graceful exist problem even with $N$ large, with clear benefits for the perturbativity of the 5D gravitational theory.

In the present paper we provide a method to deal with the large back-reaction issue. This method is a generalization of the superpotential procedure~\cite{DeWolfe:1999cp}, and to show its capabilities, we apply it to analyze a class of theories where conformality is strongly broken at the IR brane. We dub these theories soft-wall models as they generate a naked singularity in the 5D metric beyond the location of the IR brane. Although the singularity is outside  the physical interval, between the two branes, the distance of the singularity from the IR brane is important because it controls the breaking of conformality. This kind of models were introduced as minimal ultraviolet (UV) completions with no tension with EW precision data~\cite{Cabrer:2009we,Cabrer:2010si,Cabrer:2011fb,Cabrer:2011vu,Cabrer:2011mw,Carmona:2011ib,Cabrer:2011qb,deBlas:2012qf,Quiros:2013yaa,Megias:2015ory}, as an alternative to models with extended (custodial) gauge symmetry~\cite{Agashe:2003zs}. Recently, the same models were also considered to accommodate the $(g_\mu-2)$~\cite{Megias:2017dzd} and $B$-meson anomalies~\cite{Megias:2016bde,Megias:2015qqh,Megias:2016jcw,Megias:2017ove,Megias:2017isd,Megias:2017vdg,Megias:2017mll}, in agreement with the quark mass and mixing angle spectra, and the natural generation of lepton flavor universality violation.

The outline of the paper is as follows. In Sec.~\ref{sec:general}  we introduce the general formalism for the 5D action, including the Gibbons-Hawking-York (GHY) boundary term. We also review the Equations of Motion (EoM) and Boundary Conditions (BCs) of the theory, and show that solving the EoM is equivalent to applying the superpotential procedure~\cite{DeWolfe:1999cp}.

In Sec.~\ref{sec:effective_potential} we develop a novel method to employ the superpotential formalism in the presence of mistuned BCs. This allows to calculate the effective potential between the two branes as a function of their distance without major problems with the back-reaction. It hence opens up the possibility of studying warped models without imposing tight upper bounds on the amount of back-reaction.

In Sec.~\ref{sec:model} we introduce the particular soft-wall model and we apply the generalized superpotential method to it. Since the method needs to be carried out numerically, we focus on some benchmark scenarios with different degrees of back-reaction (up to $N\simeq 25$). In all cases, the relevant parameters are set to solve the hierarchy problem. 

The relation between the UV and IR brane distance and the canonically normalized radion field is analyzed in Sec.~\ref{sec:radion}. The effective potential for the brane distance, previously obtained, can then be reinterpreted as a function of the physical radion field. This in particular allows to ensure that in our benchmark scenarios the KK gravitons are much heavier than the radion. For this reason the radion phase transition can be analysed within an Effective Field Theory (EFT) where the SM-like particles and the radion are the only dynamical fields.

The EFT at finite temperature of the soft-wall model is computed in Sec.~\ref{sec:effective_potential_finiteT}. We obtain that, depending on the amount of back-reaction, the free energy difference between the confined and deconfined phases can span several orders of magnitude. This of course has relevant effects on the value of the nucleation temperature and, in turn, on the phenomenology of the model.

In Sec.~\ref{sec:phase-transition} we analyze the phase transition of the radion in detail. We find that, in agreement with precedent analyses~\cite{Creminelli:2001th,Randall:2006py,Kaplan:2006yi}, for tiny back-reaction the nucleation rate tends to be too small to overcome the Hubble expansion rate, and hence the universe is stuck in an eternal inflationary phase. Instead, for scenarios with large back-reaction, the universe inflates by (at most) a few e-folds and eventually completes the transition. In these cases the nucleation temperature is typically of the order of the EW scale, contrarily to what happens in most of the (small-back-reaction) frameworks considered in the literature~\cite{Nardini:2007me,Hassanain:2007js,Konstandin:2010cd,Bunk:2017fic,Dillon:2017ctw}. Moreover, depending on the benchmark choice, the transition can end up with a reheating temperature smaller or larger than the nucleation temperature of the EW phase transition in the SM.  We highlight the implication of this feature in Sec.~\ref{sec:EWPT}, with some remarks about the feasibility of EW baryogenesis. 

In Sec.~\ref{sec:GW} we discuss the prospects for detecting the stochastic gravitational wave (GW) background that the radion phase transition induces. Interestingly enough, the signal is so strong that both the Laser Interferometer Space Antenna (LISA) and Einstein Telescope (ET) will have very good chances to detect it. 

We observe that the large-back-reaction regime favors the radion mass to be large, typically around the TeV scale. The corresponding collider phenomenology is studied in Sec.~\ref{sec:phenomenology}.
No tension with present LHC data is found for the benchmark scenarios although, for future integrated luminosity, the radion decay into $W^+W^-$ and $ZZ$ might lead to detectable signatures. 

Finally general conclusions are drawn in Sec.~\ref{sec:conclusions}.

\section{General formalism}
\label{sec:general}

We follow the notation and conventions of Ref.~\cite{Konstandin:2010cd}~\footnote{Except for a global change in the sign of the metric exponent as $e^{A(r)}\to e^{-A(r)}$.}. We consider a slice of 5D spacetime between two branes at values $r=r_0$, the UV brane, and $r=r_1$, the IR brane. The 5D action of the model, including the stabilizing bulk scalar $\phi(x,r)$, reads as
\begin{eqnarray}
S &=& \int d^5x \sqrt{|\det g_{MN}|} \left[ -\frac{1}{2\kappa^2} R + \frac{1}{2} g^{MN}(\partial_M \phi)(\partial_N \phi) - V(\phi) \right]\nonumber \\
&-& \sum_{\alpha} \int_{B_\alpha} d^4x \sqrt{|\det \bar g_{\mu\nu}|} \Lambda_\alpha(\phi)  
 -\frac{1}{\kappa^2} \sum_{\alpha} \int_{B_\alpha} d^4x \sqrt{|\det \bar g_{\mu\nu}|} K_\alpha  \,, \label{eq:action}
\end{eqnarray}
where $V(\phi)$ and $\Lambda_\alpha(\phi)$ are the bulk and brane potentials of the scalar field $\phi$, and the index $\alpha=0 \; (\alpha=1)$ refers to the UV (IR)  brane. 
The parameter $\kappa^2=1/(2M^3)$, with $M$ being the 5D Planck scale, can be traded by the parameter $N$ in the holographic theory by the relation~\cite{Gubser:1999vj}
\be
N^2\simeq \frac{8\pi^2\ell^3}{\kappa^2}\,,
\label{eq:N}
\ee
where $\ell$ is a constant parameter of the order of the Planck length, which determines the value of the 5D curvature. 
The metric $g_{MN}$ is defined in proper coordinates by
\begin{eqnarray}
ds^2 &=&g_{MN}dx^M dx^N\equiv e^{-2A(r)} \eta_{\mu\nu} dx^\mu dx^\nu-dr^2 \,,  \label{eq:metric}  
\end{eqnarray}
so that in Eq.~(\ref{eq:action}) the 4D induced metric is $ \bar g_{\mu\nu}=e^{-2A(r)}\eta_{\mu\nu}$, where the Minkowski metric is given by $\eta_{\mu\nu} =\diag(+1,-1,-1,-1)$. 
 The last term in Eq.~(\ref{eq:action}) is the usual GHY boundary term~\cite{York:1972sj,Gibbons:1976ue}, where $K_{\alpha}$ are the extrinsic UV and IR curvatures. In terms of the metric of Eq.~(\ref{eq:metric}) the extrinsic curvature tensor reads as
\begin{equation}
K_{\mu\nu} = \frac{1}{2} \frac{d}{dr} \left(  \bar g_{\mu\nu} \right) = - e^{-2A} A^\prime \eta_{\mu\nu} \,,
\end{equation}
with trace 
\begin{equation}
K = e^{2A} \eta^{\mu\nu} K_{\mu\nu} = -A^\prime \eta^{\mu\nu} \eta_{\mu\nu} = -4 A^\prime \,,
\end{equation}
so that $K_{0,1} = \mp 4 A^\prime(r_{0,1})$.

The EoM read then as~\footnote{From here on the prime symbol $(\,{}^\prime\,)$ will stand for the derivative of a function with respect to its argument.}
\begin{eqnarray}
&&A^{\prime\prime} 
= \frac{\kappa^2}{3} \phi^{\prime \, 2}   \,, \label{eq:eom1}\\
&&A^{\prime\, 2} 
= -\frac{\kappa^2}{6} V(\phi) + \frac{\kappa^2}{12} \phi^{\prime\, 2} \,,  \label{eq:eom2}\\
&&\phi^{\prime\prime} - 4 A^\prime \phi^\prime = V^\prime(\phi) \,, \label{eq:eom3}
\end{eqnarray}
and, assuming a $\mathbb Z_2$ symmetry across the branes, the localized terms impose the constraints
\begin{align}
A^\prime(r_\alpha)&=\frac{\kappa^2}{6}(-1)^\alpha \Lambda_\alpha(\phi(r_\alpha)) \,, 
\label{eq:BCA}\\
\phi^\prime(r_\alpha)&= \frac{1}{2}(-1)^\alpha \frac{\partial\Lambda_\alpha(\phi(r_\alpha))}{\partial\phi} \,.
\label{eq:BCphi}
\end{align}
The EoM can then be written in terms of the superpotential $W(\phi)$ as~\cite{DeWolfe:1999cp}
\begin{equation}
\phi^\prime = \frac{1}{2} \frac{\partial W}{\partial \phi} \,, \qquad A^\prime = \frac{\kappa^2}{6} W \,, \label{eq:phiA}
\end{equation}
and
\begin{equation}
V(\phi) = \frac{1}{8} \left( \frac{\partial W}{\partial \phi} \right)^2 - \frac{\kappa^2}{6} W^2(\phi) \,,\label{eq:V}
\end{equation}
while the BCs read as
\begin{align}
\Lambda_\alpha(\phi(r_\alpha))&=(-1)^\alpha W(\phi(r_\alpha)) \,,
\label{eq:BCW}\\
\frac{\partial\Lambda_\alpha(\phi(r_\alpha))}{\partial\phi}&=(-1)^\alpha \frac{\partial W(\phi(r_\alpha))}{\partial \phi} \,.
\label{eq:BCWprime}
\end{align}
Note that the EoM in Eqs.~(\ref{eq:eom1})-(\ref{eq:BCphi}) and Eqs.~(\ref{eq:phiA})-(\ref{eq:BCWprime}) are completely equivalent, having both sets three integration constants. In particular one of the integration constants appears in Eq.~(\ref{eq:V}). 

Starting from a potential $V$ and integrating Eq.~(\ref{eq:V}) is usually a very complicated task which normally cannot be accomplished analytically. On the other hand starting from a superpotential function $W$, and computing the potential $V$ from Eq.~(\ref{eq:V}), amounts to fixing the corresponding integration constant to zero, and no radion potential can be generated using this method. To circumvent this problem (for details see the next section)  we propose an alternative procedure: we determine the effective potential  by integrating the action over the solutions of the EoM with the scalar 
BC (\ref{eq:BCphi}) (or equivalently~(\ref{eq:BCWprime})) imposed at both branes, but we mistune the BC (\ref{eq:BCA}) [or equivalently~(\ref{eq:BCW})] while finely adjusting the potential $\Lambda_0$~\footnote{See e.g.~the thorough discussion in Ref.~\cite{Bellazzini:2013fga}.}. In this way, by means of the mistuning we 
break the flatness of the radion potential, and by means of the $\Lambda_0$ adjustment we achieve a zero cosmological constant at the minimum of the potential.

For concreteness we consider for the brane potentials the form
\be
\Lambda_\alpha(\phi)=\Lambda_\alpha+\frac{\gamma_\alpha}{2}(\phi-v_\alpha)^2
\label{eq:brane-potentials}
\ee
where $\Lambda_\alpha$ is a constant, hereafter considered as a free parameter as it does not enter in Eqs.~(\ref{eq:BCphi}) and (\ref{eq:BCWprime}), and $\gamma_\alpha$ is a dimensionful parameter.
Using Eq.~(\ref{eq:brane-potentials}) for the brane potentials, the BCs in Eq.~(\ref{eq:BCWprime}) can be written as
\be
\phi(r_\alpha)-v_\alpha=\frac{(-1)^\alpha}{\gamma_\alpha}\frac{\partial W(\phi(r_\alpha))}{\partial\phi} \,,
\label{eq:BCWprime2}
\ee
which fixes two integration constants, from the first equality of Eq.~(\ref{eq:phiA}) and Eq.~(\ref{eq:V}), in terms of the parameters $v_\alpha$. Using now Eq.~(\ref{eq:BCWprime2}) the brane potentials can be written as
\be
\Lambda_\alpha(\phi(r_\alpha))=\Lambda_\alpha+\frac{1}{2\gamma_\alpha}\left(\frac{\partial W(\phi(r_\alpha))}{\partial\phi}\right)^2 \,.
\label{eq:brane-potentials2}
\ee 
As we will see in Sec.~\ref{sec:effective_potential} the effective potential will depend through $\Lambda_\alpha(\phi(r_\alpha))$ on the $\gamma_\alpha$ parameters. 

In the simple (stiff wall) limit where $\gamma_\alpha\to\infty$, the BCs (\ref{eq:BCWprime2}) and the potential (\ref{eq:brane-potentials2}) simplify to
\be
\phi(r_\alpha)=v_\alpha,\quad \Lambda_\alpha(\phi(r_\alpha))=\Lambda_\alpha
\label{eq:finalBC}
\ee 
in which case $\Delta_\alpha\equiv\Lambda_\alpha-(-1)^\alpha W(v_\alpha)$ measures the mistuning we are doing, while the parameters $\gamma_\alpha$ have introduced a dynamical mechanism by which $\phi(r_\alpha)=v_\alpha$. In fact the condition $\phi(0)=v_0$ is enforced by fixing the integration constant of the first equality in Eq.~(\ref{eq:phiA}), while the condition $\phi(r_1)=v_1$ is enforced by fixing the integration constant appearing in Eq.~(\ref{eq:V}) as we will see in Sec.~\ref{sec:effective_potential}. In the generic case of finite $\gamma_\alpha$, an analytic solution to the BCs (\ref{eq:BCWprime}) does in general not exist but still numerical solutions can be worked out, as we will see in Sec.~\ref{sec:effective_potential}. In the following, and unless explicit mention, we work in the limit $\gamma_\alpha\to\infty$.  

\section{The effective potential}
\label{sec:effective_potential}

By using Eqs.~(\ref{eq:eom1})-(\ref{eq:eom3}), the action~(\ref{eq:action}) can be written as
\begin{equation}
S = S_{\rm bulk}  + S_{\rm br} + S_{\rm GHY}  \,,
\end{equation}
with
\begin{eqnarray}
S_{\rm bulk} &=&  2 \int d^4x \int_{r_0}^{r_1} dr \sqrt{|\det g_{MN}|} \left[ -M^3 R + \frac{1}{2} (\partial \phi)^2 - V(\phi) \right]  \nonumber \\
&=&  \int d^4x \frac{1}{3} \left( \left[ e^{-4A}W \right]_{r_1} -  \left[ e^{-4A}W \right]_{r_0} \right) \,, \label{eq:S_bulk}  \\
S_{\rm br} =& -& \sum_\alpha \int_{B_\alpha} d^4x \sqrt{|\det \bar g_{MN}|} \Lambda_\alpha(\phi) 
=- \int d^4x  \left( \left[ e^{-4A} \Lambda_1 \right]_{r_1} + \left[ e^{-4A} \Lambda_0 \right]_{r_0} \right)   \\
S_{\rm GHY} &=& -\frac{1}{\kappa^2} \sum_\alpha \int_{B_\alpha} d^4x \sqrt{|\det \bar g_{MN}|} K_\alpha  \nonumber \\
 &=&\int d^4x \left( -\frac{4}{3}  \right) \left( \left[ e^{-4A}W \right]_{r_1} -  \left[ e^{-4A}W \right]_{r_0} \right)  \,,  \label{eq:S_GH}
\end{eqnarray}
where we have included a factor of $2$ in $S_{\rm bulk}$ and $S_{\rm GHY}$  from orbifolding, as we are integrating over $S^1/\mathbb Z_2$. By joining all these terms together we get
\begin{equation}
S \equiv - \int d^4x \,V_{eff} 
\end{equation}
with
\be
V_{eff}=  \left[ e^{-4A} \left( W + \Lambda_1 \right) \right]_{r_1}   +  \left[ e^{-4A} \left( - W + \Lambda_0 \right) \right]_{r_0}  \,, \label{eq:Veff}
\ee
where we are using the EoM degrees of freedom to fix $r_0=0$ and $A(0)=0$. The variable $r_1$ is thus the branes distance and establishes the relationship between $\kappa^2$ and the 4D rationalized Planck mass, $M_{P}=2.4\times 10^{18}$ GeV, via the expression
\be
\kappa^2 M_{P}^2=2\ell\int_0^{\bar r_1}d\bar r e^{-2A(\bar r)}\,,
\label{eq:kappa2}
\ee
where $\bar r\equiv r/\ell$ is dimensionless. In particular, for some given $N$ and $\bar r_1$,  Eqs.~(\ref{eq:N}) and (\ref{eq:kappa2}) fix the value of $\ell$.

In the limiting case $\gamma_\alpha\to\infty$, using the superpotential formalism, the first equation in (\ref{eq:phiA}) has just one integration constant and thus only the value of the field at, say the UV brane, is fixed (thus~$v_0$ is fixed). Therefore within the superpotential formalism, if we start from a superpotential $W_0$ from which the bulk potential $V$ is deduced, we fix to zero the integration constant that should have appeared in Eq.~(\ref{eq:V}). We have then lost the freedom to choose the value of $\phi$ at the IR brane ($v_1$), in particular we cannot set $v_1$ at the value for  which $r_1$ solves the hierarchy problem [cf.~Eq.~\eqref{eq:finalBC}].
However, as we now explain, there exists a way of reintroducing such a freedom. Let us call the ``lost" integration constant $s$. 

We consider a potential $V$ that is expressed in terms of a zero-order superpotential $W_0$ via Eq.~(\ref{eq:V}), with 
\be
W=\sum_{n=0}^\infty s^n W_n
\label{expansion}
\ee
being solution of Eq.~(\ref{eq:V}) to all orders. This means that Eq.~(\ref{eq:V}) does not fix the integration constant $s$, which should then be fixed from the BC $\phi(r_1)=v_1$. An explicit solution is given for $n=1$ by~\cite{Papadimitriou:2007sj} (see also discussion in~\cite{Megias:2014iwa,Megias:2015nya})
\be
W_1(\phi)=\frac{1}{\ell\kappa^2} 
\exp\left(\frac{4\kappa^2}{3}\int^\phi \frac{W_0(\bar\phi)}{W'_0(\bar\phi)} d\bar\phi \right) \,,
\ee
while for $n>1$ it can be iteratively defined as
\be
W_n(\phi)=W_1(\phi) \int^\phi \frac{Q_n(\bar\phi)}{W_0^\prime(\bar\phi) W_1(\bar\phi) }\,d\bar\phi
\ee
with
\be
Q_n=-\frac{1}{2}\sum_{m=1}^{n-1}\left[ W_m^\prime W_{n-m}^\prime - \frac{4\kappa^2}{3}
W_m W_{n-m}\right] \,.
\ee

From now on we assume $s W_1\ll W_0$, so that we can keep the expansion in Eq.~(\ref{expansion}) to linear order, which corresponds to use $W=W_0+s W_1+\mathcal O(s^2)$, an approximation that should be verified a posteriori. We can similarly expand the field $\phi$ and metric $A$ as~\footnote{Notice that the mass dimensions are $[W]=4$, $[s]=0$ and $[\phi]=3/2$.}
\begin{align}
\phi(r)&=\phi_0(r)+s\phi_1(r)+\mathcal O(s^2) \,, \\
A(r)&=A_0(r)+s A_1(r)+\mathcal O(s^2) \,.
\end{align}
As we are solving Eq.~(\ref{eq:phiA}) order by order perturbatively, condition $sW_1\ll W_0$ also implies $s\phi_1\ll \phi_0$ and $sA_1\ll A_0$.
The corresponding expansion of $W$ then reads
\be
W(\phi)=W_0(\phi_0)+s\left[ \phi_1 W^{\prime}_0(\phi_0)+W_1(\phi_0)  \right]+\mathcal O(s^2) \,.
\label{dos}
\ee
Using now the first expression in Eq.~(\ref{eq:phiA}) we get
\begin{align}
\phi_0^\prime(r)&=\frac{1}{2}W^\prime_0(\phi_0),\quad \phi_1^\prime(r)=\frac{1}{2}\left[ \phi_1 W^{\prime\prime}_0(\phi_0)+W^\prime_1(\phi_0)  \right] \,,\label{usual}\\
\phi_1(r)&\equiv\phi_1[\phi_0(r)]= W_0^\prime(\phi_0)  \int_{C_1}^{\phi_0} \frac{W_1^\prime(\bar\phi)}{[ W_0^\prime(\bar\phi) ]^2}  d\bar\phi \,,  \label{eq:phi1A}
\end{align}
where Eq.~(\ref{eq:phi1A}) defines the field $\phi_1(r)$, while the first relation in Eq.~(\ref{usual}) is the usual equation for $\phi_0(r)$ [cf.~Eq.~(\ref{eq:phiA})]. The integration constants have been chosen to fulfill the BCs
\begin{equation}
\phi(0) = v_0  \,, \qquad \phi(r_1) = v_1 \,,
\end{equation}
corresponding to the values of $\phi(r)$ in the UV and IR branes, respectively. In particular one can fix $C_1=v_0$ such that $\phi_0(0)=v_0$ and $\phi_1(0)=0$. Then the condition $\phi(r_1)=v_1$ leads to fixing the integration constant $s$ as~\footnote{For the case of finite $\gamma_\alpha$, Eq.~(\ref{eq:s-fixing}) has $\mathcal O(1/\gamma_\alpha)$ corrections.}
\be
s(r_1)=\frac{v_1-\phi_0(r_1)}{\phi_1[\phi_0(r_1)]} \,.
\label{eq:s-fixing}
\ee
Therefore the superpotential in Eq.~(\ref{dos}) gets an explicit dependence on the brane distance, $W(r_1)$, which in turn creates a non-trivial dependence on $r_1$ of the effective potential of Eq.~(\ref{eq:Veff}). 
As the latter only gets contributions  from the branes, one can then expand the superpotential on the branes as
\be
W(v_\alpha)=W_0(v_\alpha)+s(r_1) W_1(v_\alpha)
\ee
so that the effective potential can be expanded to first order in $s(r_1)$:
\begin{align}
&V_{eff}(r_1)=\Lambda_0-W_0(v_0)\label{potencialefectivo}\\
&+e^{-4 A_0(r_1)}
\left\{\left[\Lambda_1+W_0(v_1)\right][1-4 A_1 s(r_1)]+s(r_1)\left[W_1(v_1)-e^{4A_0(r_1)}W_1(v_0) \right] \right\}  \,.
\nonumber
\end{align}

Eq.~(\ref{potencialefectivo}) involves several key parameters that
play a relevant role in our analysis. The second line, and in particular the function $s(r_1)$, provides a non-trivial dependence on the brane distance $r_1$. We anticipate that $r_1$ can be interpreted as the constant background value of the (canonically unnormalized) radion/dilaton field. Consequently, the cosmological constant at the minimum of the radion potential can be set to zero by an accurate choice of  the terms in the first line, which are independent of  $r_1$. We fine-tune $\Lambda_0$ for such a purpose~\footnote{This one is the cosmological constant fine-tuning of the theory.}. 

Similarly, from Eq.~(\ref{dos}) and the second expression in Eq.~(\ref{eq:phiA}) one finds
\begin{align}
A_0^\prime(r)&=\frac{\kappa^2}{6}W_0(\phi_0) \,, \nonumber\\
A_1^\prime(r)&=\frac{\kappa^2}{6}\left(  \phi_1W^\prime_0(\phi_0)+W_1(\phi_0)  \right) \,.
\label{eq:A}
\end{align}
After solving Eqs.~(\ref{usual}) and (\ref{eq:phi1A}), we have to integrate Eqs.~(\ref{eq:A}) to obtain the metric. This yields
\begin{align}
A_0(r) &= \frac{1}{4} \log \left[\frac{W_1(\phi_0(r))}{W_1(v_0)} \right]  \,, \label{eq:AA0} \\
A_1(r) &= \frac{\kappa^2}{3} \int_{v_0}^{\phi_0(r)} d\bar\phi \left[ \frac{W_1(\bar\phi)}{W_0^\prime(\bar\phi)} + \phi_1(\bar\phi) \right] \,, %
\label{eq:AA1}
\end{align}
where $\phi_1(\bar\phi)$ is given by Eq.~(\ref{eq:phi1A}) with the substitution $\phi_0 \to \bar\phi$. The integration constants  in Eqs.~(\ref{eq:AA0}) and~(\ref{eq:AA1}) have been chosen to fix $A(0)=A_0(0)=0$. Given that $\phi_0=\phi+\mathcal O(s\phi_1)$, and since $sA_1\ll A_0$,  we can keep the zero order $\phi_0\simeq \phi$ in the definition of $A_1$ in Eq.~(\ref{eq:AA1}). This, together with the BC $\phi(r_1)=v_1$, leads to
\be
A_1(r_1)= \frac{\kappa^2}{3} \int_{v_0}^{v_1} d\bar\phi \left[ \frac{W_1(\bar\phi)}{W_0^\prime(\bar\phi)} + \phi_1(\bar\phi) \right] \,.
\label{eq:A1app}
\ee
As we see, $A_1(r_1)$ does not explicitly depend on $r_1$, it only depends on $v_\alpha$ and the superpotential parameters.

To conclude this section we want to stress here that the method we have developed to compute the effective potential, and simultaneously take into account the back reaction on the gravitational metric, is completely general and can be applied to any model defined by any superpotential. However, since the method relies on the perturbative expansion given in Eq.~(\ref{expansion}), one has to restrict the values of the free parameters of the model (e.g.~the values of $v_\alpha$, superpotential parameters, \dots) such that the perturbative expansion makes sense. This restricts the range of validity of the method for general physical conditions.

\section{The soft-wall metric}
\label{sec:model}

We consider the exponential superpotential used in soft-wall phenomenological models~\cite{Cabrer:2009we}:
\begin{equation}
W_0(\phi) = \frac{6}{\ell \kappa^2} \left(1 +  e^{\gamma \phi} \right) \,.
\end{equation}
This function $W_0(\phi)$ is an exact solution of the EoM involving the scalar potential
\begin{equation}
V(\phi) = -\frac{6}{\ell^2 \kappa^2} \left[  1  + 2 e^{\gamma\phi} + \left( 1 - \frac{3\gamma^2}{4\kappa^2} \right)  \, e^{2\gamma\phi} \right] \,.
\end{equation}
Following the general procedure described in Sec.~\ref{sec:effective_potential}, we find
\begin{equation} 
W_1(\phi) = \frac{1}{\ell\kappa^2}\exp\left[ \frac{4\kappa^2}{3\gamma^2} \left( \gamma\phi - e^{-\gamma\phi} \right) \right]  \,.
\end{equation}

The scalar field $\phi = \phi_0 + s\phi_1$ turns out to be given by
\begin{align}
\phi_0(r) &=  v_0 -\frac{1}{\gamma} \log\left( 1 - \frac{r}{r_S} \right) \,, \label{eq:phi0E} \\
\phi(r) &= \phi_0(r) + s\ \frac{2}{\gamma \left( r_S - r \right)} \int_{v_0}^{\phi_0(r)} \frac{W_1^\prime(\bar\phi)}{[W_0^\prime(\bar\phi)]^2} d\bar\phi  \,, \label{eq:phiE}
\end{align}
where the location of the naked singularity, $r_S$,  is given by
\be
r_S=\frac{\kappa^2 \ell}{3\gamma^2}e^{-\gamma v_0}\,.
\ee
Note that integration constants have been fixed such that $\phi(0) = \phi_0(0)=v_0$. 
From the condition $\phi(r_1) = v_1$ we get
\begin{align}
s(r_1) &=  \frac{\gamma(r_S-r_1)\left[v_1 - \phi_0(r_1)\right]}{  2 \int_{v_0}^{\phi_0(r_1)} \frac{W_1^\prime(\bar\phi)}{[W_0^\prime(\bar\phi)]^2} d\bar\phi  } \,, \label{eq:se}
\end{align}
in which the integrand is
\begin{equation}
\frac{W_1^\prime(\phi)}{[W_0^\prime(\phi)]^2} = \frac{\ell \kappa^4}{27 \gamma^3} \left( 1 +  e^{\gamma\phi} \right) \exp \left[ -\frac{4\kappa^2}{3\gamma^2} e^{-\gamma\phi} - 3\gamma \left( 1 - \frac{4\kappa^2}{9\gamma^2} \right) \phi \right] \,.
\end{equation}
The integrals in Eqs.~(\ref{eq:phiE}) and (\ref{eq:se}) cannot be computed analytically in general and
therefore all calculations of the effective potential will be performed numerically.

For the warp factor $A = A_0 + s A_1$, we can determine $A_0$ as
\begin{align}
A_0(r)&= \frac{r}{\ell} + \frac{\kappa^2}{3\gamma} \left( \phi_0(r) - v_0 \right) = \frac{r}{\ell} - \frac{\kappa^2}{3\gamma^2} \log\left( 1 - \frac{r}{r_S} \right) \,. \label{eq:Ae0}
\end{align}
Instead $A_1$ cannot be given in terms of an analytic solution and we have thus to determined it numerically. In particular for $A_1(r_1)$ we use the general expression provided in Eq.~(\ref{eq:A1app}).

In order to solve the hierarchy problem we have to fix $A(r_1)\simeq 35$. This can be done by conveniently choosing the brane parameters $v_\alpha$ and $\gamma$ in the superpotential, as well as $\kappa^2$, which provides the physical KK scale $\rho_1\equiv\ell^{-1}\exp[-A(r_1)]$~\footnote{The scale $\rho_1$ is $\mathcal O$(TeV) for $\ell^{-1}\simeq M_P=2.4\times 10^{18}$ GeV and $A(r_1)\simeq 35$. In the numerical calculations we will work in units where $\ell=1$. }. Moreover, by fixing the parameter $\kappa^2$ and the metric $A(r)$, the value of $\ell$ is established from the 4D Planck mass value as in Eq.~(\ref{eq:kappa2}).
Since $s(r_1)A_1(r_1)\ll A_0(r_1)$, to solve the hierarchy problem it is enough to work to zero order in the $s$ expansion, which means $A_0(r_1)\simeq 35$.  Then, from $A_0(r_1)\simeq 35$ and assigning some values to $\gamma$ and $r_S$ (i.e.~$v_0$), one can find $r_1$. Moreover using the approximation $\phi_0(r_1)\simeq v_1$ one can roughly estimate $v_1$ from 
\be
r_1\simeq\frac{\kappa^2}{3\gamma^2}\left(e^{-\gamma v_0}-e^{-\gamma v_1} \right) \,.
\label{eq:appr1}
\ee
This simple approximation is useful to guide the eye although the correct value of $r_1$ has to eventually be computed numerically. Eq.~(\ref{eq:appr1}) also highlights  
that the IR brane is shielding the singularity since $r_1<r_S$.

The amount of back-reaction in our solution can be read off from comparing the size of the two terms in the right hand side of the approximation
\be
A_0(r_1)\simeq \frac{r_1}{\ell}+\frac{\kappa^2}{3\gamma}(v_1-v_0)
\,.
\ee
Two extreme possibilities arise for a fixed value of $\kappa^2$: \textit{i)} For $\ell^{-3/2}\gamma\gtrsim1$ and $\ell^{3/2}|v_1-v_0|\lesssim  1$, the second term is small compared with the first one and the hierarchy problem is mainly solved by the first term. In this case there is little back-reaction on the metric and the length of the extra dimension is comparable to the AdS case, i.e.~$r_1\simeq A_0(r_1)\,\ell$. \textit{ii)} For $\ell^{-3/2}\gamma\ll 1$ and/or $\ell^{3/2}|v_1-v_0|\gg  1$ the second term can be comparable to the first term and the length of the extra dimension is smaller than in the AdS case, i.e.~$r_1<A(r_1)\,\ell$. This case is also characterized by the fact that the IR brane is close enough to the singularity, i.e.~$(r_S-r_1)\ll r_1$. For different values of $\kappa^2$ ($N^2$) the amount of back-reaction decreases with decreasing (increasing) values of $\kappa^2$ ($N^2$).

Since the superpotential formalism does not permit an analytic approach, in the present paper we carry out our investigation by concentrating in a few concrete benchmark scenarios. They cover parameter configurations with   large or small back-reactions on the metric~\footnote{Models with bulk potentials quadratic in $\phi$ as those originally considered by Goldberger and Wise~\cite{Goldberger:1999uk}, would fall in our formalism into the small back-reaction class.}, and are expected to give some qualitative insight on a vast class of plausible models where the hierarchy problem is solved. Our benchmark scenarios belong to the following classes: 
\begin{description}
\item{\textit{ - Small back-reaction (class A)}}
\begin{align}
&\gamma=0.55\,\ell^{3/2},\ v_0=-9.35\,\ell^{-3/2},\ v_1=-6.79\,\ell^{-3/2}, \gamma_1\to\infty\,,\nonumber\\
 &\kappa^2=\frac{1}{4}\ell^3\ (N\simeq 18),\  r_S=47.1 \, \ell,\  \langle  r_1\rangle =34.6\, \ell \,.
\label{eq:param-small}
\end{align}
\item{\textit{ - Large back-reaction (class B)}}
\begin{align}
&\gamma=0.1\,\ell^{3/2},\ v_0=-15\,\ell^{-3/2},\ v_1=-3.3\,\ell^{-3/2}, \gamma_1\to\infty\,,\nonumber\\ 
&\kappa^2=\frac{1}{4}\ell^3\ (N\simeq 18),\ r_S=37.3\, \ell,\  \langle r_1\rangle = 25.4\,\ell 
\,.
\label{eq:param-large}
\end{align}
\item{\textit{ - Large back-reaction \& larger N (class C)}}
\begin{align}
&\gamma=0.1\,\ell^{3/2},\ v_0=-20\,\ell^{-3/2},\ v_1 = 0.7 \,\ell^{-3/2} , \gamma_1\to\infty\,,\nonumber\\ 
& \kappa^2=\frac{1}{8}\ell^3\ (N\simeq 25),\ r_S=30.8\, \ell,\  \langle r_1\rangle = 26.7\,\ell 
\,.
\label{eq:N-large}
\end{align}

\item{\textit{ - Large back-reaction \& smaller N (class D)}}
\begin{align}
&\gamma=0.1\,\ell^{3/2},\ v_0=2\,\ell^{-3/2},\ v_1 = 8.9 \,\ell^{-3/2} , \gamma_1\to\infty \,,  \nonumber\\ 
& \kappa^2=\ell^3\ (N\simeq 9),\ r_S=27.3\, \ell,\  \langle r_1\rangle = 13.6\,\ell \,.
\label{eq:N-small}
\end{align}

\item{\textit{- Finite $\gamma_1$ (class E)}}
\begin{align}
&\gamma=0.1\,\ell^{3/2},\ v_0=-15\,\ell^{-3/2},\ v_1=-2.6\,\ell^{-3/2}, \gamma_1=10\,\ell^{-1}\,,\nonumber\\ 
&\kappa^2=\frac{1}{4}\ell^3\ (N\simeq 18),\ r_S=37.3\, \ell,\  \langle r_1\rangle = 25.4\,\ell   \,.
\label{eq:finitegamma1}
\end{align}
\end{description}
For concreteness, in all of them we choose the remaining free parameters to obtain $A(r_1)\simeq35$. We also rescale the brane tensions $\Lambda_\alpha$ as 
\be
\Lambda_\alpha\equiv\frac{6}{\ell\kappa^2}\lambda_\alpha
\ee
where $\lambda_\alpha$ are dimensionless constants and the value of $\lambda_0$, which is negative, is used to fine-tune the cosmological constant to zero at the minimum of the potential. At this point, $\lambda_1$ is a free parameter, which, as we will see, will determine the shape of the effective potential.

\begin{figure}[htb]
\centering
\includegraphics[width=7cm]{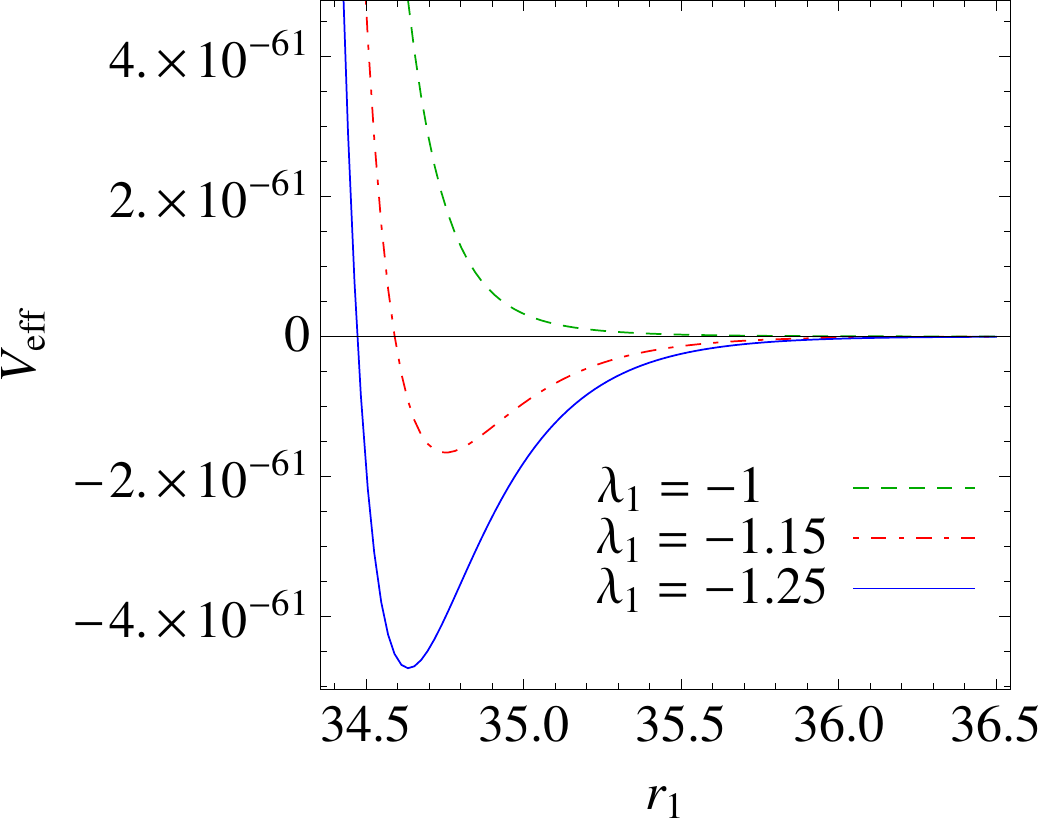} \hspace{0.5cm}
\includegraphics[width=5.9cm]{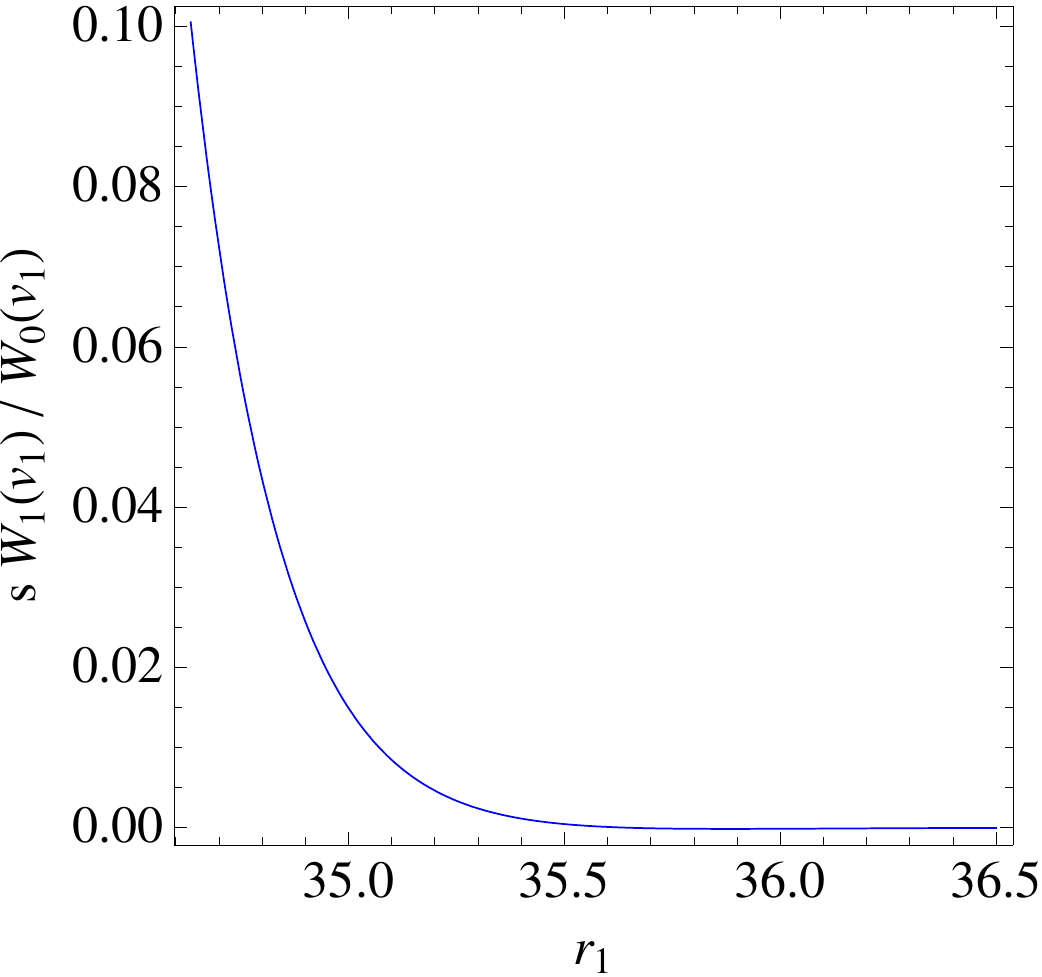} 
\includegraphics[width=7cm]{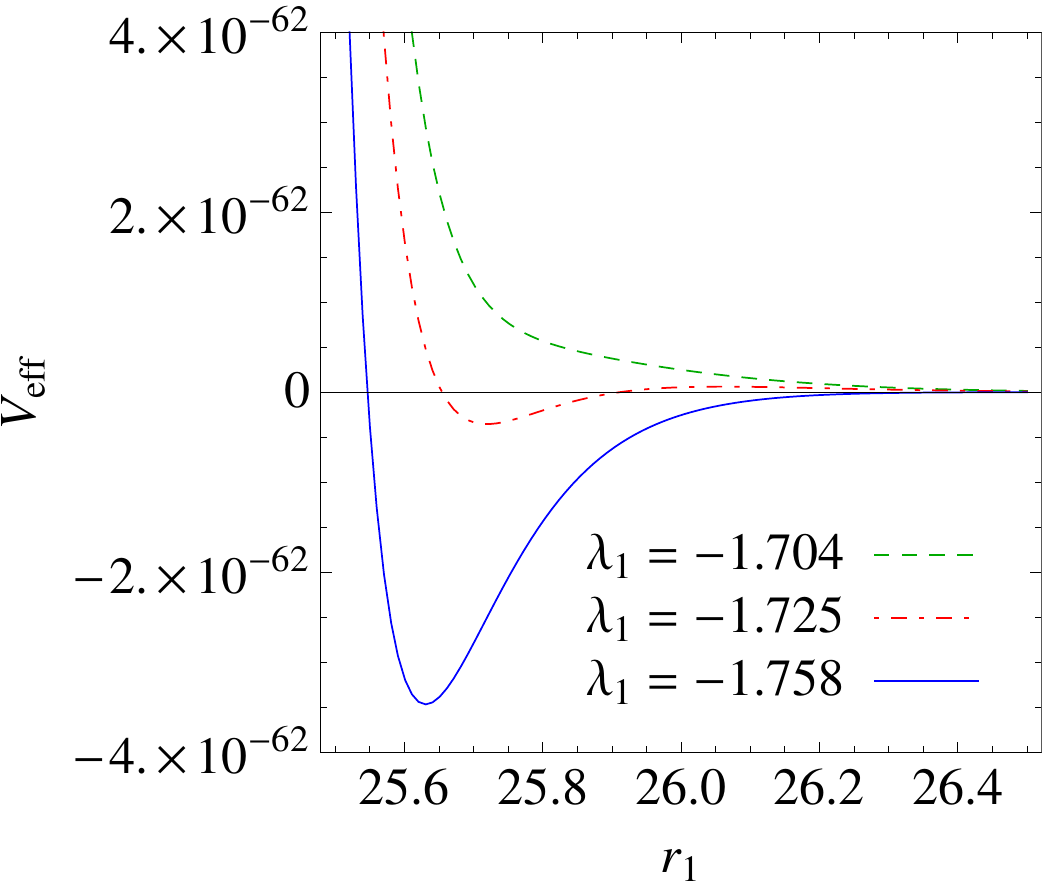} \hspace{0.5cm}
\includegraphics[width=5.7cm]{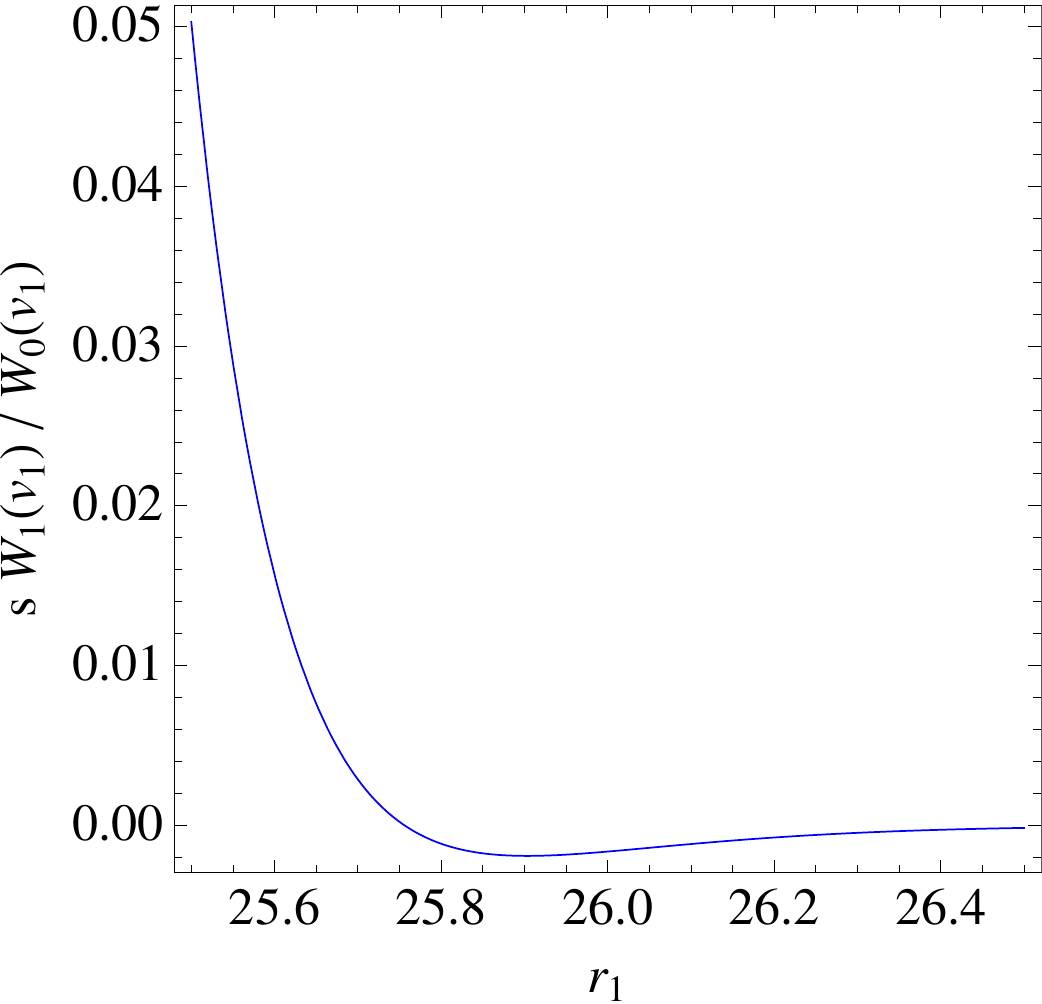} 

\caption{\it Left panels: Effective potential for different values of $\lambda_1$ in units of $\ell$. Only the relevant regime $ r_1>\langle r_1 \rangle $ is considered. Right panels: The relative correction to the superpotential $sW_1(v_1)/W_0(v_1)$ as a function of $r_1$.  The panels on the top correspond to `small back-reaction scenario (class A)', while the panels on the bottom correspond to `large back-reaction scenario (class B)'.
}
\label{fig:potential}
\end{figure}

In the left panels of Fig.~\ref{fig:potential} we show some numerical result of $V_{\rm eff}$ for several values of $\lambda_1$ in classes A and B scenarios (upper and lower panels, respectively). The potentials are normalized to zero at $r_1\to\infty$, where there is a minimum in all cases. The plots highlight how the parameter $\lambda_1$ controls the shape of the potential. For $|\lambda_1| \gg 1$ (with $\lambda_1<0$), the absolute minimum at $\langle r_1\rangle$ is very deep, and the maximum between the absolute minimum and the local minimum at $r_1\to\infty$ is tiny. Moreover there is a critical value of $|\lambda_1|$ for which the absolute minimum becomes degenerate with the minimum at $r_1\to\infty$, and even disappears (becomes a saddle point) for smaller values of $|\lambda_1|$. 

In the right panels of Fig.~\ref{fig:potential} we also show the relative size of the $\mathcal O(s)$ terms in the superpotential expansion, displayed as $sW_1(v_\alpha)/W_0(v_\alpha)$. In the upper panel we present the results for the class A scenarios (small back-reaction) while in lower panel we do it for the class B scenarios (large back-reaction).  
Notice that within one given class the ratio $sW_1(v_\alpha)/W_0(v_\alpha)$ does not depend on the particular $\lambda_1$ value.
As we can see, in the region $r_1>\langle r_1\rangle$ relevant for the study of the phase transition, the ratio is small enough to guarantee the validity of the $s$-expansion, as assumed in the analysis.  

In view of this behavior of  $V_{\rm eff}$, in the rest of the paper we restrict ourselves to configurations with potentials having two minima (and reliable $s$ expansion). Specifically, for each class we take some generic set of values for $\lambda_1$. Such values are  provided in Tab.~\ref{tab:table} (for the color code of $\lambda_1$ in the table, see Section~\ref{sec:phase-transition}). Within each class, the choice of $\lambda_1$ unequivocally define benchmark scenarios. The scenarios A$_1$, B$_{1}$, \dots, B$_{11}$, C$_{1}$, C$_{2}$, D$_1$ and E$_1$ are those we investigate numerically in the next sections.  Tab.~\ref{tab:table} also includes the value of $\ell$ in units of $M_P$ that we obtain via Eq.~(\ref{eq:kappa2}).  As expected,  $\ell^{-1}$ results very close to $M_{P}$.

\begin{table}[htb!]
\hspace{-1.0cm}
\begin{tabular}{||c|c|c|c|c|c|c||c|c|c||}
\hline\hline
Scen. & $\lambda_1$ & $\ell^{-1}/M_{P}$&$m_{\rm rad}/m_G$& $\rho_1$/TeV&$m_{\rm rad}$/TeV&$\langle\mu\rangle$/TeV&$\mu_0/\langle\mu\rangle$ &$T_c/\langle\mu\rangle$& $T_n/\langle\mu\rangle$\\ \hline
A$_1$ &  -1.250   & 0.501  & 0.0645 & 0.758 & 0.1998  &0.750 & - & 0.305 & - \\  \hline
B$_1$  &\textcolor{red}{-3.000}& 0.554 &  0.1969 & 1.085 & 1.018  & 0.828 & 0.9995  & 0.903 & 0.609 \\
B$_2$  &\textcolor{red}{-2.583}& 0.554 &  0.1905 & 1.007 & 0.915 & 0.767 & 0.989 & 0.825 & 0.428 \\
B$_3$  &\textcolor{red}{-2.500} & 0.554 & 0.1888  & 0.989  & 0.890 & 0.752 & 0.974 & 0.806 & 0.367 \\
B$_4$  &\textcolor{red}{-2.438} & 0.554 & 0.1874 & 0.973 & 0.870 & 0.741 & 0.937 & 0.790 & 0.297 \\
B$_5$  &\textcolor{blue}{-2.375} & 0.554 & 0.1859  & 0.957 & 0.849 & 0.728 & 0.982 & 0.774 & 0.193 \\
B$_6$  &\textcolor{blue}{-2.292} & 0.554 & 0.1836  & 0.934 & 0.818 & 0.710 & 0.971 & 0.750 & 0.149 \\
B$_7$  &\textcolor{blue}{-2.208} & 0.554 & 0.1809  & 0.908 & 0.784 & 0.690 & 0.949 & 0.724 & 0.0990 \\
B$_8$  &\textcolor{blue}{-2.125} & 0.554 & 0.1776  & 0.879 & 0.745 & 0.667 & 0.890 & 0.694 & 0.0388 \\
B$_9$  &\textcolor{blue}{-2.096} & 0.554 & 0.1763  & 0.8675 & 0.7303 & 0.6585 & 0.827 & 0.682 & 0.0122 \\
B$_{10}$  &\textcolor{blue}{-2.092} & 0.554 & 0.1761  & 0.8658 & 0.7281 & 0.6572 & 0.808  & 0.680 & 0.0073 \\
B$_{11}$ &     \textcolor{blue}{-2.090} & 0.554  & 0.1760 & 0.8650 & 0.7270 & 0.6565 & 0.793 &0.679 & 0.0039  \\ \hline
 C$_1$   &\textcolor{red}{-3.125} & 0.377 & 0.289 & 0.554 & 0.890 & 0.378 & 0.989  & 1.123 & 0.601 \\
 C$_2$    &  \textcolor{blue}{-2.604} & 0.377   &  0.271 & 0.496 & 0.751 & 0.336 & 0.937 &0.976 & 0.098  \\   \hline 
D$_1$  & \textcolor{red}{-3.462} & 1.49 & 0.106 & 0.468 & 0.477  &  0.250 & 0.9996& 1.007  & 0.445 \\ \hline
E$_1$  & \textcolor{blue}{-2.429} & 0.554 & 0.155 & 0.877 &  0.643 & 0.667 &0.895 & 0.694 & 0.142 \\
\hline\hline
\end{tabular}
\caption{\it List of benchmark scenarios defined by the classes in Eqs.~\eqref{eq:param-small}--\eqref{eq:finitegamma1} and the input values of $\lambda_1$ (second column). The outputs obtained in each scenario are presented from the third column on. The foreground \textcolor{red}{red} [\textcolor{blue}{blue}] color on the value of $\lambda_1$ indicates that the corresponding phase transition is driven by $O(3)$ [$O(4)$] symmetric bounce solutions. In scenario A$_1$ there is no phase transition.}
\label{tab:table}
\end{table}

\section{The radion field}
\label{sec:radion}

We now introduce the radion field as a perturbation of the metric whose definition is
\begin{eqnarray}
ds^2 &=&  -[1+2F(x,r)]^2 dr^2 + e^{-2[A+F(x,r)]} \bar g_{MN} dx^M dx^N \,, \\
\phi(x,r) &=& \phi_0(x) + \varphi(x,r) \,.
\end{eqnarray}
The Einstein EoM can be solved with the radion ansatz $F(x,r) = F(r) R(x)$ such that the excitation of the field $\phi$, $\varphi(x,r)$, can be reparametrized as~\cite{Csaki:2000zn}
\be
 \varphi(x,r)=\frac{3}{\kappa^2}\frac{F^\prime(r)-2A^\prime(r) F(r)}{\phi_0^\prime(r)}\,R(x) \,,
 \label{eq:relacion}
 \ee
so that the only remaining degree of freedom is the radion field $R(x)$. In particular we adopt the ansatz $F(r)\simeq e^{2A}$ which is appropriate for a light radion/dilaton~\footnote{We have checked numerically that this ansatz remains a good approximation for a not so light (sub-TeV) radion as far as its mass remains sufficiently smaller than the mass of KK excitations.}. In this case Eq.~(\ref{eq:relacion}) leads to $\varphi(x,r)\simeq 0$. Moreover the geodesic distance between the branes can be written as~\cite{Konstandin:2010cd} 
\begin{equation}
r(x) \equiv \int ds = \int_0^{r_1} dr \bigg[ 1+ 2 F(r)R(x) \bigg] = r_1 + X_F(r_1) R(x) \,, \label{eq:geodesic}
\end{equation}
with
\begin{equation}
X_F(r) = 2  \int_0^{r} dr  \, e^{2A(r)}\,,
\end{equation}
by which $R(x)$ can be interpreted as the excitation of the (unnormalized) radion field $r(x)$ with background value $r_1$. This provides the functional dependence of the effective potential $V_{eff}(r_1)$ we consider in Eq.~(\ref{potencialefectivo}) and subsequently.

The kinetic term of the action is given by~\cite{Megias:2015ory}
\begin{align}
&2\int d^4x \int_0^{r_1} dr \sqrt{|\det g_{MN}|} \left( -M^3R + \frac{1}{2}(\partial\phi)^2 \right) \nonumber\\
&=6 M^3 \int d^4x \sqrt{|\det \bar g_{\mu\nu} | } (\partial r)^2 X_F^{-1} (r_1)+\dots\,,  \label{eq:Kinetic1}
\end{align}
from where we can see that the field $r(x)$ is not canonically normalized. One uses to define the canonically normalized~\footnote{As it is conventional, we leave aside the action the global constant factor $12(M\ell)^3=6\ell^3/\kappa^2$.} field $\mu(x)$ with kinetic and mass terms as
\be
\mathcal L_{\textrm{rad}}=
 \frac{6\ell^3}{\kappa^2}\int d^4x \sqrt{| \det \bar g_{\mu\nu} |} \left(\frac{1}{2} (\partial \mu)^2-\frac{1}{2}m_{\rm rad}^2\,\mu^2\right) \,,
\ee
with $m_{\rm rad}$ being the mass of the normalized radion. The field $\mu(x)$ is related to $r(x)$ by 
\begin{equation}
\partial \mu(x) = -  \ell^{-3/2}X_F^{-1/2}(r_1)\partial r(x)\simeq -  \ell^{-3/2}X_F^{-1/2}[r(x)]\partial r(x) \,, \label{eq:partials}
\end{equation}
where in the last step the background field $r_1$ is approximated by the whole field configuration $r(x)$. The formal solution to Eq.~(\ref{eq:partials}) is
\begin{equation}
\mu(r) = \ell^{-3/2} \int_r^{r_S} d\bar r X_F[\bar r]^{-1/2} \,, \label{eq:mu2}
\end{equation}
which ensures $\mu(r=r_S) = 0$~\footnote{In the standard AdS scenarios the value of $\mu=0$ is achieved in the limit $r\to\infty$. Here this condition is replaced by $r\to r_S$ which is the location of the singularity and where the space is cutoff.}. If $r = r(x)$, Eq.~\eqref{eq:mu2} provides $\mu(x) \equiv \mu[r(x)]$. In this case the effective potential is given by the function
\be
V_{eff}(\mu)\equiv V_{eff}[r(\mu)]\,,
\ee
where $r(\mu)$ is the inverse function provided by Eq.~(\ref{eq:mu2}).

In general the relationship between the fields $\mu$ and $r$ can only be obtained numerically. However  the relation can be easily solved analytically in the particular regime of no back-reaction, e.g.~in the AdS scenario. In that case it turns out that $A(r)=r/\ell$ and $X_F(r)=\ell \exp(2 r/\ell)$ so that also Eq.~(\ref{eq:partials}) can be solved analytically leading to $\mu(r)=\ell^{-1}e^{-r/\ell}$ or, equivalently, $r=\ell\,\log(1/\mu\ell)$, which is the usual expression obtained in the Randall-Sundrum theory.

\begin{figure}[htb]
\centering
\includegraphics[width=7cm]{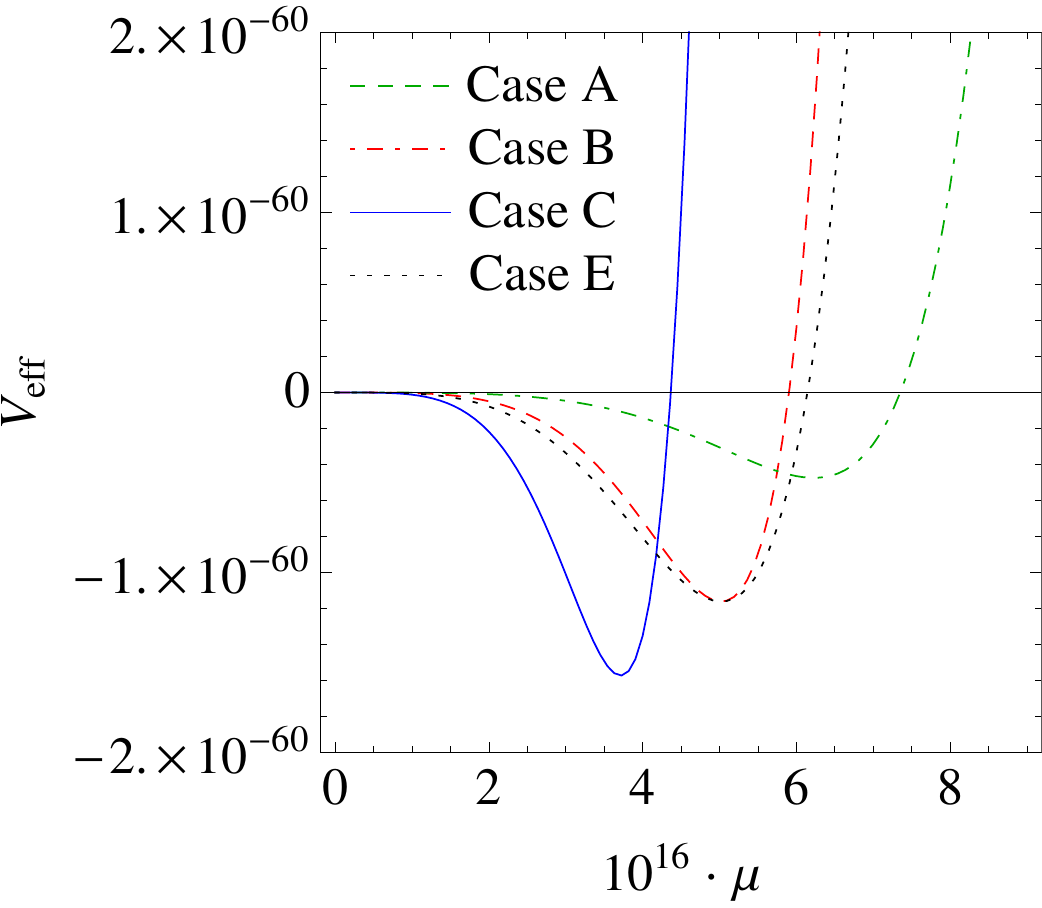} \hspace{0.5cm}
\includegraphics[width=7cm]{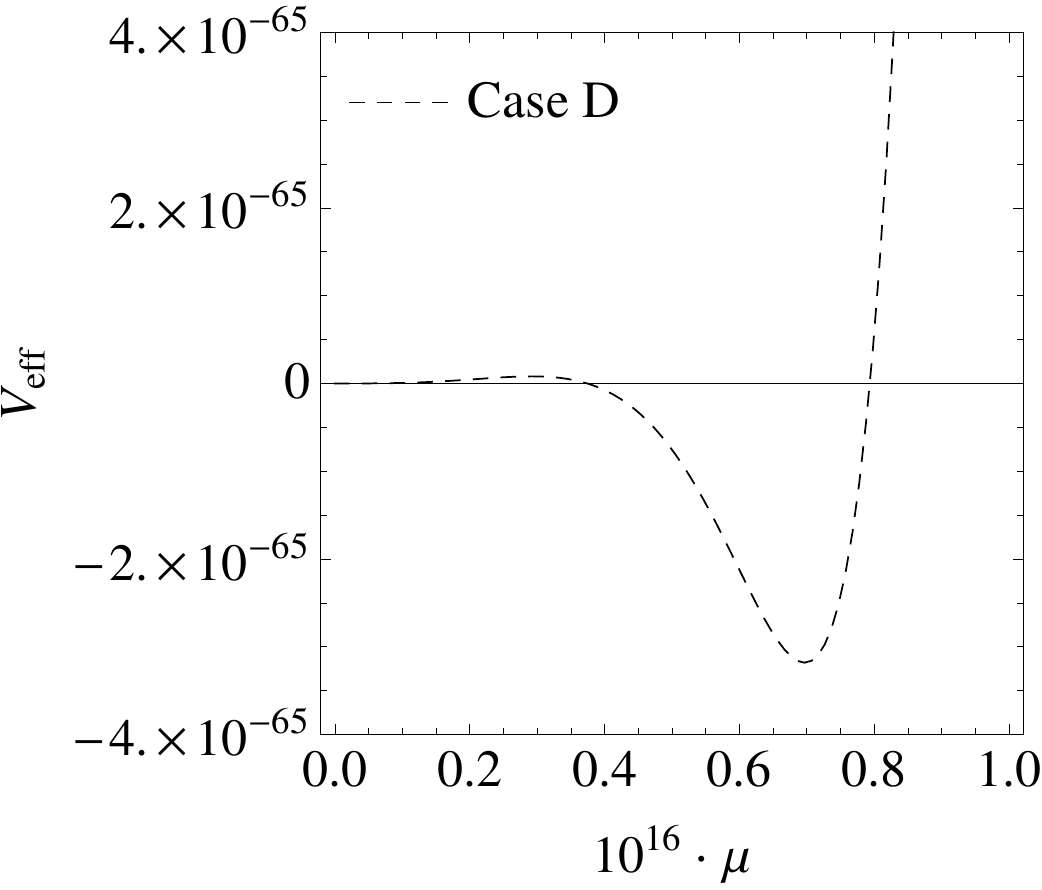} \hspace{0.5cm}
\caption{\it The effective potential as a function of $\mu$, in units of $\ell$, in scenarios A$_1$, B$_8$, C$_2$, $E_1$ (left panel) and D$_1$ (right panel).}
\label{fig:potentialmu}
\end{figure}

The effective potential for the cases of small and large back-reaction (and thus $N$ large) are shown in Fig.~\ref{fig:potentialmu}. We observe that the shape of the potential in every case, i.e.~the depth and location of the minimum, has important consequences for the dilaton phase transition. The flatter the potential, the slower the way to the false minimum, the bigger the euclidean action (as we will see) and the more difficult (if not impossible) the phase transition. The flatness of the potential is associated with the amount of back-reaction~\footnote{Notice that the case of a completely flat potential as in the Randall Sundrum model corresponds to the case where there is no back-reaction on the metric}. This happens for the potentials in classes A, B, C and E in the left panel of Fig.~\ref{fig:potentialmu}, as we can see. In fact we will see that in class A, unlike in classes B, C and E, the euclidean action is so large that the transition rate never overcomes the expansion rate of the universe. Moreover, the location of the minimum is also important for the phase transition. In fact the smaller the value of $\langle\mu\rangle$, the shorter the road along the potential to the false minimum, and thus the smaller the euclidean action. This fact is exemplified in the right panel of Fig.~\ref{fig:potentialmu} where the potential for the class D scenario is shown. Even if the potential is flatter than in case A, the value of $\langle\mu\rangle$ for case D is one order of magnitude smaller than in case A, and then the euclidean action is also smaller and allows the phase transition, as we will see.

For the validity of the 4D treatment of the radion field it is necessary that the KK graviton modes are significantly heavier than the radion and thus can be integrated out in the EFT. In that case, it is energetically expensive for the
KK fields to move and the transition can be studied in an effective theory where the only extra (with respect to the SM) dynamical degree of freedom is the radion. 
To check such a hierarchy, the following analytical approximate formulas turn out to be useful~\cite{Megias:2015ory}:
\be
m^2_{\rm rad}\simeq\frac{\rho^2_1}{\Pi_{\rm rad}(r_1)},\quad \rho_1\equiv (1/\ell) e^{-A(r_1)}\,,
\label{eq:rad-mass}
\ee
with
\begin{align}
\Pi_{\rm rad}(r_1)&=\frac{1}{\ell^2}\int_0^{r_1}dr e^{4(A-A_1)}\left(\frac{W}{W^\prime} \right)^2
\left[
\frac{2}{W[\phi(r_1)]}+\int_r^{r_1}d\bar r e^{-2(A-A_1)}\left(\frac{W^\prime}{W} \right)^2 
\right]\nonumber\\
&+\frac{4W[\phi(r_1)]}{\ell^2 W^{\prime 2}[\phi(r_1)]\left(\gamma_1 +W^{\prime\prime}[\phi(r_1)]\right)} \,,
\label{eq:Pirad}
\end{align}
in which the last term is negligible for strict stiff wall boundary potentials ($\gamma_\alpha\to\infty$). 
Similarly, the mass of graviton KK modes can be approximated as~\cite{Megias:2015ory}
\be
m_{G}^2\simeq\frac{\rho^2_1}{\Pi_{G}(r_1)}
\label{eq:KK-mass}
\ee
with
\be
\Pi_{G}(r_1)=\frac{1}{\ell^2}\frac{\int_0^{r_1}dr e^{-2(A-A_1)}\int_r^{r_1}dr^\prime e^{4(A-A_1)}
\int_{r^\prime}^{r_1} d r^{\prime\prime}e^{-2(A-A_1)}}{\int_0^{r_1}dr e^{-2(A-A_1)}
} \,.
\label{eq:PiG}
\ee
Therefore the validity of the EFT requires the ratio 
\be
\frac{m_{\rm rad}^2}{m_G^2}=\frac{\Pi_{G}(r_1)}{\Pi_{\rm rad}(r_1)} \,
\label{eq:radGratio}
\ee
to be small.

Using Eqs.~(\ref{eq:Pirad}), (\ref{eq:PiG}) and \eqref{eq:radGratio}, for our benchmark scenarios we obtain
\begin{align}
\textrm{(class A):}&\quad m_{\rm rad}\simeq\, 0.2\, \rho_1,\quad m_G\simeq 2.9\, \rho_1\,,\\
\textrm{(class B):}&\quad  m_{\rm rad}\simeq 0.9 \rho_1,\quad\,  m_G\simeq 4.8\,\rho_1\,,\\
\textrm{(class C):}&\quad  m_{\rm rad}\simeq  1.6\rho_1,\quad\,  m_G\simeq 5.6\,\rho_1\,,\\
\textrm{(class D):}&\quad  m_{\rm rad}\simeq  1.0\rho_1,\quad\,  m_G\simeq 9.6\,\rho_1\,,\\
\textrm{(class E):}&\quad m_{\rm rad}\simeq 0.7\,\rho_1,\quad m_G\simeq 4.7\,\rho_1\,,
\label{eq:masas}
\end{align}
although the more precise values depend on the specific value of $\lambda_1$ of each scenario~\footnote{As expected from Eq.~(\ref{eq:Pirad}), a way of decreasing the radion mass is to make the value of $\gamma_1$ finite and thus decrease the denominator in $\Pi_{\rm rad}$. See class D scenario.}. 

It then turns out that the radion is not very light in the scenarios of the class B and C because of the large back-reaction and the strong departure from conformality near the IR brane. Still there is enough hierarchy between the radion and the KK graviton masses to justify the use of the EFT effective potential for the analysis of the phase transition in most of the benchmark scenarios although class C might be borderline~\footnote{Our numerical results for C$_1$ and C$_2$ might hence be inaccurate.}. However this does not happen for scenarios A, B, D and E where the radion is lighter, as compared with the corresponding KK graviton mass. The precise values of the mass ratios for the different benchmarks are shown in the Tab.~\ref{tab:table} which also includes the scale $\rho_1$, defined in Eq.~(\ref{eq:rad-mass}), and the radion VEV $\langle\mu\rangle$, corresponding to the minima of the potentials.

\section{The effective potential at finite temperature}
\label{sec:effective_potential_finiteT}

At finite temperature the system allows for an additional gravitational solution with a Black Hole (BH) singularity located in the bulk. In the AdS/CFT correspondence this  BH metric describes the high temperature phase of the system where the dilaton is sent to the symmetric phase $\langle\mu\rangle=0$ and thus the condensate evaporates~\cite{Creminelli:2001th}. 

Let us consider a BH metric of the form
\begin{equation}
ds_{BH}^2 = -\frac{1}{h(r)}dr^2 + e^{-2A(r)} (h(r) dt^2 -  d\vec{x}^{\,2} ) \,,  \label{eq:metricBH}  
\end{equation}
where $h(r)$ is a blackening factor which vanishes at the position of the event horizon, $r=r_h$. The EoM with this metric read 
\begin{eqnarray}
&& \frac{h^{\prime\prime}}{h^\prime} - 4 A^\prime=0 \,, \label{eq:eomTh} \\
&&A^{\prime\prime}  - \frac{\kappa^2}{3} \phi^{\prime \, 2}=0   \,, \label{eq:eomT1}\\
&&A^{\prime\, 2} - \frac{h^\prime}{4h} A^\prime +\frac{\kappa^2}{6} \frac{V(\phi)}{h} - \frac{\kappa^2}{12} \phi^{\prime\, 2}=0 \,,  \label{eq:eomT2}\\
&&\phi^{\prime\prime} +\left( \frac{h^\prime(r)}{h(r)} - 4 A^\prime \right) \phi^\prime - \frac{1}{h(r)} \frac{\partial V}{\partial\phi}=0 \,. \label{eq:eomT3}
\end{eqnarray}
Eq.~(\ref{eq:eomT3}) can be eliminated in favor of (\ref{eq:eomTh})-(\ref{eq:eomT2}) by means of the identity
\be
\kappa^2 h\phi^\prime \cdot [\ref{eq:eomT3}]=-\frac{3}{2}A^\prime h^\prime \cdot [\ref{eq:eomTh}]+\frac{3}{2}(8A^\prime h-h^\prime) \cdot [\ref{eq:eomT1}]
-6\left(h^\prime + h \frac{d}{dr}\right) \cdot [\ref{eq:eomT2}] \,,
\ee
so that we have three differential equations with five integration constants which can be fixed by imposing BCs at the UV brane $r=0$, and regularity conditions at the singularity $r=r_h$. Four of these integration constants are then set as
\be
h(0)=1\,,\quad h(r_h)=0\,,\quad \phi(0)=v_0\,\quad A(0)=0 \,,
\label{eq:BCs}
\ee
while the fifth one is $A(r_h)$ and is traded for the physical parameter $T_h$ representing the Hawking temperature of the system.
Indeed, from Eq.~\eqref{eq:metricBH} it can be derived that the temperature $T_h$ and the entropy $S$ of the BH can be expressed as~\cite{Gibbons:1976ue,Carlip:2008wv}~\footnote{We have included in the definition of the entropy a factor of two coming from the integration over the orbifold.}
\begin{equation}
T_h = \frac{1}{4\pi} e^{-A(r_h)} \big| h^\prime(r) \big|_{r=r_h}  \,, \qquad S = \frac{4\pi}{\kappa^2} e^{-3A(r_h)} \,. \label{eq:Ths}
\end{equation}

The quantity $T_h$ has a key role in the phase transition. To appreciate this, it is useful to consider the thermodynamics relations for the internal and free energies
\bea
U(T_h)=T_h S(T_h)-\int_0^{T_h}S(\bar T_h)d\bar T_h
\label{eq:internal}\,,\\
F(T_h)=(T_h-T)S(T_h)-\int_0^{T_h}S(\bar T_h)d\bar T_h \,,
\label{eq:free}
\eea
with $U$ and $F= U-TS$ being the internal energy and the free energy, respectively. In fact Eq.~(\ref{eq:free}) makes manifest  that $F(T_h)$ has a minimum at $T_h=T$ that amounts to 
\be
F_{\rm min}=F(T)=-\int_0^T S(\bar T_h)d\bar T_h=-\frac{4\pi^4\ell^3}{\kappa^2}\int_0^Ta_h \bar T_h^3 d\bar T_h \,,
\label{eq:freemin}
\ee
where we have employed Eq.~(\ref{eq:Ths}) and the definition 
\be
a_h(T_h)=\left|\frac{4}{\ell h^\prime(r_h)}\right|^3 \,.
\label{eq:ah}
\ee
In particular, under the assumption of $a_h$ being a smooth function of $T$, we can approximate the free energy as
\be
F_{\rm min}^{\rm app}=-\frac{\pi^4\ell^3}{\kappa^2} a_h(T) T^4 \,.
\label{eq:freeminapp}
\ee

\subsection{The case of small back-reaction}

The regime of small back-reaction has been broadly studied~\cite{Creminelli:2001th, Randall:2006py,Kaplan:2006yi}. In this case the constant part of the bulk potential $V(\phi)\simeq -6/(\kappa^2\ell^2)$ 
dominates, and  neglecting the back-reaction of the scalar field on the metric $A$ is a good approximation. Thus the solutions to Eqs.~(\ref{eq:eomTh})-(\ref{eq:eomT3}) read
\be
\phi(r)\simeq v_0 \,, \quad A(r)\simeq r/\ell\,, \quad h(r)\simeq 1-e^{4(r-r_h)/\ell} \,.
\label{eq:phiAdS}
\ee
Moreover, from Eqs.~(\ref{eq:Ths}) and (\ref{eq:phiAdS}) one recovers the usual expressions
\be
T_h=\frac{e^{-r_h/\ell}}{\pi\ell},\quad r_h=1 \,,
\label{eq:ThAdS}
\ee
leading to the standard expression for the free energy in AdS space~\cite{Creminelli:2001th}:
\be
F_{\rm min}^{\rm AdS}=-\frac{\pi^4\ell^3}{\kappa^2} T^4 \,.
\label{eq:freeminAdS}
\ee

\subsection{The case of large back-reaction}
In the case of large back-reaction, the blackening factor $h(r)$ has to be obtained by solving Eqs.~(\ref{eq:eomTh})-(\ref{eq:eomT3}) numerically, from where one can easily deduce $a_h(T)$ and $F_{\rm min}$. 

\begin{figure}[htb]
\centering
\includegraphics[width=7.2cm]{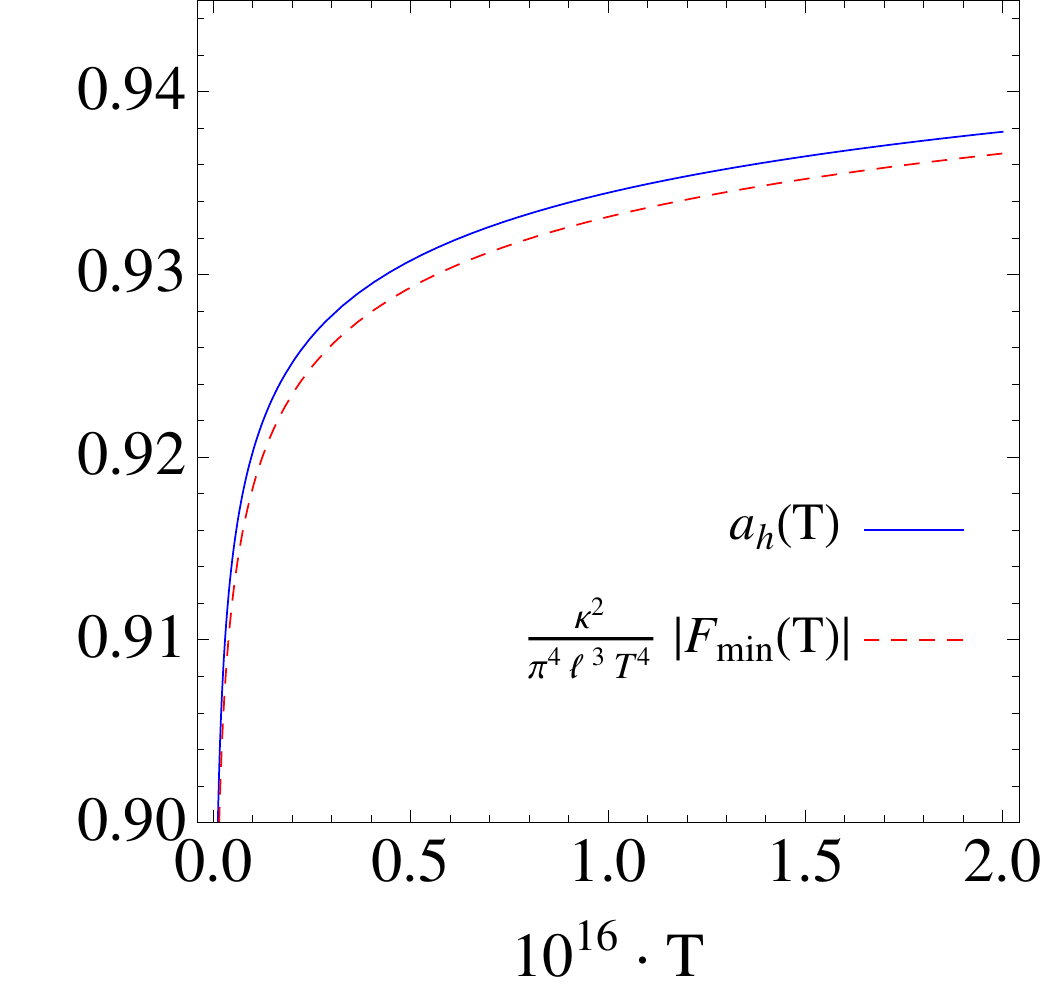} \hspace{0.5cm}
\includegraphics[width=7.2cm]{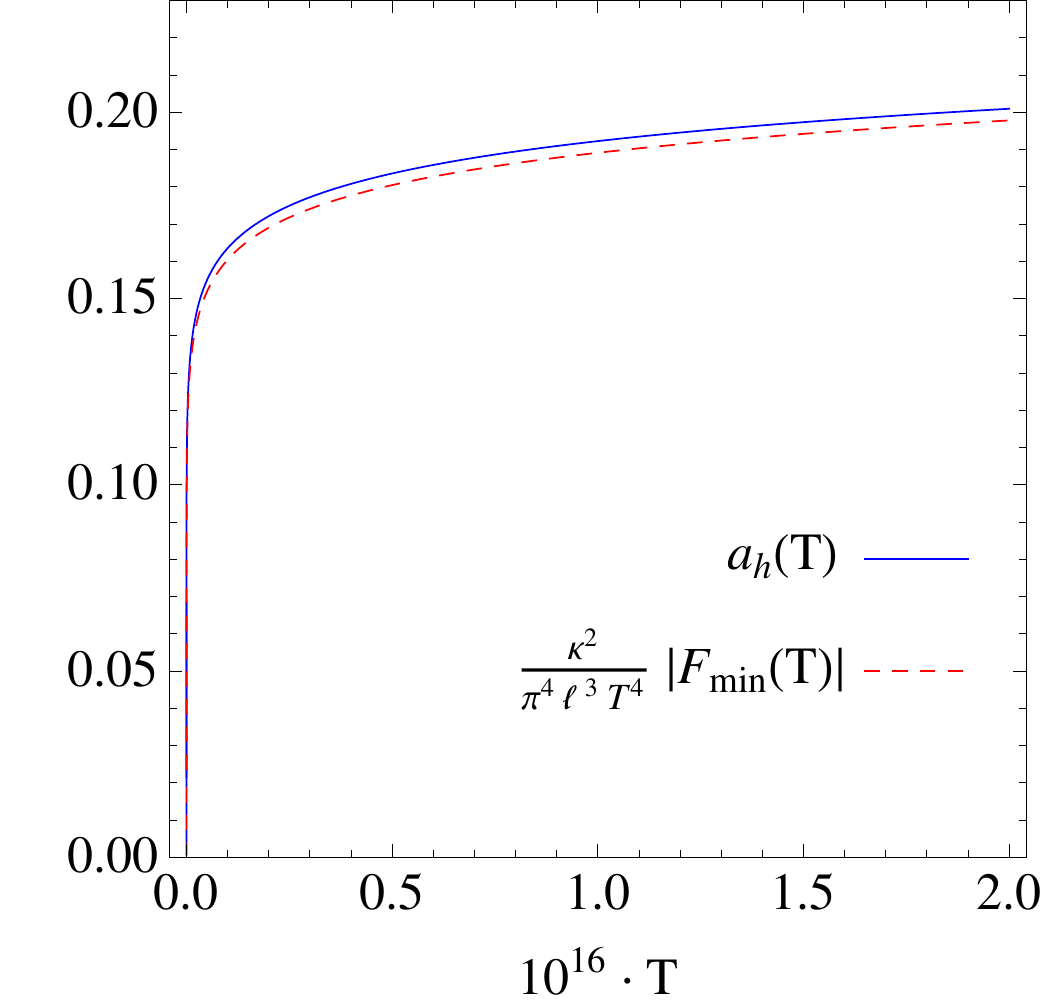} \hspace{0.5cm}
\caption{\it The quantities $a_h(T)$ (blue solid line) and $\kappa^2F_{\rm min}/(\pi^4\ell^3T^4)$ (red dashed line) as a function of $T$ in the scenarios of the class A (left panel) and class B (right panel).
}
\label{fig:aT}
\end{figure}

We show the result of this procedure in Fig.~\ref{fig:aT} whose left and right panels deal, respectively, with the class A (i.e.~small back-reaction) and  B (i.e.~large back-reaction) scenarios. The resulting function $a_h(T)$ is marked in (blue) solid, while the quantity $\kappa^2F_{\rm min}/(\pi^4\ell^3T^4)$ is marked in (red) dashed.
We see that, as anticipated, for small-back reaction $a_h(T)$ basically reproduces the case of pure AdS (i.e.~$a_h(T)= 1$), whereas for large back-reaction it results $a_h(T)\ll 1$. This effect has important phenomenological implications since it strongly influences the nucleation temperature of the phase transition, as we discuss in Sec.~\ref{sec:phase-transition}. The comparison between $a_h$ and $\kappa^2F_{\rm min}/(\pi^4\ell^3T^4)$  highlights the fact that Eq.~(\ref{eq:freeminapp}) is a very good approximation of $F_{\rm min}$  for all practical purposes. 

We have checked that these features do not depend on the specific benchmark scenarios we have considered. In particular the behavior of $a_h(T)$ is generic and only depends, in all cases, on the amount of back-reaction on the gravitational metric.

\section{The dilaton phase transition}
\label{sec:phase-transition}
The phase transition can start when the free energy of the BH deconfined phase, $F_{d}$, becomes smaller than the free energy in the soft-wall confined phase, $F_{c}$. Those free energies are defined by
\be
F_{d}(T)=E_0+F_{min}-\frac{\pi^2}{90}g_d^{eff} T^4 \,,
\ee
\be
F_{c}(T)=-\frac{\pi^2}{90}g_c^{eff} T^4 \,,
\ee
where $g_c^{eff}$ ($g_d^{eff}$) is the number of SM-like degrees of freedom in the confined (deconfined) phase, $F_{min}$ is given in (\ref{eq:freemin}) and finally $E_0$ is defined as $E_0=V_{eff}(\mu=0)-V_{eff}(\mu=\langle \mu \rangle)>0$~\footnote{For numerical purposes we need to focus on a given particle setup: we  assume that at low energy the confined phase does not contain BSM fields besides the radion. In this phase, at $T$ much below the mass scale of the  $n=1$ modes, $g_{B(F)}(T)$ matches the SM number of bosonic (fermionic) degrees of freedom. It follows that $g^{eff}=g_B(T)+\frac{7}{8}g_F(T)=106.75$ at $172\,{\rm GeV}\lesssim T\ll m_G$. 
On the other hand, at very high temperatures, in the deconfined phase only the elementary degrees of freedom will contribute to the free-energy $F_d$, which we will then assume to be contributed by most of the SM degrees of freedom, as we will only consider, as we will see later on in this section, a few (composite) states (as the right-handed top quark and the Higgs scalar) living in the IR brane.
Under this reasonable assumption, the contribution to the free energies coming from the SM degrees of freedom is balanced between the confined and deconfined phases, and can be neglected. This approximation will be justified in Secs.~\ref{sec:inflation} and \ref{sec:EWPT}.}. In this way the critical temperature $T_c$ at which the phase transition starts being allowed (the nucleation temperature $T_n$ is indeed below it) is given by
\be
F_{d}(T_c)=F_{c}(T_c) \,.
\ee
The values of $T_c$ for the different considered benchmark scenarios are shown in Tab.~\ref{tab:table}.

To study in detail the dilaton/radion phase transition we have to consider the bounce solution of the Euclidean action, as described in Refs.~\cite{Coleman:1977py,Callan:1977pt,Linde:1980tt}.
For the canonically normalized field $\mu$, the Euclidean action driven by thermal fluctuation is $O(3)$ symmetric  and given by~\cite{Linde:1980tt,Quiros:1999jp}
\be
S_3= 4\pi\int d\rho \rho^2 \frac{6\ell^3}{\kappa^2} \left(\frac{1}{2}\mu^{\prime 2}+V_{\textrm{rad}}(\mu)\right)\qquad \textrm{with}\quad V_{\textrm{rad}}\equiv\frac{\kappa^2}{6\ell^3} V_{eff} \,.
\label{eq:bounceaction3}
\ee
The corresponding bounce equation is
\begin{equation}
\frac{\partial^2 \mu}{\partial\rho^2}  + \frac{2}{\rho} \frac{\partial \mu}{\partial\rho} - \frac{\partial V_{\textrm{rad}}}{\partial\mu} = 0 \,, \label{eq:bouncesol3}
\end{equation}
with $\rho=\sqrt{\vec x^{\,2}}$ 
and BCs~\footnote{Notice that the BC at $\mu=0$ is not the standard one which fixes the behaviour of the solution at $\rho\to \infty$. Here we exploit the fact that $\mu(\rho)$ reaches $\mu=0$ at very large values of $\rho$, so that at even larger $\rho$ the friction term in Eq.~(\ref{eq:bouncesol3}) is negligible. Our BC at $\mu=0$ is thus equivalent to the standard one due to approximate energy conservation (i.e.~approximate lack of friction) in the subsequent evolution of the bounce.}
\begin{equation}
\left. \frac{3\ell^3}{\kappa^2} \mu^{\prime\, 2}(\rho)\right|_{\mu=0}=\left|F_{\rm min}(T)\right|\,, \quad \mu(0)=\mu_0 \,, \quad \frac{d\mu}{d\rho} \bigg|_{\rho=0} = 0 \,.
\label{eq:bc3}
\end{equation}

Thermal fluctuations are not the only way of overcoming the barrier between the false and true vacua. At low enough temperatures it can also occur via quantum fluctuations. In this case the bounce solution is $O(4)$ symmetric, with Euclidean action $S_4$ provided by~\cite{Coleman:1977py,Quiros:1999jp}
\be
S_4= 2\pi^2\int d\rho \rho^3 \frac{6\ell^3}{\kappa^2}\left(\frac{1}{2}\mu^{\prime 2}+V_{\textrm{rad}}(\mu)\right),\ 
\label{eq:bounceaction4}
\ee
where $\rho=\sqrt{\vec x^2+\tau^2}$ (with $\tau$ being the Euclidean time), and satisfies the differential equation
\begin{equation}
\frac{\partial^2 \mu}{\partial\rho^2}  + \frac{3}{\rho} \frac{\partial \mu}{\partial\rho} - \frac{\partial V_{\textrm{rad}}}{\partial\mu} = 0 \,, \label{eq:bouncesol4}
\end{equation}
with BCs given in Eq.~(\ref{eq:bc3}).

As we are normalizing the potential to zero at the origin $\mu=0$, instead of  normalizing it at the (fake) BH minimum as  in the original calculations~\cite{Coleman:1977py,Linde:1980tt}, we have to add the omitted contribution to the Euclidean action. In a suitable approximation this is given for the $O(3)$ solution by
\be
\frac{\Delta S_3}{T}=\frac{|F_{\rm min}|}{T}\,\frac{4}{3}\pi \bar\rho^3 \,,
\ee
and for the $O(4)$ solution by
\be
\Delta S_4=|F_{\rm min}|\frac{\pi^2}{2}\, \bar\rho^4 \,.
\ee
Here $\bar\rho$ (the bubble `radius') is calculated assuming a simple step approximation for the bubble profile, namely $\mu=\mu_0$ inside the bubble  and $\mu=0$ outside. Specifically it results
\be
\int_0^{\rho_0}\rho^{n-1} \mu(\rho) d\rho\equiv \mu_0\int_0^{\bar\rho}\rho^{n-1} d\rho=\mu_0\frac{\bar\rho^{n}}{n} \,,
\ee
for the $O(n)$ solution ($n=3,4$), with $\rho_0$ being the value of the `time' $\rho$ when $\mu$ reaches zero.

Once $S_3$ and $S_4$ are known, the bubble nucleation rate per unit volume per unit time from the false BH minimum to the true vacuum is given by the sum over configurations
\be
\Gamma/\mathcal V = \mathcal A\,e^{-S_E}
\label{eq:nucl-rate}
\ee
with 
\be
e^{-S_E} = c_3 \, e^{-(S_3+\Delta S_3)/T} + c_4 \, e^{-(S_4+\Delta S_4)} \,, \label{eq:eSE}
\ee
where, in practice, we can take $c_3 \simeq c_4 \simeq  1$ and $\mathcal A\simeq T^4_c$ (changing these values has negligible impact on the results of this paper). 
Then $\Gamma/\mathcal V$ is dominated by the least action such that in non-pathological regimes we can assume $S_E=(S_3+\Delta S_3)/T$ for $O(3)$ bubbles and $S_E=S_4+\Delta S_4$ for $O(4)$ bubbles. Only when the first and second terms in the right hand side of Eq.~(\ref{eq:eSE}) are very close to each other, one should take care of the full expression of Eq.~(\ref{eq:eSE}). 

The onset of nucleation then happens at the temperature $T_n$ such that the probability for a single bubble to be nucleated within one horizon volume 
is $\mathcal O(1)$. A simple estimate translates into the upper bound on the Euclidean action~\cite{Konstandin:2010cd,Bunk:2017fic} 
\be
S_E\lesssim 4 \log\left(M_P/\langle\mu\rangle\right)\approx 140 \,,
\label{eq:SE_bound}
\ee
%
%
which will be considered throughout the forthcoming numerical analysis.

\subsection{Small back-reaction}

\begin{figure}[htb]
\centering
\includegraphics[width=6.3cm]{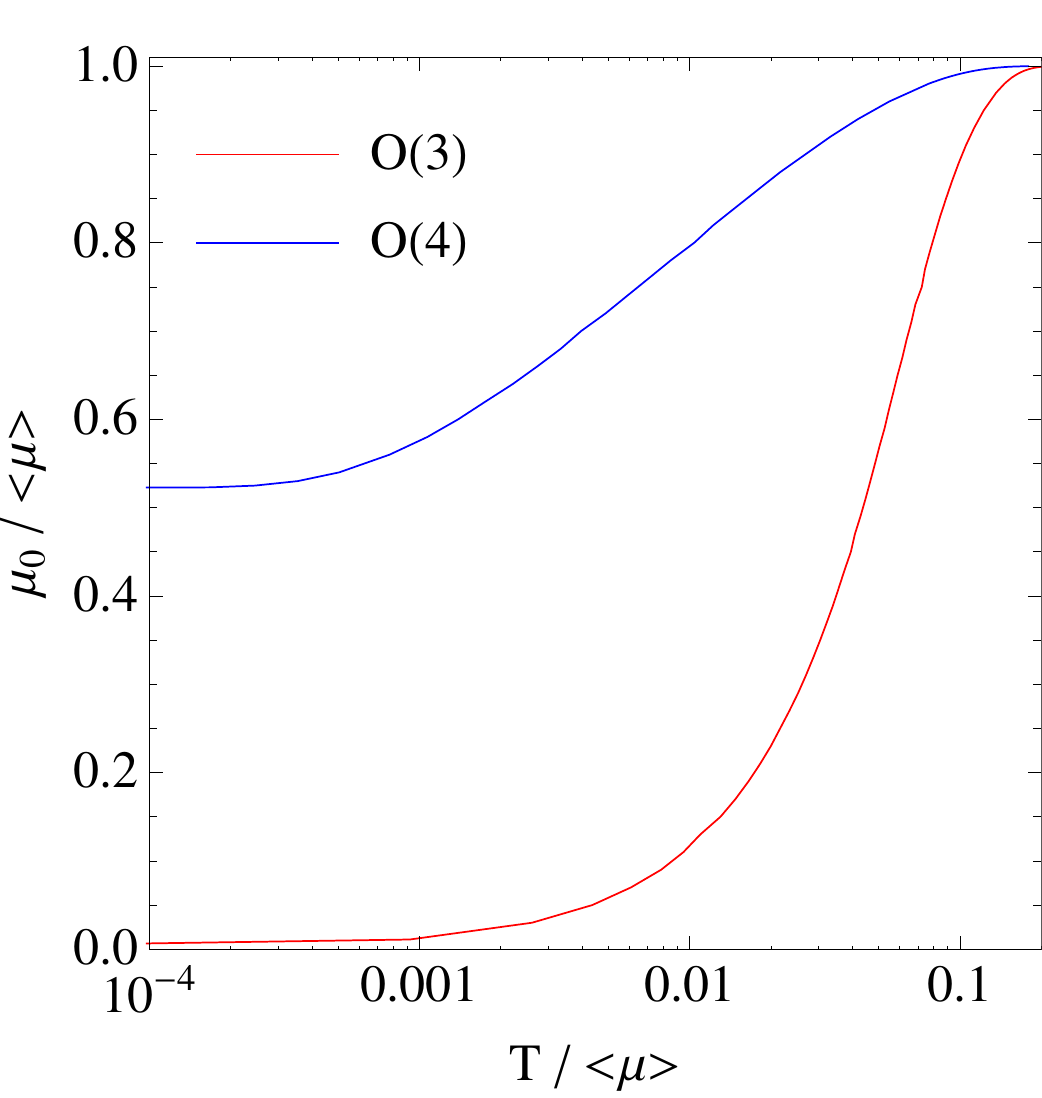} \hspace{0.5cm}
\includegraphics[width=6.6cm]{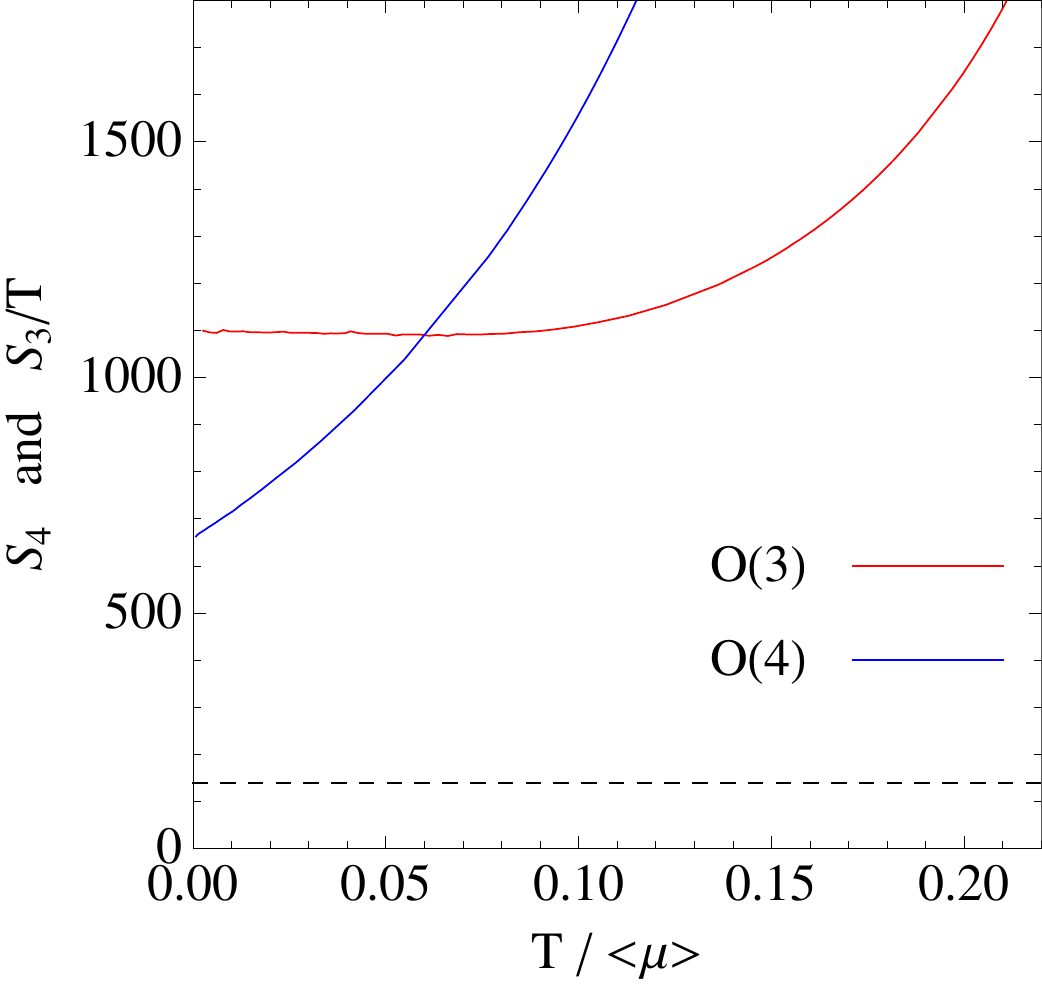} \hspace{0.5cm}
\caption{\it  $\mu_0$ (left panel) and $S_4$ and $S_3/T$ (right panel) as a function of the temperature in the benchmark scenario A$_1$ where the back-reaction is small. Dimensional quantities are in units of $\langle \mu \rangle$ with values quoted in Tab.~\ref{tab:table}.}
\label{fig:S3S4Small}
\end{figure} 

Fig.~\ref{fig:S3S4Small} presents the numerical results on the analysis of the phase transition in the (small back-reaction) scenario A$_1$. The figure displays the values of the bounce solution $\mu_0$ and the action $S_E$, as a function of the temperature, under the assumption that only the $O(3)$ or $O(4)$ \textit{ans\"atze} are valid.

As we can see, at low (high) temperatures the $O(4)$ ($O(3)$) solution dominates, as expected. Remarkably, neither $S_4$ nor $S_3/T$, and therefore $S_E$, ever reach the threshold 140. This happens because the free-energy in the BH solution is large, i.e.~$a_h\simeq\mathcal O(1)$, and the system tries to cool down as much as possible to minimize the energy barrier between the confined and deconfined phases. Nevertheless, due to the flatness of the potential, the barrier is still too big even at zero temperature. As a consequence the bubble nucleation is always too suppressed to compete with the Hubble expansion of the universe, and the bubbles of the confined phase $\langle\mu\rangle$ never percolate. This leads to a universe where a (huge) portion of the space remains in an inflationary phase (see Sec.~\ref{sec:inflation}). The viability of the scenario A$_1$ is then quite debatable and we do not further investigate it hereafter.

\subsection{Large back-reaction}

\begin{figure}[htb]
\centering
\includegraphics[width=5.8cm]{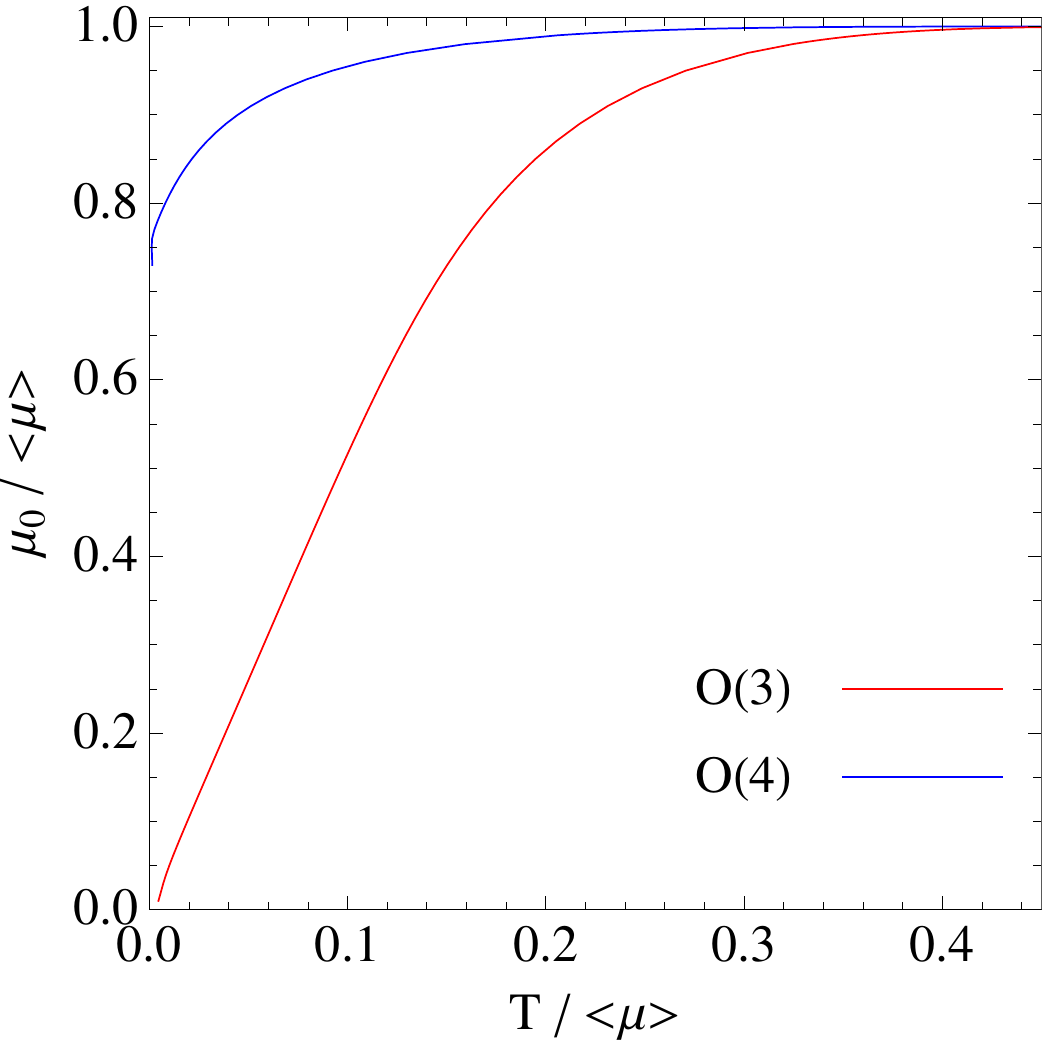} \hspace{0.5cm}
\includegraphics[width=5.9cm]{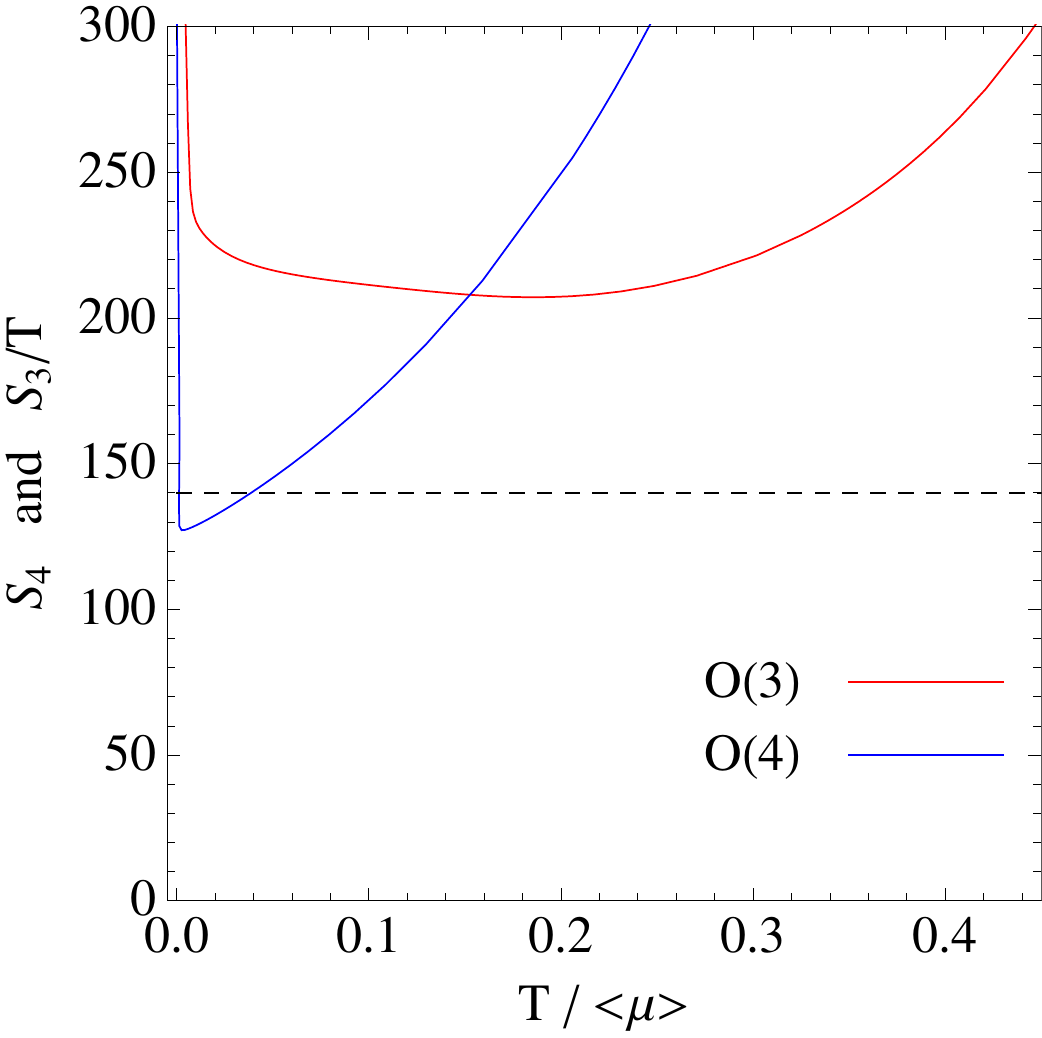} 
\includegraphics[width=5.8cm]{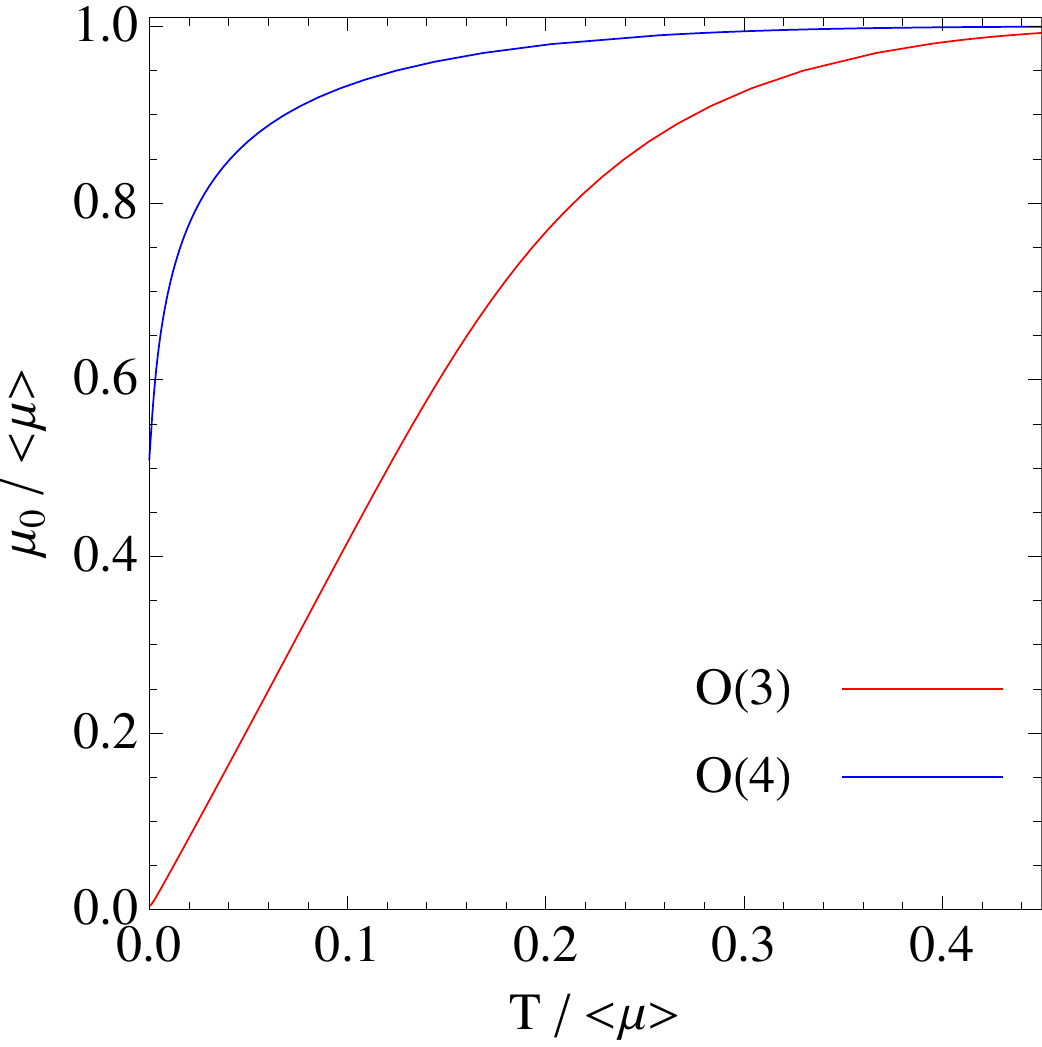} \hspace{0.5cm}
\includegraphics[width=6.2cm]{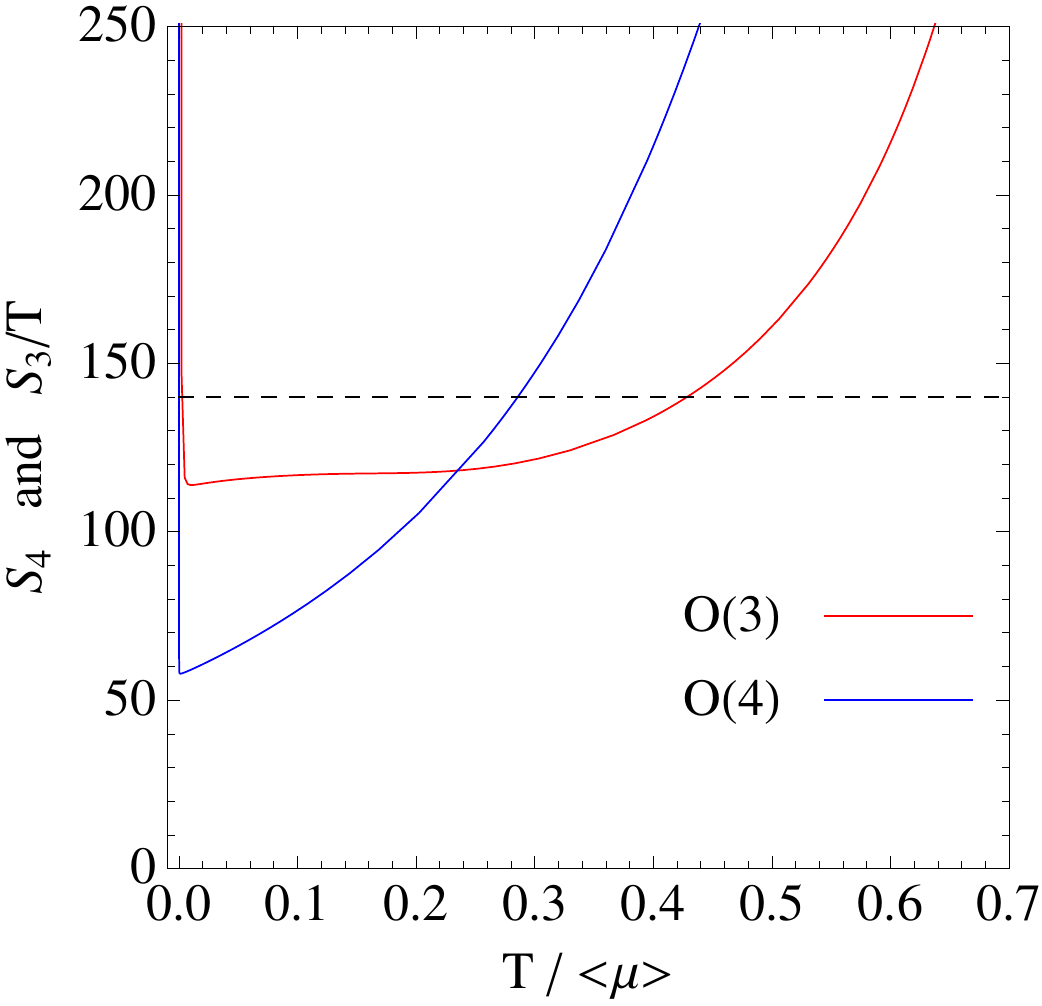} 
\caption{\it Upper panels: As in Fig.~\ref{fig:S3S4Small} but for scenario B$_8$. Lower panels: As in Fig.~\ref{fig:S3S4Small} but for scenario B$_2$. The findings are qualitatively similar to those arising in the most common  parameter scenarios where the back-reaction is large (cf.~Fig.~\ref{fig:S3S4smallkappa}).
}
\label{fig:S3S4Large}
\end{figure} 
\begin{figure}[htb]
\centering
\includegraphics[width=5.8cm]{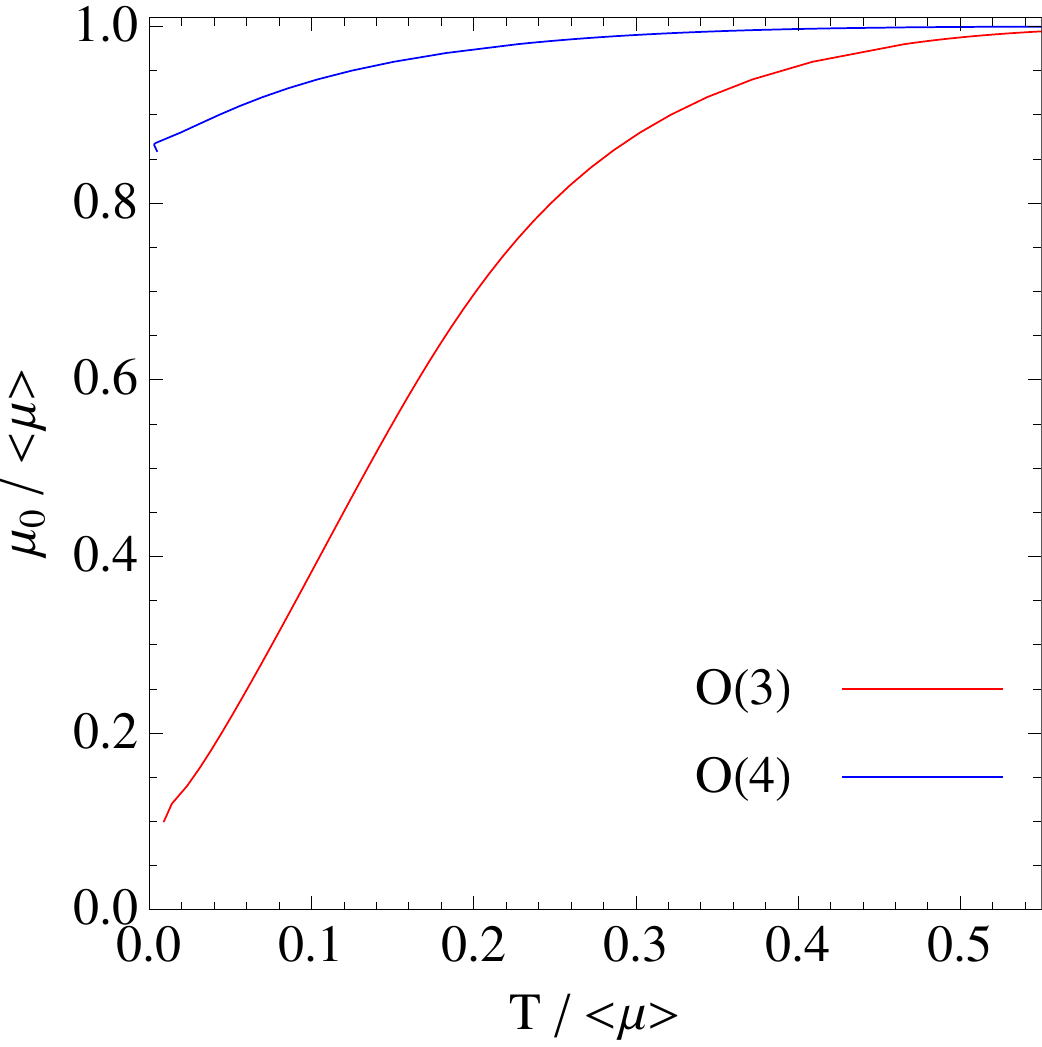} \hspace{0.8cm}
\includegraphics[width=5.9cm]{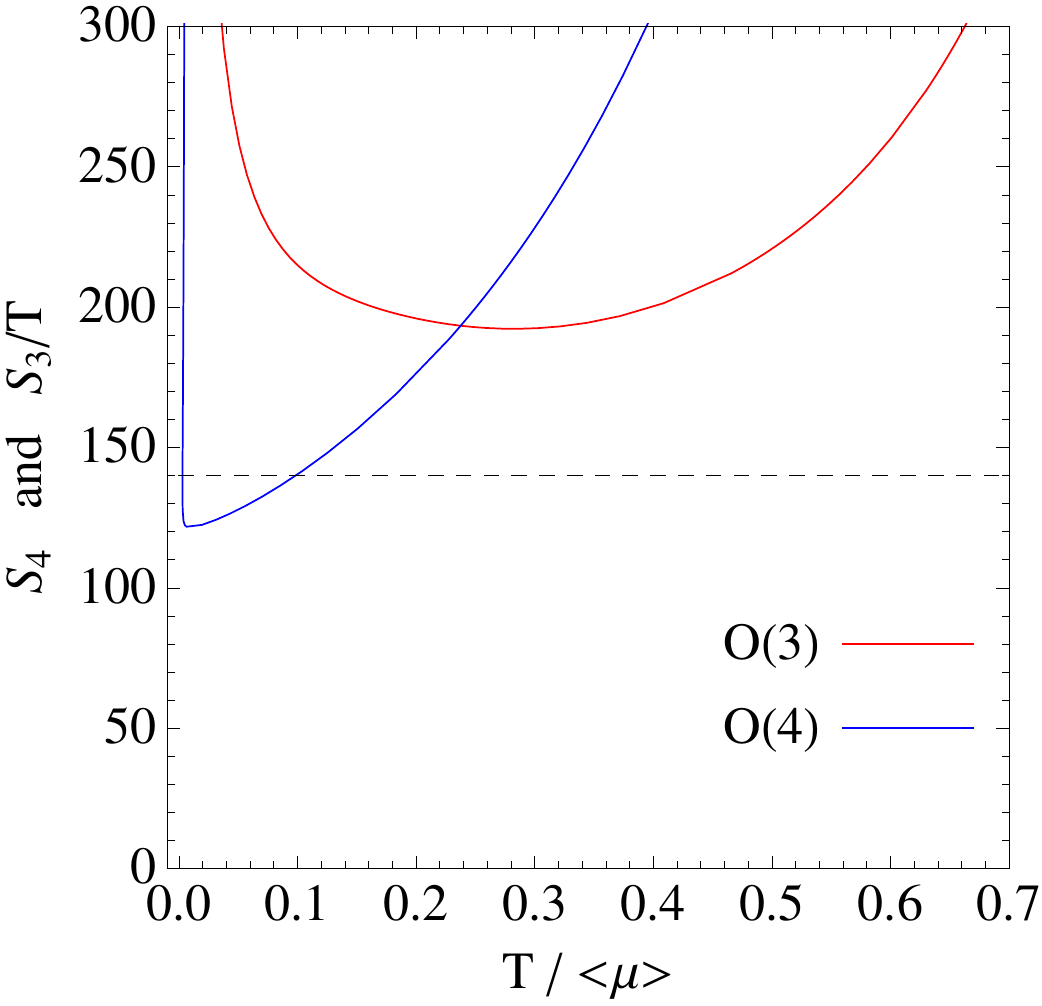} 
\includegraphics[width=5.8cm]{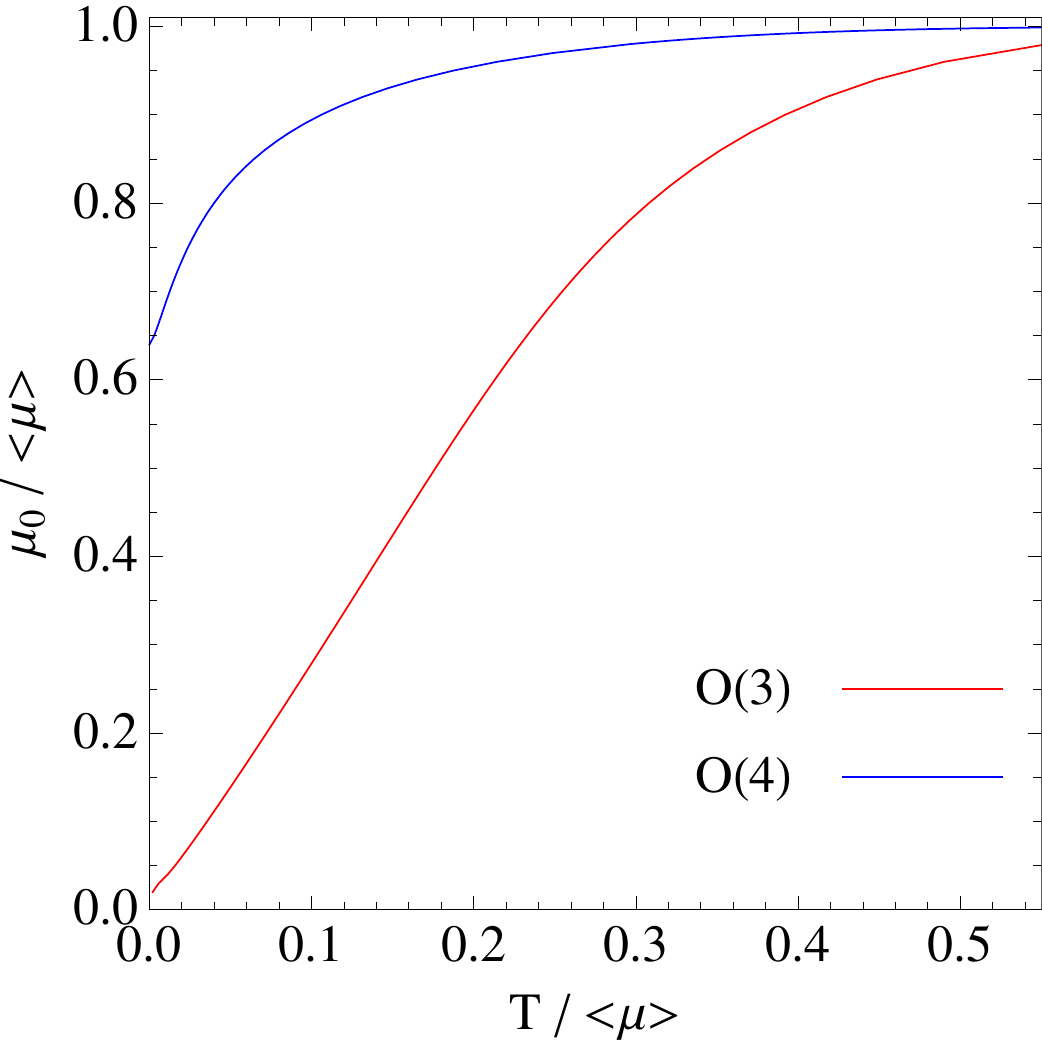} \hspace{0.5cm}
\includegraphics[width=5.7cm]{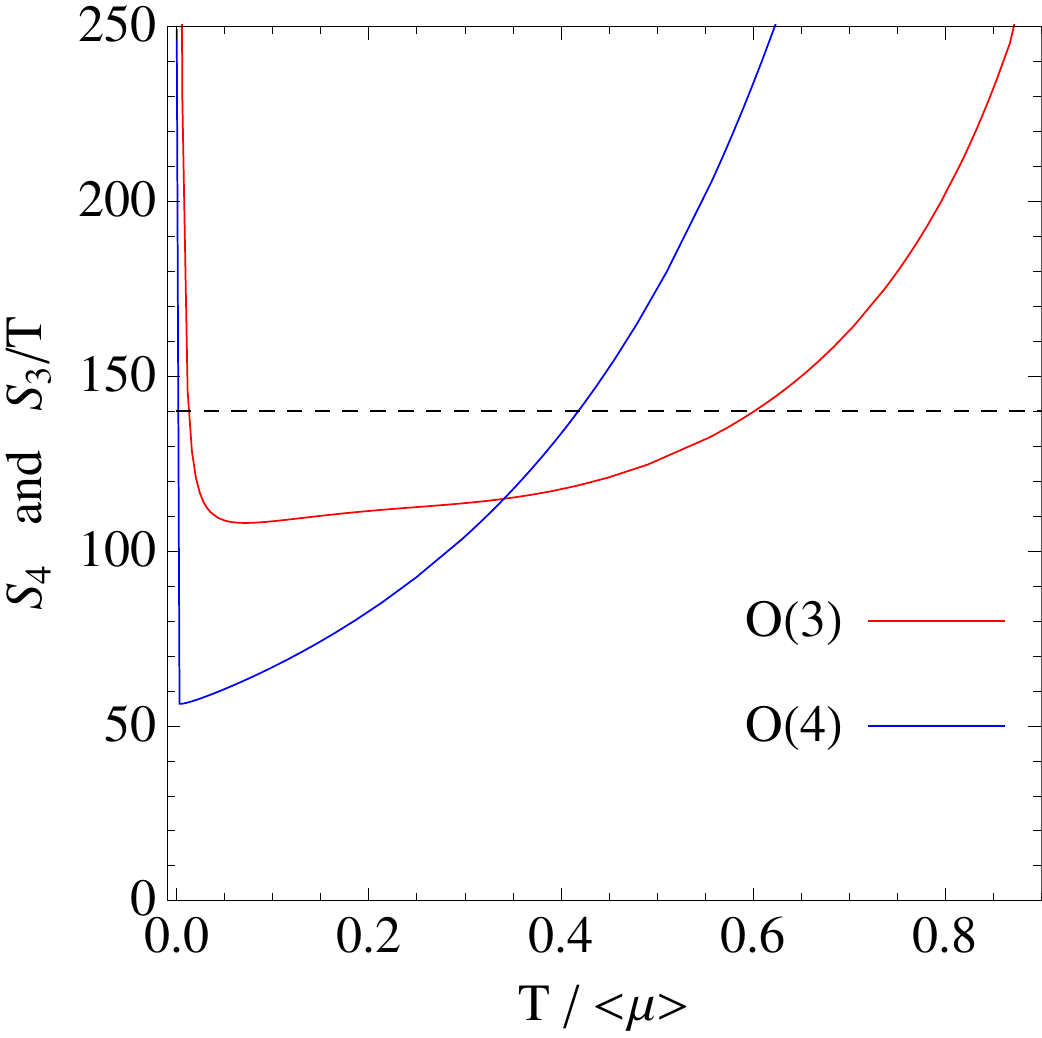} 
\caption{\it Upper panels: As in Fig.~\ref{fig:S3S4Small} but for scenario C$_2$. Lower panels: As in Fig.~\ref{fig:S3S4Small} but for scenario C$_1$. The findings are qualitatively similar to those arising in the most common  parameter scenarios where the back-reaction is large (cf.~Fig.~\ref{fig:S3S4Large}).}
\label{fig:S3S4smallkappa}
\end{figure} 

To describe the behavior of the radion phase transition in the regime of large back-reaction, we first focus on classes B and C, and then comment on the remaining parameter configurations.

The upper (lower) panel of Fig.~\ref{fig:S3S4Large} shows the numerical results for the bounce in scenario B$_8$ (B$_2$). Similarly, the upper (lower) panel of Fig.~\ref{fig:S3S4smallkappa} deals with scenario C$_2$ (C$_1$). The plots illustrate that for large values of $|\lambda_1|$ (lower panels) the phase transition is dominated by the $O(3)$ bounce, while for lower values of $|\lambda_1|$ it is dominated by the $O(4)$ bounce (upper panels). The plots moreover highlight that $\mu_0$ and $T_n$, which is the largest temperature where $S_4$ or $S_3/T$ crosses the horizontal dashed line, are of the same order of $\langle \mu \rangle$. This happens due to the fact that $a_h\ll 1$:  the temperature in the free energy  has to be substantially increased to compensate the smallness of the prefactor $a_h$, in comparison to what happens in $a_h\approx 1$ configurations (remember that $T$ appears only in $F_d$ once the SM-like plasma is neglected, as previously stressed).

As expected, Figs.~\ref{fig:S3S4Large} and \ref{fig:S3S4smallkappa} also show that the nucleation temperature provided by the $O(3)$ ansatz, if it exists, is higher than the one arising in the $O(4)$ case. In particular the nucleation temperature of the latter case is small enough not to jeopardize the correctness of the $O(4)$ action calculation. In fact, the $O(4)$ ansatz assumes a space topology that is a good approximation of the (compactified) finite-temperature space only when $\bar\rho \,T_n \ll 1$~\cite{Creminelli:2001th, Nardini:2007me}. We have checked that our solutions fulfill such a condition.

The numerical results obtained for other benchmark scenarios with large back-reaction are qualitatively similar to those just described. We then simply report our findings in Tab.~\ref{tab:table}, together with those above. Besides quoting the results, we display the value of $\lambda_1$ in blue (red) when the phase transition occurs via the $O(4)$ ($O(3)$) solution. Overall, all the considered benchmark configurations hint at the fact that in the ballpark of the large back-reaction parameter space, the transition is possible and occurs with $T_n/\langle \mu \rangle$ of the order of between one or one tenth.  Much smaller values of $T_n$ are of course feasible by tuning the parameters.

\subsection{Inflation and reheating}
\label{sec:inflation}

As Tab.~\ref{tab:table} shows, when the radion phase transition happens,  $T_n$ is smaller than the value of $\mu$ inside the nucleated bubble, $\mu_0$, or the value of  $\mu$ at the radion potential minimum, $\langle\mu\rangle$ (of course $\mu_0 < \langle\mu(T_n)\rangle$ with $\mu_0 \simeq \langle\mu(T_n)\rangle$ in our scenarios). The considered scenarios thus exhibit a quite large order parameter $\langle\mu\rangle /T_n$, namely $2 \lesssim\langle\mu\rangle /T_n\lesssim 25$, signaling the presence of a strong first order phase transition. This is a consequence of the cooling in the initial (BH) phase, which also triggers a (very brief) inflationary stage just before the onset of the phase transition.

The energy density  $\rho=F-T dF/dT$ in the two phases is given by
\be
\label{eq:E0}
\rho_{d}=E_0+\frac{3\pi^4\ell^3}{\kappa^2}a_h T^4+\frac{\pi^2}{30}g_d^{eff}T^4\,,
\ee
\be
\rho_{c}=\frac{\pi^2}{30}g_c^{eff} T^4\ .
\ee
Inflation in the deconfined phase happens provided that $E_0$ dominates the value of $\rho_d$ over the thermal corrections. So inflation in the deconfined phase starts at the temperature
\be
T_i\approx \left(\frac{30\kappa^2 E_0}{90\pi^4\ell^3 a_h+\pi^2\kappa^2 g_d^{eff}}\right)^{1/4} \,,
\label{inftemp}
\ee
and finishes everywhere when bubbles percolate, which is expected to occur at a temperature very closed to $T_n$ (for details, see e.g.~Ref.~\cite{Nardini:2007me}). So, the amount of e-folds of inflation occurring just before the radion phase transition is $N_e\approx \log(T_i/T_n)$.

The precise values of $T_i$ and $N_e$ for the different benchmark scenarios depend on the matter content in the different confinement/deconfinement phases, i.e.~the values of $g_c^{eff}$ and $g_d^{eff}$. As previously stated, we assume that in the confined phase, at low energy, the dynamical degrees of freedom are the SM fields plus the massive radion, i.e. $g_c^{eff}=106.75$~\footnote{We are not counting here the radion/dilaton, which is highly localized towards the IR brane and thus composite in the dual theory, whose mass in the confined phase is larger than the nucleation temperature. Its contribution $\Delta g_c^{eff}\propto \exp(-m_{rad}/T_n)$ is Boltzmann suppressed, as it decouples from the thermal plasma.}. Among these, only the Higgs and the right-handed top quark are localized towards the IR brane, so that $g_d^{eff}=97.5$. The consequent values of $T_i$ and $N_e$ in the considered benchmark scenarios are shown in Tab.~\ref{tab:table2}. Notice that the scenario D$_1$, 
Eq.~(\ref{inftemp}), yields $T_i<T_n$ and thus there is no inflationary period before nucleation. In the the other scenarios, instead, a brief inflationary epoch exists, so that the plasma contribution due to the SM-like degrees of freedom is subdominant at the onset of the radion phase transition. This proves a posteriori that our calculation of $T_n$ by disregarding the thermal contribution proportional to $g_d^{eff}$ is fully justified.

\begin{table}[htb]
\centering
\begin{tabular}{||c|c|c|c|c||c|c||}
\hline\hline
Scen. &   $T_i/\langle\mu\rangle$& $N_e$&$T_R/\langle\mu\rangle$&$T_R/GeV$ &$\alpha$& $\log_{10}(\beta/H_\star)$\\ \hline
B$_1$      &  0.663 & 0.09 & 1.272 & 1053 & $1.60$  & 2.36  \\ 
  B$_2$   & 0.605 &0.35&1.071& 821.8 & 4.61  & 1.99\\
  B$_3$      &  0.591 & 0.48 & 1.024 & 770.4 & $7.86$  & 1.79  \\   
  B$_4$      & 0.580 & 0.67 & 0.986 & 730.6 & $17.1$  & 1.48  \\   
 B$_5$      &   0.568 & 1.08 & 0.953 & 694.0 & $90.1$  & 1.97  \\ 
 B$_6$      &   0.551 & 1.31 & 0.921 & 654.2 & $228$  & 1.86  \\  
B$_7$       &   0.531 & 1.68 & 0.887 & 612.0 & $1047$  & 1.67  \\  
 B$_8$      &   0.509 & 2.57&0.849 & 566.4 & $4.0\cdot 10^4$  & 1.23  \\ 
  B$_9$     & 0.5004 & 3.71 & 0.834 & 549.3 & $4.1\cdot 10^6$  & 0.64  \\  
 B$_{10}$      & 0.4991 & 4.22 & 0.832 & 546.8 & $3.3\cdot 10^7$  & 0.34  \\  
 B$_{11}$      & 0.4985 & 4.86 & 0.831 & 545.6 & $4.5\cdot 10^8$  & -0.32  \\   \hline           
 C$_1$   &   0.828 &0.32&1.531 & 578.4  & 4.3  & 2.03 \\
 C$_2$       &    0.718 & 1.99&1.239 & 416.2 & $5.0\cdot 10^3$ & 1.45  \\   \hline 
D$_1$  &  --  & -- &0.535 &133.7 & 5.0  & 1.05 \\ \hline
E$_1$  & 0.509 & 1.28 &0.850 &567.2 & 203  & 1.89 \\
\hline\hline
\end{tabular}
\caption{\it Some physical parameters for the cases $B_i$, $C_i$, $D$ and $E$ considered in the text.}
\label{tab:table2}
\end{table}

Under the approximation that the percolation temperature is very similar to $T_n$, during the phase transition the energy density is approximately conserved. At the end of the phase transition the universe then ends up in the confined phase at the reheating temperature $T_{R}$ given by
 \be
 \rho_c(T_{R})=\rho_d(T_n) \,,
 \ee
 or, equivalently,
 \be
\frac{\pi^2}{30} g_c^{eff}T_R^4=E_0+\left(\frac{3\pi^4\ell^3}{\kappa^2}a_h+\frac{\pi^2}{30}g_d^{eff}\right)T_n^4  \,.  
\label{igualdad}
 \ee

The value of $T_R$ for the different benchmark scenarios is shown in Tab.~\ref{tab:table2}. It turn out that in most of the cases $T_R$ is quite close to the TeV scale, nevertheless a parameter window with $T_R$ at the EW scale exists (e.g.~scenario D$_1$). We will comment on the consequences of this observation in the next section. 
 
\section{The electroweak phase transition}
\label{sec:EWPT}

Depending on the particle setup and the embedding of the Higgs field in the model, the confinement/deconfinement phase transition can be tightly connected to the EW phase transition. This is the case in our setup (specified in Sec.~\ref{sec:inflation}) where the Higgs, the radion, and the right-handed top are localized towards the IR brane and hence only exist in the confined phase.

All SM-like fields propagating in the bulk, as well as those localized at the branes, are present in thermal plasma of the confined phase. Their contribution to the free energy is $\Delta F_{c}=-\pi^2g^{eff}_{c}T^4/90$, with $g^{eff}_{c}\simeq 106.75$ at $100\,\textrm{GeV} \lesssim T \lesssim m_{\rm rad}$. Instead, the fields localized near the IR brane are beyond the BH horizon and, being outside the physical space, they are not present in the deconfined phase. Within our particle setup, the thermal plasma before the radion phase transition contributes to the free energy as $\Delta F_{d}=-\pi^2g^{eff}_{d}T^4/90$, with $g^{eff}_{d}\simeq 97.5$ at any EW-scale temperature for all SM-like fields being massless. 
In view of this, the (model dependent) quantity $\Delta g^{eff}=g^{eff}_{c}-g^{eff}_{d}=9.25$ effectively shifts $F_{\rm min}$ in Eq.~(\ref{eq:freemin}) by
\be
\Delta F_{\rm min}=\frac{\pi^2}{90}\Delta g^{eff}T^4
\label{eq:Deltafreemin}\,,
\ee 
which corresponds to $\left|\Delta F_{min}/F_{min}\right|\simeq 0.01$. Therefore the nucleation temperature of the radion phase transition is essentially unaffected by the presence of the SM-like degrees of freedom in the plasma. Disregarding them in the calculation of $T_n$ is hence fully justified, even when the phase transition does not start in an inflationary epoch, as in our scenario D$_1$. 

On the other hand the SM-like particles do not contribute to the free energy only via the plasma term: when the BH horizon moves beyond the IR brane during the phase transition, the Higgs field ($\mathcal H$) appears and there is an extra dynamical field besides the radion. The effective potential  becomes a function of both fields and can be written as~\cite{Nardini:2007me}
\be
V(\mu,\mathcal H)=V_{eff}(\mu)+\left(\frac{\mu}{\langle\mu\rangle}\right)^4V_{\rm SM}(\mathcal H,T) \,,
\label{eq:totalpotential}
\ee
while the SM potential $V_{\rm SM}$ in the effective theory, after integrating the extra dimension, is given by
\be
V_{\rm SM}(\mathcal H,T)=-\frac{1}{2}m^2\mathcal H^2+\frac{\lambda}{4}\mathcal H^4+\Delta V_{\rm SM}(\mathcal H,T) \,,
\label{eq:SMpotential}
\ee
where the Higgs mass is $m_\mathcal H^2=2\lambda v^2\simeq(125\,\textrm{GeV})^2$ with $\lambda=v^2/m^2\simeq0.123$ and $v=246$ GeV, and the term $\Delta V_{\rm SM}(\mathcal H,T)$ contains the Higgs field dependent loop corrections both at zero and at finite temperature. $V_{\rm SM}(\mathcal H,T)$ has its absolute minimum at $\langle\mathcal  H(T)\rangle=v(T)$ whose value, in the first (leading) approximation for the thermal corrections, turns out to be ~\cite{Quiros:1994dr,Quiros:1999jp}
\be
v(T)=\left\{
\begin{tabular}{cl}
0 & for $T>T_{EW}$  \\
$v\sqrt{1-T^2/T^2_{EW}}$ & for $T\leq T_{EW}$ 
\end{tabular}
\right. 
\ee
where $T_{EW}$, the temperature at which the SM minimum at the origin turns into a maximum, is given by
\be
T_{EW}\simeq m_\mathcal H/\left(m_W^2/v^2+m_Z^2/2 v^2+m_t^2/v^2\right)^{1/2}\simeq 150\ \textrm{GeV} \,.
\ee

In principle, the analysis of the radion phase transition should also take into account the $\mathcal H$ degree of freedom.  However, in practice, this is not necessary. In fact,  $V_{\rm SM}$ provides a contribution $\mathcal O(\lambda v^4(T)/4)$, so that it  effectively shifts the $\mu^4$ term in $V_{eff}(\mu)$  by the amount $\mathcal O(\lambda v^4(T)/(4\langle\mu \rangle^4))$, which is vanishing for $T_n>T_{EW}$ and is +$\mathcal O( 10^{-4})$ otherwise. Such a correction is therefore too small to substantially affect the results of the radion phase transition, obtained without including $V_{\rm SM}$ (cf.~Fig.~\ref{fig:potentialmu}). The calculations in Sec.~\ref{sec:phase-transition} turn out to be justified a posteriori.

We can see from Tab.~\ref{tab:table} that some scenarios lead to $T_n<T_{EW}$, so that the EW symmetry is broken at the same time that the confinement/deconfinement phase transition,  while other scenarios yield $T_n>T_{EW}$ and the EW symmetry remains unbroken during the radion phase transition~\footnote{For $T_n<T_{EW}$ we have $\mu=\langle \mu \rangle$ and $\mathcal H=v(T_n)$ deep inside the bubbles (the confined phase), while far outside the bubble walls, in the sea of the deconfined phase, we have $\mu=\mathcal H=0$. For $T_n>T_{EW}$ we instead have the same behaviour for $\mu$ but the $\mathcal H$ profile is zero both outside and inside the radion bubbles.}. Nevertheless, it ultimately depends on $T_R$ whether the universe really ends up in the EW broken phase after the deconfined/confined bubble percolation or, in other words, whether the dilaton and the EW phase transitions are sequential or simultaneous. This has consequences for electroweak baryogenesis~\cite{Kuzmin:1985mm,Quiros:1994dr},  as we now discuss.

\subsection{Sequential phase transitions: $T_{R}>T_{EW}$.}

Models with $T_{R}>T_{EW}$ are exhibited by the scenarios of classes B, C and E (see Tab.~\ref{tab:table2}). In those cases, even when $T_n<T_{EW}$, at the end of the reheating process the Higgs field is in its symmetric phase and the universe evolves along a radiation dominated era. Within the particle setup we have assumed so far,  the EW symmetry breaking would occur as in the SM, that is, via a crossover that prevents the phenomenon of electroweak baryogenesis~\cite{Kajantie:1996mn,Rummukainen:1998as}.  Had we chosen a low energy particle content rich of new BSM degrees of freedom, the dynamics of the EW symmetry breaking would have been the one corresponding to the chosen low energy setup (while the radion phase transition would have been basically unchanged). In this sense, when $T_{R}>T_{EW}$, the implementation of electroweak baryogenesis remains a puzzle for which the UV soft-wall framework is not helpful.

\subsection{Simultaneous phase transitions: $T_{R}<T_{EW}$.}
For $T_{R}<T_{EW}$ the reheating does not restore the EW symmetry and eventually the Higgs lies at the minimum of $V_{\rm}(\mathcal H,T_R)$.  The value of its minimum, $v(T_R)$, can be considered as the upper bound of the Higgs VEV during the (simultaneous) EW and deconfined/confined phase transitions. Taking this upper bound, it results that the EW baryogenesis condition~\footnote{The SM at finite temperature has an IR singularity at the origin such that perturbative calculations in this region are unreliable. In fact lattice calculations point toward an extremely weak phase transition, or cross-over, for Higgs masses around the experimental value. However for temperatures low enough condition (\ref{eq:condicion}) is fulfilled, and the perturbative potential near the minimum can be approximately trusted.}
\be
\frac{v(T_R)}{T_R}\gtrsim 1 
\label{eq:condicion}
\ee
is fulfilled in the presence of a SM-like low energy particle content (and  $m_\mathcal H\simeq 125$\,GeV) when $T_R$ satisfies the bound~\cite{Nardini:2007me} (see also~Ref.~\cite{DOnofrio:2014rug})
\be
T_R\lesssim T_\mathcal H\simeq 140\ \textrm{GeV}  \,.
\label{eq:temperature}
\ee
To summarize, in scenarios with $T_R<T_{EW}$ the nature of the EW phase transition is then entirely dependent on the radion reheating temperature. More specifically:
\begin{itemize}
\item
If the reheating temperature is $T_{EW}\gtrsim T_{R}>T_\mathcal H$, the EW phase transition is too weak (i.e. it does not satisfy Eq.~\eqref{eq:condicion}) and the sphalerons inside  the bubble wipe out any previously created baryon asymmetry. 
\item
If the reheating temperature is below $T_\mathcal H$, then the sphalerons inside the bubble do not erase the possible baryon asymmetry accumulated inside the bubble during their expansion. Therefore EW baryogenesis can take place if there is a strong enough source of CP violation in the theory. However, the radion phase transition in the generic scenarios leading to $T_{R}\ll T_\mathcal H$ should be studied paying particular attention to the bounce procedure. In fact the vacuum energy $E_0$ might not have dominated the energy density prior to the transition (see Eq.~(\ref{igualdad})), as the dilaton and Higgs potentials might be of the same order of magnitude. The precise bounce solution would then need to be solved in the two-field space $(\mu,\mathcal H)$, as in Ref.~\cite{Bruggisser:2018mus}~\footnote{The precise evaluation of such bounce solutions goes beyond the scope of the present paper whose main aim is more to stress new possibilities than providing refined results.}.  
 
\end{itemize}

A parameter configuration leading to $T_{R}<T_{\mathcal H}$ is provided by scenario D$_1$. In this case the dilaton and EW phase transitions happen simultaneously at $T=T_n\simeq 112$ GeV, ending up with $T=T_{R}=133.7\,\textrm{GeV}<T_{EW}$, so that both the radion and the Higgs acquire a VEV. Before and after the reheating, the bound of Eq.~(\ref{eq:temperature}) is fulfilled, and the condition of strong-enough first order phase transition for EW baryogenesis is satisfied~\footnote{
For a recent analysis see Refs.~\cite{Bruggisser:2018mus,Bruggisser:2018mrt}.}.

\section{Gravitational waves}
\label{sec:GW}

A cosmological first-order phase transition generates a stochastic gravitational waves background (SGWB)~\cite{Witten:1984rs, Kosowsky:1991ua, Kosowsky:1992vn, Kamionkowski:1993fg, Hogan:1986qda, Caprini:2006jb, Caprini:2007xq, Huber:2008hg, Kahniashvili:2008pe, Kahniashvili:2008pf, Caprini:2009yp, Kahniashvili:2009mf, Hindmarsh:2013xza, Giblin:2013kea, Giblin:2014qia, Kisslinger:2015hua, Hindmarsh:2015qta, Weir:2016tov, Jinno:2016vai, Jinno:2017fby, Konstandin:2017sat, Cutting:2018tjt}~\footnote{It has been recently observed that the SGWB from first order phase transitions can contain anisotropies, correlated to
those of the cosmic microwave background of photons, which may be within the reach
of the forthcoming gravitational wave detectors~\cite{Geller:2018mwu}.}. The corresponding GW power spectrum depends on several quantities that characterize the phase transition~\cite{Caprini:2015zlo}. Determining accurately all of them is challenging even in the simplest setups. Hereafter we discuss the main uncertainties and assumptions influencing our estimate of the SGWB sourced by the radion phase transition.

A key quantity is the velocity $v_w$ at which the bubble walls are expanding at the moment of their collisions. In standard cases this would be determined as the asymptotic solution of the EoM of the field driving the phase transition~\cite{Moore:1995si,John:2000zq,Konstandin:2014zta}:
\begin{equation}
\label{eq:friction}
\Box \tilde\mu + \frac{\partial V(\tilde\mu,T)}{\partial \tilde\mu}+ 
\sum_j \frac{\partial m_j^2(\tilde \mu)}{\partial \tilde \mu} \int \frac{d^3p}{(2\pi)^3}  \frac{\delta f_j(\vec{p},E)}{2 E} 
=0~,
\end{equation}
where $\delta f_j$ is the {\it small} deviation from the Boltzmann distribution of the species $j$ with mass $m_j$. However in our case, where $\tilde \mu  =\mu$ for $\tilde \mu\ge 0$ and $\tilde \mu =-T_h$ for $\tilde \mu<0$, not all $\delta f_j$ are  small~\footnote{For instance, fields exactly localized on the IR brane are degrees of freedom that do not exist in the deconfined phase and suddenly appear when the BH horizon crosses the IR brane (at $\tilde \mu =0$). This abrupt change implies $\delta f_j$ to be of the same order of the Boltzmann distribution $f_j$, i.e.~the species $j$ is far away from the thermal equilibrium. By continuity, large deviations are also expected for fields non-exactly localized.  For these, it is manifest that their non-trivial prefactor $\partial m_j^2/ \partial \tilde \mu$ is not sufficient to enforce the sum in Eq.~\eqref{eq:friction} to be a small perturbation.}. Thus Eq.~\eqref{eq:friction} does not capture the complex dynamics of the confined/deconfined phase transition, and strongly-coupled techniques, still under development, should be applied; see~e.g.~Refs.~\cite{Fukushima:2010bq, Andersen:2014xxa}. In any case, it seems reasonable to expect supersonic walls, even reaching $v_w\approx 1$ in the extremely supercooled scenarios (i.e.~very strong phase transitions, in practice). For concreteness we thus discuss in detail two reasonable options for $v_w$, namely  $v_w = v_1 \equiv 0.70$ and $v_w = v_2 \equiv 0.95$.

A further critical feature is the behavior of the plasma during, and after, the bubble collisions.   Besides the energy stored in the bubble walls, the turbulent or coherent motions of the plasma, excited by the bubble expansion, can contribute to the SGWB spectrum too. Including them would enhance not only the amplitude of the GW frequency spectrum but even the shape of the spectrum at high frequencies. Unfortunately, no robust result on the plasma effects exists for the subtle case of a deconfined/confined phase transition. We thus refrain ourselves from including plasma effects in the subsequent analysis. 

In view of the above considerations, in our analysis we employ the envelope approximation results~\cite{ Kosowsky:1992vn,Steinhardt:1981ct, Caprini:2007xq, Huber:2008hg,  Konstandin:2017sat, Cutting:2018tjt}. In such a regime, the frequency power spectrum of the SGWB is given by~\cite{Caprini:2015zlo}
\begin{eqnarray}
h^2\Omega_{\rm GW}(f)&\simeq&  
h^2\overline \Omega_{\rm GW} ~\frac{3.8(f/ f_p)^{2.8}}{1+2.8 (f/ f_p)^{3.8}}  \,,
\label{eq:OmGW}
\end{eqnarray}
with
\be
h^2\overline \Omega_{\rm GW}\simeq 0.80\times 10^{-4} 
\left( \frac{H_\star}{\beta}   \frac{\alpha}{\alpha+1} \right)^2 \frac{\xi(v_w)}{\sqrt[3]{g_c(T_R)}} \,, ~
\ee
\begin{equation}
f_p\simeq 7.7\times 10^{-5} \textrm{Hz} ~\tilde{\xi}(v_w) \left(\frac{\beta}{H_\star}\right)
\frac{T_R\sqrt[6]{g_c(T_R)}}{100\, \textrm{GeV}}~,
\label{eq:fp}
\end{equation}
\begin{equation}
\xi(v_w)=\frac{0.11 v_w^3}{0.42+v_w^2} ~, \qquad  \tilde \xi(v_w) = \frac{0.62}{1.8-0.1+v_w^2}~,
\end{equation}
\be
\alpha \simeq\frac{E_0 
}{3(\pi^4\ell^3/\kappa^2) a_h(T_n) T_n^4} ~,
\label{eq:alphaGW}
\ee
\be
\frac{\beta}{H_\star}\simeq 
T_n\left. \frac{dS_E}{dT}\right|_{T=T_n}
\label{eq:betaGW}~.
\ee
In particular for the chosen velocities $v_1$ and $v_2$ it turns out that $\xi(v_1)\simeq 0.04$, $\xi(v_2)\simeq 0.07$, $\tilde\xi(v_1)\simeq 0.28$, $\tilde \xi(v_2)\simeq 0.24$.

The size of the peak of the power spectrum, $f_p$, can span many orders of magnitudes, and strongly depends on $\beta/H_\star$ and $T_R$. The latter is basically set by $\sqrt[4]{E_0}$ (see Eq.~\eqref{eq:E0}). Had we not bothered about the solution to the hierarchy problem, values of $T_R$ differing  from the TeV scale by orders of magnitude~\footnote{Even though $\sqrt[4]{E_0}$ much below the TeV scale might be in tension with LHC data; see Sec.~\ref{sec:phenomenology}.} (in particular for $\sqrt[4]{E_0}$ much larger than TeV~\footnote{For some theories with large $T_R$, see e.g.~\cite{Dev:2016feu}.}) would have been consistent with the theoretical framework~\footnote{For the production of GW from the QCD phase transition, see e.g.~\cite{Ahmadvand:2017tue}.}. Also $\beta/H_\star$ can span many order of magnitude and radically modify $f_p$. Its lower bound is set by Big Bang Nucleosynthesis (BBN), which provides an upper bound on the number of relativistic species during nucleosynthesis that can be converted into the constraint $\int_0^\infty df f^{-1}h^2\Omega_{\rm GW}(f) \lesssim 1.12\times 10^{-6}$~\cite{Cyburt:2004yc, Caprini:2018mtu}. For the spectrum in Eq.~\eqref{eq:OmGW} this constraint implies $h^2\overline \Omega_{\rm GW}\lesssim 5.6\times 10^{-7}$, corresponding to $\log_{10}(\beta/H_\star) \gtrsim 0.045$ for $v_w=v_1$ and $\log_{10}(\beta/H_\star) \gtrsim 0.16$ for $v_w=v_2$.

\begin{figure}[t]
\centering
\includegraphics[width=12cm]{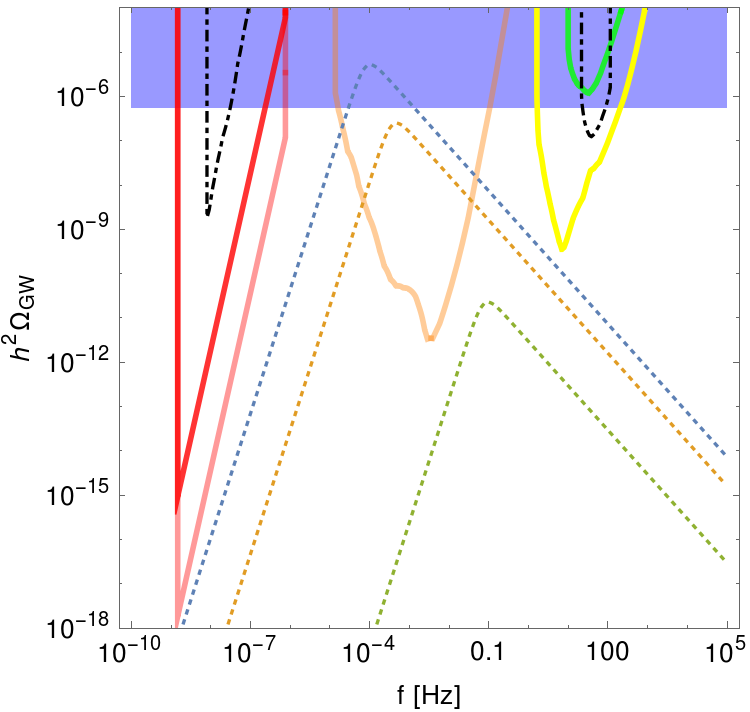} 
\caption{\it
SGWB signals of the benchmark scenarios $B_1$ (lower dashed curve), $B_2$ (middle dashed curve) and $B_{11}$ (upper dashed curve), and the current and forthcoming GW experiments able to test them. The dotted-dashed lines correspond to the power-law sensitivity curves $h^2\Omega_{\rm pls}$ of  PPTA \& EPTA \& NANOGRAV (at frequencies $f\sim$nHz) and aLIGO O1 (at frequencies $f\sim$100\,Hz); the solid lines correspond to the sensitivity curves $\Omega_{\rm sens}(f)$ of SKA observing 100 milli-second pulsars (dark red), SKA observing 2000 milli-second pulsars (light red), LISA (orange), aLIGO at its final design (green) and ET (yellow).  
}
\label{fig:BetaVsTRleft}
\end{figure}

In the next two decades several GW observatories will have the potential to observe, or constrain, the SGWB produced in our benchmark models.  Fig.~\ref{fig:BetaVsTRleft} highlights the sensitivity curves of the main existing and forthcoming GW experiments. The dashed-dotted lines at $f \sim 1$\,nHz  and $\sim10$\,Hz are the {\it power-law} sensitivity curves $h^2\Omega_{\rm pls,NANO}$ and $h^2\Omega_{\rm pls,LIGO\, O1}$~\cite{Thrane:2013oya, Moore:2014eua} reached by the NANOGRAV and aLIGO collaborations, respectively~\cite{ Arzoumanian:2018saf, TheLIGOScientific:2016dpb}. These collaborations do not find any SGWB in their data and consequently rule out any spectrum $h^2\Omega_{\rm GW}(f)$ that intersects one of the two dashed-dotted curves and behaves as a power law inside  them (EPTA and PPTA also achieve a bound similar to the NANOGRAV's one~\cite{Lentati:2015qwp, Shannon:2015ect}). The solid lines correspond to the future sensitivity curves of SKA, LISA, ET and aLIGO at its design sensitivity. Since for SKA, LISA and ET there exists no official and/or updated power-law sensitivity curve, for all future detectors we perform our analysis starting from the ``standard"  sensitivity curves. Specifically, for SKA we determine $h^2 \Omega_{\rm sens, SKA (100)}$ and  $h^2 \Omega_{\rm sens, SKA (2000)}$ from Ref.~\cite{Moore:2014lga, GWplotter}, assuming observation of respectively 100 and 2000 milli-second pulsars (light and dark red lines respectively) during 20 years with 14 days of cadence and $3\times 10^{-8}$ timing precision. For LISA (orange line) we take the sensitivity curve $h^2 \Omega_{\rm sens,LISA}$ from Ref.~\cite{Audley:2017drz}, while for aLIGO at its design sensitivity (green line) we obtain $h^2 \Omega_{\rm sens,LIGO design}$ by joining the sensitivity curves of Virgo, LIGO and KAGRA of Ref.~\cite{Aasi:2013wya}. For ET (yellow line) we use the ``ET-D" sensitivity curve presented in Ref.~\cite{Sathyaprakash:2012jk}.
The dashed lines display the SGWBs $h^2 \Omega_{\rm GW}$ corresponding to the benchmark scenarios B$_1$, B$_2$ and B$_{11}$  summarized in Tab.~\ref{tab:table2} (the values of $\alpha$ and $\beta/H$ from Eqs.~\eqref{eq:betaGW} and \eqref{eq:alphaGW} are also quoted in the table). The SGWB spectra touching the blue area are ruled out by the BBN bound previously discussed~\footnote{
Notice that the blue area includes the region $\beta/H_\star\ll 1$. In this limit the phase transition is so slow that our prediction of the GW spectrum should be corrected, taking into account e.g.~the expansion of the universe during the phase transition. For continuity we do not however expect such corrections to make points with $\beta/H_\star\ll 1$ compatible with BBN, while points with $\beta/H_\star\simeq 1.2$, for which our GW spectrum prediction is rather trustable, are excluded. We thank the referee for pointing out this (implicit) approximation.
 }.

\begin{figure}[t]
\centering
\includegraphics[width=12cm]{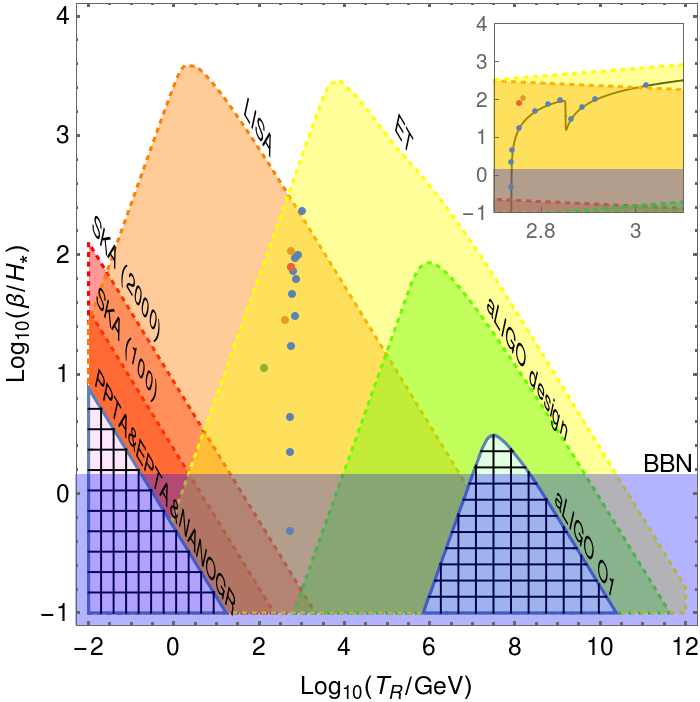}
\caption{\it
The $T_R$--$\beta/H_\star$ parameter space that exhibits  SNR $>10$ at SKA, LISA, aLIGO, and ET for $v_w=v_2$. The etched areas are in tension with present aLIGO O1, EPTA, PPTA and NANOGRAV constraints~\cite{TheLIGOScientific:2016dpb, Lentati:2015qwp, Arzoumanian:2018saf, Shannon:2015ect}. The BBN bound excludes the blue area. The considered benchmark scenarios B$_i$ (blue points), C$_i$ (orange points), D$_1$ (green point) and  E$_1$ (red point) are detectable at both LISA and ET. The stepwise behavior shown in the inserted figure, a zoom of the main one,  is a consequence of the continuous change of regime from $O(4)$ to $O(3)$ bubbles when decreasing the IR brane parameter $\lambda_1$; cf. Eqs.~(\ref{eq:nucl-rate}) and (\ref{eq:eSE}) and colors of $\lambda_1$ in Tab.~\ref{tab:table}.
}
\label{fig:BetaVsTRright}
\end{figure} 

Fig.~\ref{fig:BetaVsTRright} sketches the parameter region (hatched areas) of the plane $\beta/H_\star$--$T_R$ that NANOGRAV, EPTA, PPTA and aLIGO O1 rules out, assuming the spectrum in Eq.~\eqref{eq:OmGW} with $v_\omega=v_2$. The 
exclusion is based on the criterion that a spectrum touching the power-low sensitivity curves of these experiments would have already been detected~\footnote{We do not check that $h^2\Omega_{\rm GW}$ behaves as a power law within the full frequency band of each experiment. Were we adopting this (correct) criterion, we would not expect appreciable differences in the corresponding plot region.}. The blue area is the BBN bound above mentioned. The remaining areas sketch the $\beta/H_\star$--$T_R$ parameter regions for which $\Omega_{\rm GW}$ will yield a Signal-to-Noise Ratio (SNR) larger than 10 in the data which will be collected in the next two decades. For concreteness, for each experiment we check the condition 
\begin{equation}
{\rm SNR_i} = \sqrt{(3.16\times 10^7 s)\frac{\mathcal T_i}{\textrm{1 year}}  \int_0^\infty df \frac{\Omega_{\rm GW}^2(f)}{\Omega_{\rm sens,i}^2(f)}     } > 10~,
\end{equation}
with $\mathcal T_i = $ 20, 3, 7 and 8 years, respectively, for $i=$``SKA",``LISA", ``ET" and ``aLIGO design" (these numbers are  very indicative estimates of the amount of data that each experiment may take by 2040 including duty cycles). The parameter reach that we obtain for LISA does not substantially differ from the one previously calculated in Ref.~\cite{Caprini:2015zlo}.

We remind that Figs.~\ref{fig:BetaVsTRleft} and~\ref{fig:BetaVsTRright} assume $v_w=v_2$. The  forecast for a different bubble velocity, $v_\omega$, can be obtained from the right panel of the figure by shifting the coloured regions by $0.5\log_{10}[0.07/\xi(v_\omega)]$ and $\log_{10}[0.06/(\tilde \xi(v_\omega)\sqrt{\xi(v_\omega)})]$ along the $\log_{10}(\beta/H_\star)$ and $\log_{10}(T_R/\textrm{GeV})$ axes, respectively. Thus for the case $v_w=v_1$ the shifts are around 10\% in $\log_{10}(\beta/H_\star)$, and 1\% in $\log_{10}(T_R/\textrm{GeV})$, which are negligible with respect to the approximations on the 
spectrum we are making. Notice also that this rescaling proves that subsonic velocities, suitable for EW baryogenesis, are not incompatible with detection. For instance, within our approximations (which might not be reliable for small velocities), the ``simultaneous phase transition" of the scenario D$_1$ would be detectable at LISA, even with $v_w\gtrsim 0.02$~\footnote{See e.g.~the more complete analysis in Ref.~\cite{Figueroa:2018xtu}.} (fully consistent with the scenario of EW baryogenesis~\cite{Carena:2000id}, which is known to work only for the cases of low (subsonic) wall velocities, as said above). Unfortunately this would not hold for the ET detector, whose detection region would stay completely on the right of the point representing D$_1$.

In conclusion, for both $v_w=v_1$ and $v_w=v_2$, all our benchmark scenarios are promising for detection at both LISA and ET, whereas SKA and aLIGO, as well as present GW constraints, do not reach them. Out of our benchmarks, only scenario B$_{11}$ is ruled out due to the BBN bound.  In general, measuring the SGWB at two experiments, sensitive to very different frequencies, will allow to better understand the nature of the SGWB. We further comment on the possible implications of this result in the conclusions.

\section{Heavy radion phenomenology}
\label{sec:phenomenology}

As concluded in the previous sections, a considerable amount of back-reaction on the metric facilitates the confined/deconfined phase transition. It also typically implies that the radion is lighter than any KK resonances and has a mass around the TeV scale, at least for the parameter choices solving the  hierarchy problem. Due to this mass hierarchy, in our particle setup with only SM-like fields at the EW scale the radion can decay only into SM-like fields. In particular, since the radion couples to the trace of the energy momentum tensor, its production and decay channels are those of the SM Higgs, although with different strengths. We can thus estimate the detection prospects for the radion at the LHC by rescaling the cross sections and branching rations valid for a generic SM-like Higgs, $H$~\cite{Dittmaier:2011ti}~\footnote{In non-minimal particle setups the radion might be coupled to sectors that do not interact with the SM fields. In this case the considerations in this section would be relaxed, as all radion signal strengths would be correspondingly reduced, with benefits on the  minimal radion mass experimentally allowed and, in turn, on the range of values that are permitted for $E_0$.} with mass equal to the radion mass.

\subsection{Radion couplings}

As in our particle setup the 125-GeV Higgs boson is localized towards the IR brane to solve the hierarchy problem, hereafter we make the \textit{simplifying hypothesis} that the Higgs is exactly localized at the IR brane. This allows to avoid technicalities that  would affect the final result only marginally. The relevant 4D action for the radion, the generic Higgs $H$ and the SM fields is then
\begin{align}
&S_4=2\int_0^{r_1}dr\left[(1-F)\bar{\psi}_{L,R}i\slashed D {\psi}_{L,R}-
\left(\frac{1}{4}+\frac{F}{2}\right)\tr F_{\mu\nu}^2 - e^{-A} (1-2F)M(\phi)\bar\psi\psi\right. \label{eq:accion}\\
&+\delta(r-r_1)\left.\left\{-\frac{\ell h_f}{\sqrt{2}}(1-4F) (H\bar{{\psi}}_L\psi_R+h.c.)+(1-2F)\frac{1}{2}(D_\mu H)^2-(1-4F)V(H)\right\}\right]\nonumber
\end{align}
where all 5D fields have already been rescaled with the corresponding power of the warp factor and the 5D Dirac mass is $M(\phi)=\mp c_{L,R}W(\phi)$.  Moreover $V(H)$ has the form of $V_{\rm SM}(\mathcal H, T=0)$ in Eq.~\eqref{eq:totalpotential} but with a generic $\lambda$ ($\lambda\simeq 0.123$ only when $H$ matches the 125-GeV SM Higgs $\mathcal H$). In addition  the zero modes are defined, in terms of the 4D fields, as
\begin{align}
A_\mu(x,r)&=\frac{A_\mu(x)}{\sqrt{2r_1}} \,, \nonumber\\
\psi_{L,R}(x,r)&=\frac{e^{(1/2-c_{L,R})A}}{\left[2\,\int_0^{r_1} dr e^{(1-2 c_{L,R})A}  \right]^{1/2}} f_{L,R}(x) \,,
\label{eq:Apsi}
\end{align}
and the 5D ($g_5$) and 4D ($g_4$) gauge couplings are correspondingly related by $g_5=g_4\sqrt{2r_1}$. 

Using the radion ansatz $F(x,r)\equiv e^{2A}R(x)$ and expanding
Eq.~(\ref{eq:accion}) to first order in $R(x)$, we obtain the reparametrization (see Eq.~(\ref{eq:mu2}))
\be
\mu(x)=\ell^{-3/2}\int_{r_1}^{r_S}X_F^{-1}-\ell^{-3/2}X_F^{1/2}(r_1)R(x)+\mathcal O(R^2) \,.
\ee
This leads to the canonically normalized radion field $\mathcal R(x)$ defined, in terms of the Planck scale relation in Eq.~(\ref{eq:kappa2}), by
\be
R(x)=-\left[ \frac{\int_0^{r_1}e^{-2A}}{6\int_0^{r_1}e^{2A}} \right]^{1/2}  \frac{\mathcal R(x)}{M_P}\,.
\label{eq:redefinicion}
\ee

\subsubsection*{Couplings to massless gauge bosons}

To compare the loop-induced couplings of the radion  with those of the heavy Higgs $H$, it is useful to calculate the loop-induced couplings of both scalar fields. In the case of the heavy Higgs, the interactions to photons and gluons are given by the Lagrangians
\begin{align}
\mathcal L_{H\gamma\gamma}&=\frac{\alpha}{8\pi}\left[\sum_f N_c Q_f^2 A_{1/2}(\tau_f)+A_1(\tau_W)\right] \frac{h}{v}F_{\mu\nu}F^{\mu\nu}  \,, \\
\mathcal L_{Hgg}&=\frac{\alpha_s}{16\pi}\left[\sum_Q A_{1/2}(\tau_Q)\right] \frac{h}{v}\tr G_{\mu\nu}G^{\mu\nu} \,,
\end{align}
where $\tau_i=m_H^2/4m_i^2$ and $H=v+h$. For the functions $A_{1/2}(\tau)$ and $A_1(\tau)$ we use their generic expressions defined e.g.~in~Ref.~\cite{Djouadi:2005gi} although, in our regime of heavy Higgs with $\tau=m_H^2/(4 m_i^2)\gg 1$, they can be well approximated as
\begin{equation}
A_1(\tau)\to -2,\quad A_{1/2}(\tau)\to -\frac{[\log(4\tau)-i\pi]^2}{2\tau} \,.
\end{equation}
It follows that
\begin{align}
&\sum_fN_c Q_f^2 A_{1/2}(\tau_f)+A_1(\tau_W)=-2+\mathcal O(m_t^2/m_H^2) \,, \\
&\left|\sum_Q A_{1/2}(\tau_Q)\right|=\frac{2m_t^2}{m_H^2}\left[ \log^2(m_H^2/m_t^2)+\pi^2 \right]
+\mathcal O(m_b^2/m_H^2)\,,
\end{align}
which implies that $\mathcal L_{H\gamma\gamma}$ and $\mathcal L_{Hgg}$ are respectively dominated by diagrams with $W$-boson exchange and top exchange.

For the radion interactions with the massless gauge bosons we take the results from Ref.~\cite{Megias:2015ory}. The Lagrangian relevant for photons is given by
\be
\mathcal L_{\mathcal{R}\gamma\gamma}=-\frac{R(x)}{2}F_{\mu\nu}^2(x)\frac{\int_0^{r_1}dr e^{2A}}{r_1} \,,
\ee
and similarly for gluons. For our aim it is convenient to re-express such Lagrangians in terms of the canonically normalized radion $\mathcal R$. We find
\begin{align}
\mathcal L_{\mathcal{R}\gamma\gamma}&=\frac{\alpha}{8\pi}\left[\sum_f N_c Q_f^2 A_{1/2}(\tau_f)+A_1(\tau_W)\right]c_\gamma \frac{\mathcal R(x)}{v}F_{\mu\nu}F^{\mu\nu}  \,, \\
\mathcal L_{\mathcal{R}gg}&=\frac{\alpha_s}{16\pi}\left[\sum_Q A_{1/2}(\tau_Q)\right]c_g \frac{\mathcal R(x)}{v}\tr G_{\mu\nu}G^{\mu\nu} \,,
\end{align}
where $c_\gamma$ ($c_g$) measures the departure of the $\gamma\gamma$ ($gg$) coupling from the value that the hypothetical SM Higgs $H$ has when $m_H=m_{\rm rad}$. If the radion had couplings exactly equal to those of the SM Higgs, then $c_\gamma$ and $c_g$ would be equal to one, but  in  general they are given by
\begin{align}
&c_\gamma=-\frac{4\pi}{\alpha\sqrt{6}\left[\sum_f N_c Q_f^2A_{1/2}(\tau_f)+A_1(\tau_W)\right]}\frac{v}{kr_1e^{-A_1}M_{P}}\left[k^2\int dr\, e^{-2A}\int dr\, e^{2A-2A_1}\right]^{1/2}   \,,  \nonumber\\
&c_g=-\frac{8\pi}{\alpha_{s}\sqrt{6}\left[\sum_QA_{1/2}(\tau_Q)\right]}\frac{v}{kr_1e^{-A_1}M_{P}}\left[k^2\int dr\, e^{-2A}\int dr\, e^{2A-2A_1}  \right]^{1/2} \,.
\end{align}
Tab.~\ref{tab:tablecouplings} reports the numerical results of $c_\gamma$ and $c_g$ arising in the benchmark scenarios B$_2$, B$_8$, C$_1$, C$_2$, D$_1$ and E$_1$ introduced in Tab.~\ref{tab:table}.
\begin{table}[htb]
\centering
\begin{tabular}{||c|c|c|c|c|c|c|c||}
\hline\hline
Scen.  &  $m_{\rm rad}$/TeV& $m_{G}$/TeV& $c_\gamma$&$c_g$&$c_V$ &$c_\mathcal H$& $c_f$\\ \hline
B$_2$   & 0.915   & 4.80 & 0.472  & 0.164 & 0.0649  & 0.259  & 0.259 \\
B$_8$  & 0.745 & 4.19 & 0.542 &0.146 & $0.0744$ & $0.298$ &  0.298 \\ \hline
C$_1$  & 0.890 & 3.08  & 0.532 &0.179 & $0.0904$  & $0.362$ &  0.362 \\
C$_2$   &  0.751 & 2.77 & 0.595 & 0.162 & $0.101$ & $0.404$ &  0.404 \\   \hline 
D$_1$  & 0.477 & 4.50 & 3.791 & 0.475 & $0.397$ & $1.586$ & 1.586 \\ \hline
E$_1$  & 0.643 & 4.16  & 0.562 & 0.124 & $0.0746$ & $0.298$ & 0.298 \\
\hline\hline
\end{tabular}
\caption{\it  Masses of the radion and the $n=1$ graviton mode, and coupling coefficients of the radion interactions with the SM fields, for the scenario B$_2$, B$_8$, C$_1$, C$_2$, D$_1$ and E$_1$.}
\label{tab:tablecouplings}
\end{table}

\subsubsection*{Couplings to fermions}
After canonically normalizing the fermions, the fermion  masses are given by
\be
m_f=\frac{\ell h_f v}{\sqrt{2}}\frac{e^{(1-c_{f_L}-c_{f_R})A_1}}{2\left[\int e^{(1-2c_{f_L})A} \int e^{(1-2c_{f_R})A}  \right]^{1/2}} \,,
\ee
and their couplings to the radion are manifest in the Lagrangian interaction
\begin{equation}
\mathcal L_{r\bar ff}=-\frac{\mathcal R(x)}{v}c_f m_f \bar f f   \label{eq:Lf}\,,
\end{equation}
with 
\be
c_f=\sqrt{\frac{8}{3}}\left( \frac{\int e^{-2A}}{\int e^{2(A-A_1)}} \right)^{1/2}\, \frac{v}{e^{-A_1}M_P} \,.
\ee
As before, the coupling coefficient $c_f$ would be equal to one for a radion coupled to fermions exactly like the SM Higgs.

The coefficient $c_f$ is universal, i.e.~equal for all fermions.  However the full radion couplings to fermions depend on the fermion masses, as it happens for the Higgs; see Eq.~(\ref{eq:Lf}). The values of $c_f$ in the considered benchmark scenarios are listed in Tab.~\ref{tab:tablecouplings}.

\subsubsection*{Couplings to massive gauge bosons}

In the Lagrangian involving the radion interactions with the massive gauge bosons, the couplings can be again normalized as
\be
\mathcal L_{\mathcal{R}VV}=-\frac{\mathcal R(x)}{v}\left\{2 c_W\,m_W^2 W_\mu W^\mu+c_Z\,m_Z^2 Z_\mu Z^\mu \right\} \,,  \label{eq:LV}
\ee
with
\begin{equation}
c_V=c_W=c_Z=\frac{1}{4}c_f \,.
\end{equation}
Were these couplings of the same size of those of the SM Higgs, we would have obtained $c_W=c_Z=1$.
The values of the coefficients $c_W$ and $c_Z$ in our selected scenarios are shown in Tab.~\ref{tab:tablecouplings}. 

\subsubsection*{Coupling to the Higgs boson}

The coupling of the radion to Higgs bosons can be deduced from the interaction
\be
\mathcal L_{\mathcal{R} \mathcal H\mathcal H}=-\frac{\mathcal R(x)}{v}c_\mathcal H\frac{1}{2}m_\mathcal H^2 \mathcal H^2 \,.
\ee
The interaction would have the same size of the SM trilinear interaction for $c_\mathcal H=1$. For a generic radion it instead results
\be
c_\mathcal H=c_f\, .
\ee
The numerical values of $c_\mathcal H$ for the considered models are exhibited in Tab.~\ref{tab:tablecouplings}. 

\subsection{LHC constraints on the radion signal strengths}
The production cross section  and decays of the radion at the LHC can be calculated by manipulating the results on the productions and decays of a (heavy) SM Higgs. We concentrate on the scenarios B$_2$, B$_8$ and D$_1$ since they well represent the collider phenomenology of our scenarios.
 
\subsubsection*{Radion production}
At the LHC we can produce the heavy radion by the following main production mechanisms:
\begin{itemize}
\item
\textit{\underline{Gluon fusion}}, with a cross-section $\sigma^{ggF}(gg\to \mathcal R)$ related to the corresponding heavy SM Higgs prediction $\sigma^{ggF}_{SM}(gg\to H)$ by
\be
\sigma^{ggF}_\mathcal R\equiv\sigma^{ggF}(gg\to \mathcal R)\simeq |c_g|^2\sigma^{ggF}_{SM}(gg\to H) \,,
\ee
assuming $m_H=m_{\rm rad}$.
Taking $\sigma^{ggF}_{SM}(gg\to H)$ for $m_H=$(0.915, 0.745, 0.477)\,TeV at $\sqrt{s}=13$ TeV~\cite{Dittmaier:2011ti}, we get $\sigma^{ggF}_{SM}(gg\to H)\simeq $ (0.219, 0.685, 5.62)\,pb in B$_2$, B$_8$  and D$_1$, respectively. Using the values of $c_g$ from Tab.~\ref{tab:tablecouplings} we then obtain
\be
\sigma^{ggF}_ \mathcal R\simeq (5.88,\  14.6,\  1270) \textrm{ fb} \,
\ee
in the three considered benchmark scenarios.

\item
\textit{\underline{Vector-boson fusion}}, with a cross-section $\sigma^{VBF}(VV\to \mathcal R)$ related to $\sigma^{VBF}_{SM}(VV\to H)$ by
\be
\sigma^{VBF}_\mathcal R\equiv\sigma^{VBF}(VV\to \mathcal R)\simeq |c_V|^2\sigma^{VBF}_{SM}(VV\to H) \,,
\ee
provided $m_H=m_{\rm rad}$.
For a Higgs as heavy as the radion in B$_2$, B$_8$ or D$_1$, Ref.~\cite{Dittmaier:2011ti} provides $\sigma^{VBF}_{SM}(VV\to H)\simeq$ (0.141, 0.220, 0.546)~pb. From the values of $c_V$ in Tab.~\ref{tab:tablecouplings} we hence obtain 
\be
\sigma^{VBF}_\mathcal R\simeq (0.59,\ 1.22 ,\  86)  \textrm{ fb} \ .
\ee
\end{itemize}

Likewise there exists the \textit{{associated production with $V$}}, $\sigma(pp\to V^\ast\to \mathcal R V)$, which is proportional to $|c_V|^2$, and the \textit{{associated production with $t\bar t$}}, $\sigma(gg\to t\bar t \mathcal R)$. However they are tiny at the considered values of the radion mass so that they can be neglected as compared to the aforementioned production processes.  In conclusion our benchmark scenarios highlight that at the LHC the TeV-scale radion is mainly produced via gluon fusion, and to some extent via vector-boson fusion. 

\subsubsection*{Radion decay}
The radion decays, mimicking the (heavy) SM Higgs, have the partial widths
\be
\Gamma(\mathcal R \to X\bar X)\simeq |c_X|^2 \Gamma_{SM}(H\to X\bar X)\,,
\ee
with $X=\gamma,g,W,Z,f$.
On top of these channels, the radion can also decay into a pair of 125-GeV Higgses with partial width
\be
\Gamma(\mathcal R\to \mathcal H\mathcal H)=\frac{|c_\mathcal H|^2}{16\pi}\frac{m_\mathcal H^4}{v^2\, m_{r}}\sqrt{1-\frac{4m_\mathcal H^2}{m_{r}^2}} \,,
\ee
from which it turns out that the radion branching fraction into an $X$ pair is
\be
\mathcal B^{\, \mathcal R}_{XX}\simeq \frac{|c_X|^2\Gamma_{SM}(H\to X\bar X)}{\Gamma (\mathcal R \to \mathcal{HH})+\sum _Y|c_Y|^2\Gamma_{SM}(H\to Y\bar Y)} \,,
\ee
with $Y=\gamma,g,W,Z,f$. 
The numerical values of the radion partial widths and branching ratios in scenarios B$_1$, B$_8$ and D$_1$ are quoted in Tabs.~\ref{tab:Gammas} and~\ref{tab:BRs}. As we can see, at the TeV scale the radion mainly decays into $WW$, $ZZ$ and $t\bar t$.

From these results we observe that the radion total width is $\Gamma_{\mathcal R}\simeq $ (4.51, 3.86, 35.7) GeV in B$_1$, B$_8$ and D$_1$, respectively. The radion is therefore a narrow resonance since in these three scenarios it turns out that
\be
\frac{\Gamma_{\mathcal R}}{m_{r}}\simeq (4.9, 5.2, 75) \times 10^{-3}\, .
\ee

\begin{table}[htb]
\centering
\begin{tabular}{||c||c|c|c|c|c|c|c||}
\hline\hline
Scen.& $\Gamma_{\mathcal R\to WW}$ & $\Gamma_{\mathcal R\to ZZ}$ & $\Gamma_{\mathcal R\to hh}$& $\Gamma_{\mathcal R\to t\bar t}$& $\Gamma_{\mathcal R\to b\bar b}$& $\Gamma_{\mathcal R\to \tau\bar \tau}$& $\Gamma_{\mathcal R\to \gamma\gamma}$\\ \hline

B$_2$ & 1220  & 610   & 5.70   & 2670  & 0.825  & 0.129  &  0.0385    \\
\hline
B$_8$&  786   & 389   &  9.01   & 2680 &  0.917 &  0.138  & 0.0143    \\ \hline
D$_1$&4960 &2350 & 362    & 28000   &17.73   &2.49   & 0.378    \\
\hline\hline
\end{tabular}
\caption{\it Partial widths of the radion in the scenarios B$_2$, B$_8$ and D$_1$. All widths are in MeV units. }
\label{tab:Gammas}
\end{table}

\begin{table}[htb]
\centering
\begin{tabular}{||c||c|c|c|c|c|c|c||}
\hline\hline
Scen.& $\mathcal B^{\mathcal R}_{WW}$ & $\mathcal B^{\mathcal R}_{ZZ}$ & $\mathcal B^{\mathcal R}_{hh}$& $\mathcal B^{\mathcal R}_{t\bar t}$& $\mathcal B^{\mathcal R}_{b\bar b}$& $\mathcal B^{\mathcal R}_{\tau\bar \tau}$& $\mathcal B^{\mathcal R}_{\gamma\gamma}$\\ \hline

B$_2$ & 0.271  &  0.135   & $1.26 \cdot 10^{-3}$   & 0.592  & $1.83 \cdot 10^{-4}$  &  $2.85 \cdot 10^{-5}$  &  $8.55 \cdot 10^{-6}$   \\
\hline
B$_8$&  0.203  & 0.101    & $2.33 \cdot 10^{-3}$    & 0.693 & $2.37 \cdot 10^{-4}$   & $3.58 \cdot 10^{-5}$  & $3.70 \cdot 10^{-6}$     \\ \hline
D$_1$& 0.139 & $6.58 \cdot 10^{-2}$ & $1.01 \cdot 10^{-2}$  & $0.785$  & $4.97 \cdot 10^{-4}$   & $6.99 \cdot 10^{-5}$  & $1.06 \cdot 10^{-5}$   \\
\hline\hline
\end{tabular}
\caption{\it The radion branching fractions in the scenarios B$_2$, B$_8$ and D$_1$. }
\label{tab:BRs}
\end{table}

\subsubsection*{Experimental bounds}

\begin{table}[htb]
\centering
\hspace{-.3cm}
\begin{tabular}{||c||c|c|c|c|c|c|c||}
\hline\hline
Scen.& $\mathcal S^{ggF}_{WW}$ & $\mathcal S^{ggF}_{ZZ}$ &  $\mathcal S^{ggF}_{\tau\bar \tau} $ & $\mathcal S^{ggF}_{\gamma\gamma}$+$\mathcal S^{VBF}_{\gamma\gamma}$ &  $\mathcal S^{VBF}_{WW}$ & $\mathcal S^{VBF}_{ZZ}$ &  $\mathcal S^{VBF}_{\tau\bar \tau} $
\\ \hline

B$_2$ (predic.) & 1.59   &  0.80   & $1.7 \cdot 10^{-4}$   &  $(5.0 + 0.5)\cdot 10^{-5}$  & 0.16  & 0.080  &$1.7 \cdot 10^{-5}$  \\
\hline
B$_2$ (bound) &  52 & 14   & 11   & 0.29  & 12  & 8  &     -- \\
\hline
B$_8$ (predic.) & 2.96   & 1.47   & $5.2 \cdot 10^{-4}$  & $(5.4 + 0.5) \cdot 10^{-5}$  & 0.25  & 0.12  & $4.4 \cdot 10^{-5}$     \\
\hline
B$_8$ (bound) & 91  &  42      & 20  & 0.34  &  19 & 19  & --\\
\hline
D$_1$ (predic.) & 176 & 83 & 0.09 & 0.013+0.001  & 12   & 6   & 0.006   \\ 
\hline
D$_1$ (bound) & 1100 & 300 & 90  & 2  & 200   & 130   & --  \\ \hline
\hline\hline
\end{tabular}
\caption{\it The predictions of $S^{ggF(VBF)}_{XX}$  and their corresponding 95\% C.L.~upper bounds in the scenarios B$_2$, B$_8$ and D$_1$. The bound on the $\gamma \gamma$ channel does not distinguish between gluon and vector fusion production and then has to be compared to the sum of the two processes. No specific bound on $S^{VBF}_{\tau \bar \tau}$ is considered. All quantities are in fb units. The bounds are taken from Refs.~\cite{Aaboud:2017fgj, Aaboud:2017itg, Aaboud:2017yyg, Aaboud:2017sjh}}.
\label{tab:crossdecay}
\end{table}

Since the radion is a narrow resonance, the cross section $\mathcal S^{ggF(VBF)}_{XX} \equiv \sigma^{ggF(VBF)}(pp \to \mathcal{R} \to XX)$ can be calculated as 
\begin{equation}
\mathcal S_{XX}^{ggF(VBF)} = \sigma_\mathcal{R}^{ggF(VBF)}~\mathcal B^{\, \mathcal R}_{XX} \,.
\end{equation}
To determine whether such collider features are experimentally allowed, we consider the ATLAS searches of Refs.~\cite{Aaboud:2017fgj, Aaboud:2017itg, Aaboud:2017yyg, Aaboud:2017sjh}  constraining the $WW$, $ZZ$, $\tau \tau$ and $\gamma\gamma$  channels~\footnote{The equivalent CMS searches (see e.g.~Ref.~\cite{Khachatryan:2016yec}) tend to provide weaker bounds and therefore we do not take them into account. On the other hand, since we eventually find that our scenarios are well within the current limits, we do not expect our conclusions to depend on the particular analyses we consider.}. 
These furnish 95\% C.L.~bounds on $\mathcal S_{WW}^{ggF}$, $\mathcal S_{ZZ}^{ggF}$, $\mathcal S_{\tau \bar \tau}^{ggF}$, $\mathcal S_{\tau \bar \tau}^{VBF}$, $\mathcal S_{WW}^{VBF}$, $\mathcal S_{ZZ}^{VBF}$ and  $\mathcal S_{\gamma\gamma}^{ggF}+ \mathcal S_{\gamma\gamma}^{VBF}$  as functions of the scalar mass.  Tab.~\ref{tab:crossdecay} reports the pertinent limits  and the respective predictions of $\mathcal S_{XX}^{ggF}$ and $\mathcal S_{XX}^{VBF}$ in each of the considered scenarios. Notice that the constraint on the $\gamma\gamma$ channel does not distinguish between the gluon and the vector-boson fusion productions, and for this reason it has to be compared with the sum of the two production processes.

We conclude that the scenarios B$_2$, B$_8$ and D$_1$ are in full agreement with the current bounds~\footnote{In principle also the searches for the SM-like and graviton KK modes might be relevant. Under some model assumptions, the bounds in Ref.~\cite{Aaboud:2018mjh}, for instance, require the KK gluons to be above 4\,TeV, approximatively,  and thus should not be in tension with most of our scenarios. Moreover such bounds are extremely model dependent and can thus be circumvented by adjusting our particle setups. For instance, assuming the first and second generation of quarks localized towards the UV brane could relax the bounds from Drell-Yan production, as the KK modes are extremely localized towards the IR brane, without major changing on our main results.} 
and, given the values collected in Tabs.~\ref{tab:Gammas} and \ref{tab:BRs}, we expect the same conclusion to hold for all previously investigated benchmark configurations, as D$_1$ is the scenario with the smallest radion mass and largest coupling coefficients. In particular, among scenarios B$_2$, B$_8$ and D$_1$, only D$_1$ has some channels (i.e. the $ZZ$ and $WW$ ones) that are not far below the experimental constraints. It then results that, at least for the parameter regions our benchmark points represent,  future LHC data, with much larger integrated luminosity, will be able to probe some of the decay channels here investigated, but likely only future colliders~\cite{Golling:2016gvc} will be capable of discovering the soft-wall radion, or putting strong constraints on the model. This will probably happen in conjunction with the LISA and ET measurements, given the time schedule of future collider and GW facilities. Of course such conclusion might be not generic, as it is potentially biased by the limited number of benchmark points we have investigated. To clarify this point we should extend the above procedure to a much larger set of parameter points, an analysis that we postpone to a future publication.

\section{Conclusions}
\label{sec:conclusions}

The hierarchy problem has motivated several ultraviolet completions of the Standard Model. Among these, the frameworks of warped extra dimensions have gained popularity in the last decade. The interest in these frameworks is two-fold: on the one side, they may be the correct description of nature if the latter has a five dimensional spacetime; on the other, they may be a useful tool for understanding a strongly-coupled sector in a four dimensional nature.
The most investigated warped model is the Randall-Sundrum one, followed by scenarios where the metric is less trivial, which can show phenomenological advantages related to the description of precision electroweak observables. In the present paper we have explored technical challenges and phenomenological issues of one of these setups, the soft-wall models, with special emphasis on the so-called holographic phase transition.

Concerning the technical achievements, we have extended the application of the superpotential formalism to configurations where the mechanism stabilizing the extra dimension can have a strong back-reaction on the metric.
This formal result is remarkable because, in principle, it can be applied to any warped model, with clear advantages on the parameter space that can be investigated without losing control on the back-reaction effects. (We remind that the correct treatment of the back-reaction has strongly limited the parameter space that some of the previous studies could explore~\cite{Creminelli:2001th,Randall:2006py,Kaplan:2006yi,Nardini:2007me}).  

As a concrete application, we have applied the proposed formalism to the soft-wall model, where the potential in the bulk behaves exponentially near the IR brane. The radion phase transition is controlled, on the one hand by the free energy in the confined phase (i.e.~essentially the depth of the effective potential at its minimum), and on the other hand by the free energy in the deconfined phase. Concerning the confined phase, the depth of the effective potential is essentially controlled by the radion mass, which in turn is controlled by the amount of back-reaction on the gravitational metric~\footnote{In the extreme case of no back-reaction, the radion potential is flat, consequently the radion is massless and there is no phase transition. We have exemplified such situation in scenario A$_1$ above.}. The heavier the radion, the deeper (and steeper) the effective potential, the smaller the Euclidean action (as the Euclidean "time" to get to the true minimum is shorter), and consequently the higher the nucleation temperature. In this way, for the cases of large back-reaction considered in this paper, there is no supercooling and the nucleation temperature is usually above the electroweak temperature. On the other hand the free energy in the deconfined phase, as given by Eq.~(\ref{eq:freeminapp}), depends on the factor $a_h(T)$ which, for the cases of small back-reaction is $a_h(T)\simeq 1$, while for the cases of large back-reaction is $a_h(T)\ll 1$, a fact which makes easier the phase transition and increases the nucleation temperature by a factor $\mathcal O(a_h^{-1/4})$.
Therefore the nucleation temperature is not necessarily much below the electroweak scale, which implies that  the SM-like particles in the plasma are not Boltzmann suppressed during the bubble expansion and collision. Clearly, the presence of this rich plasma could have relevant effects on the dynamics of the conformal symmetry breaking and, in turn, on the phenomenology of the model.

Our method has allowed to determine the radion potential even in the regime of large t'Hooft coupling, e.g.~as large as $N\simeq 25$, when the back-reaction goes away if all the other parameters are fixed, and the phase transition meets more difficulties to happen. The reason the phase transition can take place in those cases is because we can still compensate the sizable back-reaction by changing the values of the other parameters, in particular the values of the field $\phi$ at the UV and IR branes, $v_0$ and $v_1$. However we have found that the radion mass increases parametrically in the cases with large values of $N$, the low energy effective theory describing the SM degrees of freedom and the radion field should not be trustable, and one instead should consider the whole set of 5D Kaluza-Klein modes in the thermal plasma, a task outside the scope of the present paper.

In summary, in the class of models we consider in this paper, where conformality is strongly broken in the IR brane, and we can keep track of the back-reaction, the nucleation temperature is higher than the electroweak temperature and thus the dilaton phase transition naturally occurs (sequentially) before (at higher temperatures than) the electroweak phase transition~\footnote{This makes a difference with respect to the class of models presented in Ref.~\cite{vonHarling:2017yew} where conformality is only weakly broken and both phase transitions occur simultaneously.}. However (less natural) solutions where both phase transitions are simultaneous can be implemented in our class of models only at the price of decreasing the value of $N$ (see the benchmark scenario D$_1$ above). 
In all the cases there is no supercooling in the deconfined phase and the amount of inflation which takes place is very marginal and does not affect at all the dynamics of the phase transition.

Together with other quantities, the reheating temperature plays a key role in the signatures of the radion phase transition. In most of the considered benchmark scenarios (cf.~scenarios of classes B, C and E) the reheating temperature is much above $150\,$GeV; thus the electroweak phase transition is subsequent to the holographic one and resembles the one of the SM. On the contrary, when this does not happen (see~e.g.~benchmark scenario D$_1$) and the Higgs is localized at the infrared brane, the electroweak phase transition turns out to be supercooled and then of first order. Electroweak baryogenesis in soft-wall models thus looks possible, although further studies would be required to better understand this issue.

We also have investigated the detectability prospects of the model at the forthcoming gravitational wave observatories. Present and future pulsar time array experiments, and the current generation of ground based interferometers, are not sensitive to the stochastic gravitational signals our benchmark scenarios lead to. The LISA and ET interferometers can instead measure all of them with a signal to noise ratio of about 10 or larger (assuming the absence of further astrophysical~\cite{Abbott:2017xzg} or cosmological~\cite{Caprini:2018mtu} sources). Simultaneous detection of the signal at both experiments is possible due to the broadband of the predicted power spectrum and the large signal amplitude that emerges when the bubbles are supersonic. Curiously,  in a corner of parameter space, the signal is so powerful that the big bang nucleosynthesis constraint rules it out. On the other hand, for subsonic bubble velocities ($v_w\gtrsim 0.02$), which are those favored by electroweak baryogenesis, the signal is weaker and redshifted, and only LISA can detect (most of) the benchmark scenarios.

We have moreover noticed that in the large-back-reaction regime the soft-wall scenario tends to provide a radion mass that is only slightly suppressed with respect to the radion vacuum  expectation value.  For this reason, once such a vacuum expectation value is fixed at the electroweak scale to alleviate  the hierarchy problem, the radion mass is not necessarily of the order of the electroweak scale or below. Thanks to this feature, the radion is not in tension with present LHC searches. Its observation may be however feasible at the future LHC runs.

In conclusion, by means of the aforementioned superpotential formalism,  we have determined some interesting features of the soft-wall models in the presence of large back-reaction. A heavy radion and a large nucleation temperature look to be the main smoking guns. Whereas measuring the former at colliders would be suitable by standard techniques, inferring the latter would need improvements in the prediction and detection of stochastic gravitational wave backgrounds. In fact, based on the envelope approximation we have followed, two phase transitions having the same reheating temperature but different nucleation temperatures would provide the same stochastic signal. Only going beyond the envelope approximation, and having well under control the plasma effects during the phase transition, would allow to disentangle scenarios with tiny nucleation temperature --- where most of the SM degrees of freedom are Boltzmann suppressed, as it typically happens in the Randall-Sundrum model --- from those with large nucleation temperature. More detailed theoretical predictions, as well as more refined phase transition simulations, are thus required in order to break this degeneracy. We look forward to knowing them in order to understand how to possibly disentangle a given warped framework from another one.  

\vspace{0.5cm}
\noindent\emph{Note added:} Before submission, the LISA CosWG preview of this paper unveiled the existence of Refs.~\cite{Axen:2018zvb, TalkPedro}, which partially overlap with Section~\ref{sec:GW}, but all done independently of this paper.

\vspace{0.5cm}
\section*{Acknowledgments}
We would like to thank Geraldine Servant for useful comments on the draft of this paper, prior to publication, concerning the main features of the confined/deconfined and electroweak phase transitions in different approaches.
MQ thanks the ICTP-South American Institute for Fundamental Research and the Instituto de F\'isica Te\'orica, Universidade Estadual Paulista, S\~ao Paulo, Brazil, where part of this work has been done, for hospitality. He also acknowledges the Albert Einstein Institute of the University of Bern for hosting him during the last stage of this work. GN is grateful to the LISA Cosmology Working Group for important discussions.  
The work of EM is supported by the Spanish MINEICO under Grant FPA2015-64041-C2-1-P and FIS2017-85053-C2-1-P, by the Junta de Andaluc\'{\i}a under Grant FQM-225,  by the Basque Government under Grant IT979-16, and by the Spanish Consolider Ingenio 2010 Programme CPAN (CSD2007-00042). The research of EM is also supported by the Ram\'on y Cajal Program of the Spanish MINEICO, and by the Universidad del Pa\'{\i}s Vasco UPV/EHU, Bilbao, Spain, as a Visiting Professor. GN is supported by the Swiss National Science
Foundation (SNF) under grant 200020-168988. 
 The work of MQ is partly supported by Spanish MINEICO
under Grant CICYT-FEDER-FPA2014-55613-P and FPA2017-88915-P, by the Severo Ochoa Excellence Program of MINEICO under  Grant  SEV-2016-0588, and by CNPq PVE fellowship project 405559/2013-5.

\bibliographystyle{JHEP}
\bibliography{refs}

\providecommand{\href}[2]{#2}\begingroup\raggedright\begin{thebibliography}{100}

\bibitem{ALEPH:2005ab}
{\scshape SLD Electroweak Group, DELPHI, ALEPH, SLD, SLD Heavy Flavour Group,
  OPAL, LEP Electroweak Working Group, L3} collaboration, S.~Schael et~al.,
  \emph{{Precision electroweak measurements on the $Z$ resonance}},
  \href{http://dx.doi.org/10.1016/j.physrep.2005.12.006}{\emph{Phys. Rept.}
  {\bfseries 427} (2006) 257--454},
  [\href{https://arxiv.org/abs/hep-ex/0509008}{{\ttfamily hep-ex/0509008}}].

\bibitem{Olive:2016xmw}
{\scshape Particle Data Group} collaboration, C.~Patrignani et~al.,
  \emph{{Review of Particle Physics}},
  \href{http://dx.doi.org/10.1088/1674-1137/40/10/100001}{\emph{Chin. Phys.}
  {\bfseries C40} (2016) 100001}.

\bibitem{Randall:1999ee}
L.~Randall and R.~Sundrum, \emph{{A Large mass hierarchy from a small extra
  dimension}}, \href{http://dx.doi.org/10.1103/PhysRevLett.83.3370}{\emph{Phys.
  Rev. Lett.} {\bfseries 83} (1999) 3370--3373},
  [\href{https://arxiv.org/abs/hep-ph/9905221}{{\ttfamily hep-ph/9905221}}].

\bibitem{Goldberger:1999uk}
W.~D. Goldberger and M.~B. Wise, \emph{{Modulus stabilization with bulk
  fields}}, \href{http://dx.doi.org/10.1103/PhysRevLett.83.4922}{\emph{Phys.
  Rev. Lett.} {\bfseries 83} (1999) 4922--4925},
  [\href{https://arxiv.org/abs/hep-ph/9907447}{{\ttfamily hep-ph/9907447}}].

\bibitem{Creminelli:2001th}
P.~Creminelli, A.~Nicolis and R.~Rattazzi, \emph{{Holography and the
  electroweak phase transition}},
  \href{http://dx.doi.org/10.1088/1126-6708/2002/03/051}{\emph{JHEP} {\bfseries
  03} (2002) 051}, [\href{https://arxiv.org/abs/hep-th/0107141}{{\ttfamily
  hep-th/0107141}}].

\bibitem{Randall:2006py}
L.~Randall and G.~Servant, \emph{{Gravitational waves from warped spacetime}},
  \href{http://dx.doi.org/10.1088/1126-6708/2007/05/054}{\emph{JHEP} {\bfseries
  05} (2007) 054}, [\href{https://arxiv.org/abs/hep-ph/0607158}{{\ttfamily
  hep-ph/0607158}}].

\bibitem{Kaplan:2006yi}
J.~Kaplan, P.~C. Schuster and N.~Toro, \emph{{Avoiding an Empty Universe in RS
  I Models and Large-N Gauge Theories}},
  \href{https://arxiv.org/abs/hep-ph/0609012}{{\ttfamily hep-ph/0609012}}.

\bibitem{Nardini:2007me}
G.~Nardini, M.~Quiros and A.~Wulzer, \emph{{A Confining Strong First-Order
  Electroweak Phase Transition}},
  \href{http://dx.doi.org/10.1088/1126-6708/2007/09/077}{\emph{JHEP} {\bfseries
  09} (2007) 077}, [\href{https://arxiv.org/abs/0706.3388}{{\ttfamily
  0706.3388}}].

\bibitem{Hassanain:2007js}
B.~Hassanain, J.~March-Russell and M.~Schvellinger, \emph{{Warped Deformed
  Throats have Faster (Electroweak) Phase Transitions}},
  \href{http://dx.doi.org/10.1088/1126-6708/2007/10/089}{\emph{JHEP} {\bfseries
  10} (2007) 089}, [\href{https://arxiv.org/abs/0708.2060}{{\ttfamily
  0708.2060}}].

\bibitem{Konstandin:2010cd}
T.~Konstandin, G.~Nardini and M.~Quiros, \emph{{Gravitational Backreaction
  Effects on the Holographic Phase Transition}},
  \href{http://dx.doi.org/10.1103/PhysRevD.82.083513}{\emph{Phys. Rev.}
  {\bfseries D82} (2010) 083513},
  [\href{https://arxiv.org/abs/1007.1468}{{\ttfamily 1007.1468}}].

\bibitem{Bunk:2017fic}
D.~Bunk, J.~Hubisz and B.~Jain, \emph{{A Perturbative RS I Cosmological Phase
  Transition}},
  \href{http://dx.doi.org/10.1140/epjc/s10052-018-5529-2}{\emph{Eur. Phys. J.}
  {\bfseries C78} (2018) 78},
  [\href{https://arxiv.org/abs/1705.00001}{{\ttfamily 1705.00001}}].

\bibitem{Dillon:2017ctw}
B.~M. Dillon, B.~K. El-Menoufi, S.~J. Huber and J.~P. Manuel, \emph{{A rapid
  holographic phase transition with brane-localized curvature}},
  \href{https://arxiv.org/abs/1708.02953}{{\ttfamily 1708.02953}}.

\bibitem{vonHarling:2017yew}
B.~von Harling and G.~Servant, \emph{{QCD-induced Electroweak Phase
  Transition}}, \href{http://dx.doi.org/10.1007/JHEP01(2018)159}{\emph{JHEP}
  {\bfseries 01} (2018) 159},
  [\href{https://arxiv.org/abs/1711.11554}{{\ttfamily 1711.11554}}].

\bibitem{DeWolfe:1999cp}
O.~DeWolfe, D.~Z. Freedman, S.~S. Gubser and A.~Karch, \emph{{Modeling the
  fifth-dimension with scalars and gravity}},
  \href{http://dx.doi.org/10.1103/PhysRevD.62.046008}{\emph{Phys. Rev.}
  {\bfseries D62} (2000) 046008},
  [\href{https://arxiv.org/abs/hep-th/9909134}{{\ttfamily hep-th/9909134}}].

\bibitem{Cabrer:2009we}
J.~A. Cabrer, G.~von Gersdorff and M.~Quiros, \emph{{Soft-Wall Stabilization}},
  \href{http://dx.doi.org/10.1088/1367-2630/12/7/075012}{\emph{New J. Phys.}
  {\bfseries 12} (2010) 075012},
  [\href{https://arxiv.org/abs/0907.5361}{{\ttfamily 0907.5361}}].

\bibitem{Cabrer:2010si}
J.~A. Cabrer, G.~von Gersdorff and M.~Quiros, \emph{{Warped Electroweak
  Breaking Without Custodial Symmetry}},
  \href{http://dx.doi.org/10.1016/j.physletb.2011.01.058}{\emph{Phys. Lett.}
  {\bfseries B697} (2011) 208--214},
  [\href{https://arxiv.org/abs/1011.2205}{{\ttfamily 1011.2205}}].

\bibitem{Cabrer:2011fb}
J.~A. Cabrer, G.~von Gersdorff and M.~Quiros, \emph{{Suppressing Electroweak
  Precision Observables in 5D Warped Models}},
  \href{http://dx.doi.org/10.1007/JHEP05(2011)083}{\emph{JHEP} {\bfseries 05}
  (2011) 083}, [\href{https://arxiv.org/abs/1103.1388}{{\ttfamily 1103.1388}}].

\bibitem{Cabrer:2011vu}
J.~A. Cabrer, G.~von Gersdorff and M.~Quiros, \emph{{Improving Naturalness in
  Warped Models with a Heavy Bulk Higgs Boson}},
  \href{http://dx.doi.org/10.1103/PhysRevD.84.035024}{\emph{Phys. Rev.}
  {\bfseries D84} (2011) 035024},
  [\href{https://arxiv.org/abs/1104.3149}{{\ttfamily 1104.3149}}].

\bibitem{Cabrer:2011mw}
J.~A. Cabrer, G.~von Gersdorff and M.~Quiros, \emph{{Warped 5D Standard Model
  Consistent with EWPT}},
  \href{http://dx.doi.org/10.1002/prop.201100054}{\emph{Fortsch. Phys.}
  {\bfseries 59} (2011) 1135--1138},
  [\href{https://arxiv.org/abs/1104.5253}{{\ttfamily 1104.5253}}].

\bibitem{Carmona:2011ib}
A.~Carmona, E.~Ponton and J.~Santiago, \emph{{Phenomenology of Non-Custodial
  Warped Models}}, \href{http://dx.doi.org/10.1007/JHEP10(2011)137}{\emph{JHEP}
  {\bfseries 10} (2011) 137},
  [\href{https://arxiv.org/abs/1107.1500}{{\ttfamily 1107.1500}}].

\bibitem{Cabrer:2011qb}
J.~A. Cabrer, G.~von Gersdorff and M.~Quiros, \emph{{Flavor Phenomenology in
  General 5D Warped Spaces}},
  \href{http://dx.doi.org/10.1007/JHEP01(2012)033}{\emph{JHEP} {\bfseries 01}
  (2012) 033}, [\href{https://arxiv.org/abs/1110.3324}{{\ttfamily 1110.3324}}].

\bibitem{deBlas:2012qf}
J.~de~Blas, A.~Delgado, B.~Ostdiek and A.~de~la Puente, \emph{{LHC Signals of
  Non-Custodial Warped 5D Models}},
  \href{http://dx.doi.org/10.1103/PhysRevD.86.015028}{\emph{Phys. Rev.}
  {\bfseries D86} (2012) 015028},
  [\href{https://arxiv.org/abs/1206.0699}{{\ttfamily 1206.0699}}].

\bibitem{Quiros:2013yaa}
M.~Quiros, \emph{{Higgs Bosons in Extra Dimensions}},
  \href{http://dx.doi.org/10.1142/S021773231540012X}{\emph{Mod. Phys. Lett.}
  {\bfseries A30} (2015) 1540012},
  [\href{https://arxiv.org/abs/1311.2824}{{\ttfamily 1311.2824}}].

\bibitem{Megias:2015ory}
E.~Megias, O.~Pujolas and M.~Quiros, \emph{{On dilatons and the LHC diphoton
  excess}}, \href{http://dx.doi.org/10.1007/JHEP05(2016)137}{\emph{JHEP}
  {\bfseries 05} (2016) 137},
  [\href{https://arxiv.org/abs/1512.06106}{{\ttfamily 1512.06106}}].

\bibitem{Agashe:2003zs}
K.~Agashe, A.~Delgado, M.~J. May and R.~Sundrum, \emph{{RS1, custodial isospin
  and precision tests}},
  \href{http://dx.doi.org/10.1088/1126-6708/2003/08/050}{\emph{JHEP} {\bfseries
  08} (2003) 050}, [\href{https://arxiv.org/abs/hep-ph/0308036}{{\ttfamily
  hep-ph/0308036}}].

\bibitem{Megias:2017dzd}
E.~Megias, M.~Quiros and L.~Salas, \emph{{$g_\mu-2$ from Vector-Like Leptons in
  Warped Space}}, \href{http://dx.doi.org/10.1007/JHEP05(2017)016}{\emph{JHEP}
  {\bfseries 05} (2017) 016},
  [\href{https://arxiv.org/abs/1701.05072}{{\ttfamily 1701.05072}}].

\bibitem{Megias:2016bde}
E.~Megias, G.~Panico, O.~Pujolas and M.~Quiros, \emph{{A Natural origin for the
  LHCb anomalies}},
  \href{http://dx.doi.org/10.1007/JHEP09(2016)118}{\emph{JHEP} {\bfseries 09}
  (2016) 118}, [\href{https://arxiv.org/abs/1608.02362}{{\ttfamily
  1608.02362}}].

\bibitem{Megias:2015qqh}
E.~Megias, O.~Pujolas and M.~Quiros, \emph{{On light dilaton extensions of the
  Standard Model}},
  \href{http://dx.doi.org/10.1051/epjconf/201612605010}{\emph{EPJ Web Conf.}
  {\bfseries 126} (2016) 05010},
  [\href{https://arxiv.org/abs/1512.06702}{{\ttfamily 1512.06702}}].

\bibitem{Megias:2016jcw}
E.~Megias, G.~Panico, O.~Pujolas and M.~Quiros, \emph{{Light dilatons in warped
  space: Higgs boson and LHCb anomalies}},
  \href{http://dx.doi.org/10.1016/j.nuclphysbps.2016.12.037}{\emph{Nucl. Part.
  Phys. Proc.} {\bfseries 282-284} (2017) 194--198},
  [\href{https://arxiv.org/abs/1609.01881}{{\ttfamily 1609.01881}}].

\bibitem{Megias:2017ove}
E.~Megias, M.~Quiros and L.~Salas, \emph{{Lepton-flavor universality violation
  in R$_{K}$ and $ {R}_{D^{{\left(\ast \right)}}} $ from warped space}},
  \href{http://dx.doi.org/10.1007/JHEP07(2017)102}{\emph{JHEP} {\bfseries 07}
  (2017) 102}, [\href{https://arxiv.org/abs/1703.06019}{{\ttfamily
  1703.06019}}].

\bibitem{Megias:2017isd}
E.~Megias, G.~Panico, O.~Pujolas and M.~Quiros, \emph{{A natural
  extra-dimensional origin for the LHCb anomalies}},  in \emph{{Proceedings,
  52nd Rencontres de Moriond on Electroweak Interactions and Unified Theories:
  La Thuile, Italy, March 18-25, 2017}}, pp.~225--232, 2017.
\newblock \href{https://arxiv.org/abs/1705.04822}{{\ttfamily 1705.04822}}.

\bibitem{Megias:2017vdg}
E.~Megias, M.~Quiros and L.~Salas, \emph{{Lepton-flavor universality limits in
  warped space}},
  \href{http://dx.doi.org/10.1103/PhysRevD.96.075030}{\emph{Phys. Rev.}
  {\bfseries D96} (2017) 075030},
  [\href{https://arxiv.org/abs/1707.08014}{{\ttfamily 1707.08014}}].

\bibitem{Megias:2017mll}
E.~Megias, M.~Quiros and L.~Salas, \emph{{Flavor anomalies from warped space}},
   in \emph{{20th High-Energy Physics International Conference in Quantum
  Chromodynamics (QCD 17) Montpellier, France, July 3-7, 2017}}, 2017.
\newblock \href{https://arxiv.org/abs/1709.05100}{{\ttfamily 1709.05100}}.

\bibitem{Gubser:1999vj}
S.~S. Gubser, \emph{{AdS / CFT and gravity}},
  \href{http://dx.doi.org/10.1103/PhysRevD.63.084017}{\emph{Phys. Rev.}
  {\bfseries D63} (2001) 084017},
  [\href{https://arxiv.org/abs/hep-th/9912001}{{\ttfamily hep-th/9912001}}].

\bibitem{York:1972sj}
J.~W. York, Jr., \emph{{Role of conformal three geometry in the dynamics of
  gravitation}},
  \href{http://dx.doi.org/10.1103/PhysRevLett.28.1082}{\emph{Phys. Rev. Lett.}
  {\bfseries 28} (1972) 1082--1085}.

\bibitem{Gibbons:1976ue}
G.~W. Gibbons and S.~W. Hawking, \emph{{Action Integrals and Partition
  Functions in Quantum Gravity}},
  \href{http://dx.doi.org/10.1103/PhysRevD.15.2752}{\emph{Phys. Rev.}
  {\bfseries D15} (1977) 2752--2756}.

\bibitem{Bellazzini:2013fga}
B.~Bellazzini, C.~Csaki, J.~Hubisz, J.~Serra and J.~Terning, \emph{{A Naturally
  Light Dilaton and a Small Cosmological Constant}},
  \href{http://dx.doi.org/10.1140/epjc/s10052-014-2790-x}{\emph{Eur. Phys. J.}
  {\bfseries C74} (2014) 2790},
  [\href{https://arxiv.org/abs/1305.3919}{{\ttfamily 1305.3919}}].

\bibitem{Papadimitriou:2007sj}
I.~Papadimitriou, \emph{{Multi-Trace Deformations in AdS/CFT: Exploring the
  Vacuum Structure of the Deformed CFT}},
  \href{http://dx.doi.org/10.1088/1126-6708/2007/05/075}{\emph{JHEP} {\bfseries
  05} (2007) 075}, [\href{https://arxiv.org/abs/hep-th/0703152}{{\ttfamily
  hep-th/0703152}}].

\bibitem{Megias:2014iwa}
E.~Megias and O.~Pujolas, \emph{{Naturally light dilatons from nearly marginal
  deformations}}, \href{http://dx.doi.org/10.1007/JHEP08(2014)081}{\emph{JHEP}
  {\bfseries 08} (2014) 081},
  [\href{https://arxiv.org/abs/1401.4998}{{\ttfamily 1401.4998}}].

\bibitem{Megias:2015nya}
E.~Megias, \emph{{A holographic realization of light dilatons}},
  \href{http://dx.doi.org/10.1088/1742-6596/670/1/012034}{\emph{J. Phys. Conf.
  Ser.} {\bfseries 670} (2016) 012034},
  [\href{https://arxiv.org/abs/1510.01990}{{\ttfamily 1510.01990}}].

\bibitem{Csaki:2000zn}
C.~Csaki, M.~L. Graesser and G.~D. Kribs, \emph{{Radion dynamics and
  electroweak physics}},
  \href{http://dx.doi.org/10.1103/PhysRevD.63.065002}{\emph{Phys. Rev.}
  {\bfseries D63} (2001) 065002},
  [\href{https://arxiv.org/abs/hep-th/0008151}{{\ttfamily hep-th/0008151}}].

\bibitem{Carlip:2008wv}
S.~Carlip, \emph{{Black Hole Thermodynamics and Statistical Mechanics}},
  \href{http://dx.doi.org/10.1007/978-3-540-88460-6_3}{\emph{Lect. Notes Phys.}
  {\bfseries 769} (2009) 89--123},
  [\href{https://arxiv.org/abs/0807.4520}{{\ttfamily 0807.4520}}].

\bibitem{Coleman:1977py}
S.~R. Coleman, \emph{{The Fate of the False Vacuum. 1. Semiclassical Theory}},
  \href{http://dx.doi.org/10.1103/PhysRevD.15.2929,
  10.1103/PhysRevD.16.1248}{\emph{Phys. Rev.} {\bfseries D15} (1977)
  2929--2936}.

\bibitem{Callan:1977pt}
C.~G. Callan, Jr. and S.~R. Coleman, \emph{{The Fate of the False Vacuum. 2.
  First Quantum Corrections}},
  \href{http://dx.doi.org/10.1103/PhysRevD.16.1762}{\emph{Phys. Rev.}
  {\bfseries D16} (1977) 1762--1768}.

\bibitem{Linde:1980tt}
A.~D. Linde, \emph{{Fate of the False Vacuum at Finite Temperature: Theory and
  Applications}},
  \href{http://dx.doi.org/10.1016/0370-2693(81)90281-1}{\emph{Phys. Lett.}
  {\bfseries 100B} (1981) 37--40}.

\bibitem{Quiros:1999jp}
M.~Quiros, \emph{{Finite temperature field theory and phase transitions}},  in
  \emph{{Proceedings, Summer School in High-energy physics and cosmology:
  Trieste, Italy, June 29-July 17, 1998}}, pp.~187--259, 1999.
\newblock \href{https://arxiv.org/abs/hep-ph/9901312}{{\ttfamily
  hep-ph/9901312}}.

\bibitem{Quiros:1994dr}
M.~Quiros, \emph{{Field theory at finite temperature and phase transitions}},
  {\emph{Helv. Phys. Acta} {\bfseries 67} (1994) 451--583}.

\bibitem{Kuzmin:1985mm}
V.~A. Kuzmin, V.~A. Rubakov and M.~E. Shaposhnikov, \emph{{On the Anomalous
  Electroweak Baryon Number Nonconservation in the Early Universe}},
  \href{http://dx.doi.org/10.1016/0370-2693(85)91028-7}{\emph{Phys. Lett.}
  {\bfseries 155B} (1985) 36}.

\bibitem{Kajantie:1996mn}
K.~Kajantie, M.~Laine, K.~Rummukainen and M.~E. Shaposhnikov, \emph{{Is there a
  hot electroweak phase transition at m(H) larger or equal to m(W)?}},
  \href{http://dx.doi.org/10.1103/PhysRevLett.77.2887}{\emph{Phys. Rev. Lett.}
  {\bfseries 77} (1996) 2887--2890},
  [\href{https://arxiv.org/abs/hep-ph/9605288}{{\ttfamily hep-ph/9605288}}].

\bibitem{Rummukainen:1998as}
K.~Rummukainen, M.~Tsypin, K.~Kajantie, M.~Laine and M.~E. Shaposhnikov,
  \emph{{The Universality class of the electroweak theory}},
  \href{http://dx.doi.org/10.1016/S0550-3213(98)00494-5}{\emph{Nucl. Phys.}
  {\bfseries B532} (1998) 283--314},
  [\href{https://arxiv.org/abs/hep-lat/9805013}{{\ttfamily hep-lat/9805013}}].

\bibitem{DOnofrio:2014rug}
M.~D'Onofrio, K.~Rummukainen and A.~Tranberg, \emph{{Sphaleron Rate in the
  Minimal Standard Model}},
  \href{http://dx.doi.org/10.1103/PhysRevLett.113.141602}{\emph{Phys. Rev.
  Lett.} {\bfseries 113} (2014) 141602},
  [\href{https://arxiv.org/abs/1404.3565}{{\ttfamily 1404.3565}}].

\bibitem{Bruggisser:2018mus}
S.~Bruggisser, B.~von Harling, O.~Matsedonskyi and G.~Servant, \emph{{The
  Baryon Asymmetry from a Composite Higgs}},
  \href{https://arxiv.org/abs/1803.08546}{{\ttfamily 1803.08546}}.

\bibitem{Bruggisser:2018mrt}
S.~Bruggisser, B.~von Harling, O.~Matsedonskyi and G.~Servant,
  \emph{{Electroweak Phase Transition and Baryogenesis in Composite Higgs
  Models}},  \href{https://arxiv.org/abs/1804.07314}{{\ttfamily 1804.07314}}.

\bibitem{Witten:1984rs}
E.~Witten, \emph{{Cosmic Separation of Phases}},
  \href{http://dx.doi.org/10.1103/PhysRevD.30.272}{\emph{Phys. Rev.} {\bfseries
  D30} (1984) 272--285}.

\bibitem{Kosowsky:1991ua}
A.~Kosowsky, M.~S. Turner and R.~Watkins, \emph{{Gravitational radiation from
  colliding vacuum bubbles}},
  \href{http://dx.doi.org/10.1103/PhysRevD.45.4514}{\emph{Phys. Rev.}
  {\bfseries D45} (1992) 4514--4535}.

\bibitem{Kosowsky:1992vn}
A.~Kosowsky and M.~S. Turner, \emph{{Gravitational radiation from colliding
  vacuum bubbles: envelope approximation to many bubble collisions}},
  \href{http://dx.doi.org/10.1103/PhysRevD.47.4372}{\emph{Phys. Rev.}
  {\bfseries D47} (1993) 4372--4391},
  [\href{https://arxiv.org/abs/astro-ph/9211004}{{\ttfamily
  astro-ph/9211004}}].

\bibitem{Kamionkowski:1993fg}
M.~Kamionkowski, A.~Kosowsky and M.~S. Turner, \emph{{Gravitational radiation
  from first order phase transitions}},
  \href{http://dx.doi.org/10.1103/PhysRevD.49.2837}{\emph{Phys. Rev.}
  {\bfseries D49} (1994) 2837--2851},
  [\href{https://arxiv.org/abs/astro-ph/9310044}{{\ttfamily
  astro-ph/9310044}}].

\bibitem{Hogan:1986qda}
C.~J. Hogan, \emph{{Gravitational radiation from cosmological phase
  transitions}}, {\emph{Mon. Not. Roy. Astron. Soc.} {\bfseries 218} (1986)
  629--636}.

\bibitem{Caprini:2006jb}
C.~Caprini and R.~Durrer, \emph{{Gravitational waves from stochastic
  relativistic sources: Primordial turbulence and magnetic fields}},
  \href{http://dx.doi.org/10.1103/PhysRevD.74.063521}{\emph{Phys. Rev.}
  {\bfseries D74} (2006) 063521},
  [\href{https://arxiv.org/abs/astro-ph/0603476}{{\ttfamily
  astro-ph/0603476}}].

\bibitem{Caprini:2007xq}
C.~Caprini, R.~Durrer and G.~Servant, \emph{{Gravitational wave generation from
  bubble collisions in first-order phase transitions: An analytic approach}},
  \href{http://dx.doi.org/10.1103/PhysRevD.77.124015}{\emph{Phys. Rev.}
  {\bfseries D77} (2008) 124015},
  [\href{https://arxiv.org/abs/0711.2593}{{\ttfamily 0711.2593}}].

\bibitem{Huber:2008hg}
S.~J. Huber and T.~Konstandin, \emph{{Gravitational Wave Production by
  Collisions: More Bubbles}},
  \href{http://dx.doi.org/10.1088/1475-7516/2008/09/022}{\emph{JCAP} {\bfseries
  0809} (2008) 022}, [\href{https://arxiv.org/abs/0806.1828}{{\ttfamily
  0806.1828}}].

\bibitem{Kahniashvili:2008pe}
T.~Kahniashvili, L.~Campanelli, G.~Gogoberidze, Y.~Maravin and B.~Ratra,
  \emph{{Gravitational Radiation from Primordial Helical Inverse Cascade MHD
  Turbulence}}, \href{http://dx.doi.org/10.1103/PhysRevD.78.123006,
  10.1103/PhysRevD.79.109901}{\emph{Phys. Rev.} {\bfseries D78} (2008) 123006},
  [\href{https://arxiv.org/abs/0809.1899}{{\ttfamily 0809.1899}}].

\bibitem{Kahniashvili:2008pf}
T.~Kahniashvili, A.~Kosowsky, G.~Gogoberidze and Y.~Maravin,
  \emph{{Detectability of Gravitational Waves from Phase Transitions}},
  \href{http://dx.doi.org/10.1103/PhysRevD.78.043003}{\emph{Phys. Rev.}
  {\bfseries D78} (2008) 043003},
  [\href{https://arxiv.org/abs/0806.0293}{{\ttfamily 0806.0293}}].

\bibitem{Caprini:2009yp}
C.~Caprini, R.~Durrer and G.~Servant, \emph{{The stochastic gravitational wave
  background from turbulence and magnetic fields generated by a first-order
  phase transition}},
  \href{http://dx.doi.org/10.1088/1475-7516/2009/12/024}{\emph{JCAP} {\bfseries
  0912} (2009) 024}, [\href{https://arxiv.org/abs/0909.0622}{{\ttfamily
  0909.0622}}].

\bibitem{Kahniashvili:2009mf}
T.~Kahniashvili, L.~Kisslinger and T.~Stevens, \emph{{Gravitational Radiation
  Generated by Magnetic Fields in Cosmological Phase Transitions}},
  \href{http://dx.doi.org/10.1103/PhysRevD.81.023004}{\emph{Phys. Rev.}
  {\bfseries D81} (2010) 023004},
  [\href{https://arxiv.org/abs/0905.0643}{{\ttfamily 0905.0643}}].

\bibitem{Hindmarsh:2013xza}
M.~Hindmarsh, S.~J. Huber, K.~Rummukainen and D.~J. Weir, \emph{{Gravitational
  waves from the sound of a first order phase transition}},
  \href{http://dx.doi.org/10.1103/PhysRevLett.112.041301}{\emph{Phys. Rev.
  Lett.} {\bfseries 112} (2014) 041301},
  [\href{https://arxiv.org/abs/1304.2433}{{\ttfamily 1304.2433}}].

\bibitem{Giblin:2013kea}
J.~T. Giblin, Jr. and J.~B. Mertens, \emph{{Vacuum Bubbles in the Presence of a
  Relativistic Fluid}},
  \href{http://dx.doi.org/10.1007/JHEP12(2013)042}{\emph{JHEP} {\bfseries 12}
  (2013) 042}, [\href{https://arxiv.org/abs/1310.2948}{{\ttfamily 1310.2948}}].

\bibitem{Giblin:2014qia}
J.~T. Giblin and J.~B. Mertens, \emph{{Gravitional radiation from first-order
  phase transitions in the presence of a fluid}},
  \href{http://dx.doi.org/10.1103/PhysRevD.90.023532}{\emph{Phys. Rev.}
  {\bfseries D90} (2014) 023532},
  [\href{https://arxiv.org/abs/1405.4005}{{\ttfamily 1405.4005}}].

\bibitem{Kisslinger:2015hua}
L.~Kisslinger and T.~Kahniashvili, \emph{{Polarized Gravitational Waves from
  Cosmological Phase Transitions}},
  \href{http://dx.doi.org/10.1103/PhysRevD.92.043006}{\emph{Phys. Rev.}
  {\bfseries D92} (2015) 043006},
  [\href{https://arxiv.org/abs/1505.03680}{{\ttfamily 1505.03680}}].

\bibitem{Hindmarsh:2015qta}
M.~Hindmarsh, S.~J. Huber, K.~Rummukainen and D.~J. Weir, \emph{{Numerical
  simulations of acoustically generated gravitational waves at a first order
  phase transition}},
  \href{http://dx.doi.org/10.1103/PhysRevD.92.123009}{\emph{Phys. Rev.}
  {\bfseries D92} (2015) 123009},
  [\href{https://arxiv.org/abs/1504.03291}{{\ttfamily 1504.03291}}].

\bibitem{Weir:2016tov}
D.~J. Weir, \emph{{Revisiting the envelope approximation: gravitational waves
  from bubble collisions}},
  \href{http://dx.doi.org/10.1103/PhysRevD.93.124037}{\emph{Phys. Rev.}
  {\bfseries D93} (2016) 124037},
  [\href{https://arxiv.org/abs/1604.08429}{{\ttfamily 1604.08429}}].

\bibitem{Jinno:2016vai}
R.~Jinno and M.~Takimoto, \emph{{Gravitational waves from bubble collisions: An
  analytic derivation}},
  \href{http://dx.doi.org/10.1103/PhysRevD.95.024009}{\emph{Phys. Rev.}
  {\bfseries D95} (2017) 024009},
  [\href{https://arxiv.org/abs/1605.01403}{{\ttfamily 1605.01403}}].

\bibitem{Jinno:2017fby}
R.~Jinno and M.~Takimoto, \emph{{Gravitational waves from bubble dynamics:
  Beyond the Envelope}},  \href{https://arxiv.org/abs/1707.03111}{{\ttfamily
  1707.03111}}.

\bibitem{Konstandin:2017sat}
T.~Konstandin, \emph{{Gravitational radiation from a bulk flow model}},
  \href{http://dx.doi.org/10.1088/1475-7516/2018/03/047}{\emph{JCAP} {\bfseries
  1803} (2018) 047}, [\href{https://arxiv.org/abs/1712.06869}{{\ttfamily
  1712.06869}}].

\bibitem{Cutting:2018tjt}
D.~Cutting, M.~Hindmarsh and D.~J. Weir, \emph{{Gravitational waves from vacuum
  first-order phase transitions: from the envelope to the lattice}},
  \href{https://arxiv.org/abs/1802.05712}{{\ttfamily 1802.05712}}.

\bibitem{Geller:2018mwu}
M.~Geller, A.~Hook, R.~Sundrum and Y.~Tsai, \emph{{Primordial Anisotropies in
  the Gravitational Wave Background from Cosmological Phase Transitions}},
  \href{https://arxiv.org/abs/1803.10780}{{\ttfamily 1803.10780}}.

\bibitem{Caprini:2015zlo}
C.~Caprini et~al., \emph{{Science with the space-based interferometer eLISA.
  II: Gravitational waves from cosmological phase transitions}},
  \href{http://dx.doi.org/10.1088/1475-7516/2016/04/001}{\emph{JCAP} {\bfseries
  1604} (2016) 001}, [\href{https://arxiv.org/abs/1512.06239}{{\ttfamily
  1512.06239}}].

\bibitem{Moore:1995si}
G.~D. Moore and T.~Prokopec, \emph{{How fast can the wall move? A Study of the
  electroweak phase transition dynamics}},
  \href{http://dx.doi.org/10.1103/PhysRevD.52.7182}{\emph{Phys. Rev.}
  {\bfseries D52} (1995) 7182--7204},
  [\href{https://arxiv.org/abs/hep-ph/9506475}{{\ttfamily hep-ph/9506475}}].

\bibitem{John:2000zq}
P.~John and M.~G. Schmidt, \emph{{Do stops slow down electroweak bubble
  walls?}}, \href{http://dx.doi.org/10.1016/S0550-3213(00)00768-9,
  10.1016/S0550-3213(02)01014-3}{\emph{Nucl. Phys.} {\bfseries B598} (2001)
  291--305}, [\href{https://arxiv.org/abs/hep-ph/0002050}{{\ttfamily
  hep-ph/0002050}}].

\bibitem{Konstandin:2014zta}
T.~Konstandin, G.~Nardini and I.~Rues, \emph{{From Boltzmann equations to
  steady wall velocities}},
  \href{http://dx.doi.org/10.1088/1475-7516/2014/09/028}{\emph{JCAP} {\bfseries
  1409} (2014) 028}, [\href{https://arxiv.org/abs/1407.3132}{{\ttfamily
  1407.3132}}].

\bibitem{Fukushima:2010bq}
K.~Fukushima and T.~Hatsuda, \emph{{The phase diagram of dense QCD}},
  \href{http://dx.doi.org/10.1088/0034-4885/74/1/014001}{\emph{Rept. Prog.
  Phys.} {\bfseries 74} (2011) 014001},
  [\href{https://arxiv.org/abs/1005.4814}{{\ttfamily 1005.4814}}].

\bibitem{Andersen:2014xxa}
J.~O. Andersen, W.~R. Naylor and A.~Tranberg, \emph{{Phase diagram of QCD in a
  magnetic field: A review}},
  \href{http://dx.doi.org/10.1103/RevModPhys.88.025001}{\emph{Rev. Mod. Phys.}
  {\bfseries 88} (2016) 025001},
  [\href{https://arxiv.org/abs/1411.7176}{{\ttfamily 1411.7176}}].

\bibitem{Steinhardt:1981ct}
P.~J. Steinhardt, \emph{{Relativistic Detonation Waves and Bubble Growth in
  False Vacuum Decay}},
  \href{http://dx.doi.org/10.1103/PhysRevD.25.2074}{\emph{Phys. Rev.}
  {\bfseries D25} (1982) 2074}.

\bibitem{Dev:2016feu}
P.~S.~B. Dev and A.~Mazumdar, \emph{{Probing the Scale of New Physics by
  Advanced LIGO/VIRGO}},
  \href{http://dx.doi.org/10.1103/PhysRevD.93.104001}{\emph{Phys. Rev.}
  {\bfseries D93} (2016) 104001},
  [\href{https://arxiv.org/abs/1602.04203}{{\ttfamily 1602.04203}}].

\bibitem{Ahmadvand:2017tue}
M.~Ahmadvand and K.~Bitaghsir~Fadafan, \emph{{The cosmic QCD phase transition
  with dense matter and its gravitational waves from holography}},
  \href{http://dx.doi.org/10.1016/j.physletb.2018.01.066}{\emph{Phys. Lett.}
  {\bfseries B779} (2018) 1--8},
  [\href{https://arxiv.org/abs/1707.05068}{{\ttfamily 1707.05068}}].

\bibitem{Cyburt:2004yc}
R.~H. Cyburt, B.~D. Fields, K.~A. Olive and E.~Skillman, \emph{{New BBN limits
  on physics beyond the standard model from $^4He$}},
  \href{http://dx.doi.org/10.1016/j.astropartphys.2005.01.005}{\emph{Astropart.
  Phys.} {\bfseries 23} (2005) 313--323},
  [\href{https://arxiv.org/abs/astro-ph/0408033}{{\ttfamily
  astro-ph/0408033}}].

\bibitem{Caprini:2018mtu}
C.~Caprini and D.~G. Figueroa, \emph{{Cosmological Backgrounds of Gravitational
  Waves}},  \href{https://arxiv.org/abs/1801.04268}{{\ttfamily 1801.04268}}.

\bibitem{Thrane:2013oya}
E.~Thrane and J.~D. Romano, \emph{{Sensitivity curves for searches for
  gravitational-wave backgrounds}},
  \href{http://dx.doi.org/10.1103/PhysRevD.88.124032}{\emph{Phys. Rev.}
  {\bfseries D88} (2013) 124032},
  [\href{https://arxiv.org/abs/1310.5300}{{\ttfamily 1310.5300}}].

\bibitem{Moore:2014eua}
C.~J. Moore, S.~R. Taylor and J.~R. Gair, \emph{{Estimating the sensitivity of
  pulsar timing arrays}},
  \href{http://dx.doi.org/10.1088/0264-9381/32/5/055004}{\emph{Class. Quant.
  Grav.} {\bfseries 32} (2015) 055004},
  [\href{https://arxiv.org/abs/1406.5199}{{\ttfamily 1406.5199}}].

\bibitem{Arzoumanian:2018saf}
{\scshape NANOGRAV} collaboration, Z.~Arzoumanian et~al., \emph{{The NANOGrav
  11-year Data Set: Pulsar-timing Constraints On The Stochastic
  Gravitational-wave Background}},
  \href{https://arxiv.org/abs/1801.02617}{{\ttfamily 1801.02617}}.

\bibitem{TheLIGOScientific:2016dpb}
{\scshape Virgo, LIGO Scientific} collaboration, B.~P. Abbott et~al.,
  \emph{{Upper Limits on the Stochastic Gravitational-Wave Background from
  Advanced LIGO's First Observing Run}},
  \href{http://dx.doi.org/10.1103/PhysRevLett.118.121101,
  10.1103/PhysRevLett.119.029901}{\emph{Phys. Rev. Lett.} {\bfseries 118}
  (2017) 121101}, [\href{https://arxiv.org/abs/1612.02029}{{\ttfamily
  1612.02029}}].

\bibitem{Lentati:2015qwp}
L.~Lentati et~al., \emph{{European Pulsar Timing Array Limits On An Isotropic
  Stochastic Gravitational-Wave Background}},
  \href{http://dx.doi.org/10.1093/mnras/stv1538}{\emph{Mon. Not. Roy. Astron.
  Soc.} {\bfseries 453} (2015) 2576--2598},
  [\href{https://arxiv.org/abs/1504.03692}{{\ttfamily 1504.03692}}].

\bibitem{Shannon:2015ect}
R.~M. Shannon et~al., \emph{{Gravitational waves from binary supermassive black
  holes missing in pulsar observations}},
  \href{http://dx.doi.org/10.1126/science.aab1910}{\emph{Science} {\bfseries
  349} (2015) 1522--1525}, [\href{https://arxiv.org/abs/1509.07320}{{\ttfamily
  1509.07320}}].

\bibitem{Moore:2014lga}
C.~J. Moore, R.~H. Cole and C.~P.~L. Berry, \emph{{Gravitational-wave
  sensitivity curves}},
  \href{http://dx.doi.org/10.1088/0264-9381/32/1/015014}{\emph{Class. Quant.
  Grav.} {\bfseries 32} (2015) 015014},
  [\href{https://arxiv.org/abs/1408.0740}{{\ttfamily 1408.0740}}].

\bibitem{GWplotter}
C.~J. Moore, R.~H. Cole and C.~P.~L. Berry, ``Gravitational wave detectors and
  sources.'' \url{http://rhcole.com/apps/GWplotter}.

\bibitem{Audley:2017drz}
H.~Audley et~al., \emph{{Laser Interferometer Space Antenna}},
  \href{https://arxiv.org/abs/1702.00786}{{\ttfamily 1702.00786}}.

\bibitem{Aasi:2013wya}
{\scshape VIRGO, LIGO Scientific} collaboration, B.~P. Abbott et~al.,
  \emph{{Prospects for Observing and Localizing Gravitational-Wave Transients
  with Advanced LIGO, Advanced Virgo and KAGRA}},
  \href{https://arxiv.org/abs/1304.0670}{{\ttfamily 1304.0670}}.

\bibitem{Sathyaprakash:2012jk}
B.~Sathyaprakash et~al., \emph{{Scientific Objectives of Einstein Telescope}},
  \href{http://dx.doi.org/10.1088/0264-9381/29/12/124013,
  10.1088/0264-9381/30/7/079501}{\emph{Class. Quant. Grav.} {\bfseries 29}
  (2012) 124013}, [\href{https://arxiv.org/abs/1206.0331}{{\ttfamily
  1206.0331}}].

\bibitem{Figueroa:2018xtu}
D.~G. Figueroa, E.~Megias, G.~Nardini, M.~Pieroni, M.~Quiros, A.~Ricciardone
  et~al., \emph{{LISA as a probe for particle physics: electroweak scale tests
  in synergy with ground-based experiments}},
  \href{https://arxiv.org/abs/1806.06463}{{\ttfamily 1806.06463}}.

\bibitem{Carena:2000id}
M.~Carena, J.~M. Moreno, M.~Quiros, M.~Seco and C.~E.~M. Wagner,
  \emph{{Supersymmetric CP violating currents and electroweak baryogenesis}},
  \href{http://dx.doi.org/10.1016/S0550-3213(01)00032-3}{\emph{Nucl. Phys.}
  {\bfseries B599} (2001) 158--184},
  [\href{https://arxiv.org/abs/hep-ph/0011055}{{\ttfamily hep-ph/0011055}}].

\bibitem{Dittmaier:2011ti}
{\scshape LHC Higgs Cross Section Working Group} collaboration, S.~Dittmaier
  et~al., \emph{{Handbook of LHC Higgs Cross Sections: 1. Inclusive
  Observables}},  \href{https://arxiv.org/abs/1101.0593}{{\ttfamily
  1101.0593}}.

\bibitem{Djouadi:2005gi}
A.~Djouadi, \emph{{The Anatomy of electro-weak symmetry breaking. I: The Higgs
  boson in the standard model}},
  \href{http://dx.doi.org/10.1016/j.physrep.2007.10.004}{\emph{Phys. Rept.}
  {\bfseries 457} (2008) 1--216},
  [\href{https://arxiv.org/abs/hep-ph/0503172}{{\ttfamily hep-ph/0503172}}].

\bibitem{Aaboud:2017fgj}
{\scshape ATLAS} collaboration, M.~Aaboud et~al., \emph{{Search for $WW/WZ$
  resonance production in $\ell \nu qq$ final states in $pp$ collisions at
  $\sqrt{s} =$ 13 TeV with the ATLAS detector}},
  \href{https://arxiv.org/abs/1710.07235}{{\ttfamily 1710.07235}}.

\bibitem{Aaboud:2017itg}
{\scshape ATLAS} collaboration, M.~Aaboud et~al., \emph{{Searches for heavy
  $ZZ$ and $ZW$ resonances in the $\ell\ell qq$ and $\nu\nu qq$ final states in
  $pp$ collisions at $\sqrt{s}=13$ TeV with the ATLAS detector}},
  \href{https://arxiv.org/abs/1708.09638}{{\ttfamily 1708.09638}}.

\bibitem{Aaboud:2017yyg}
{\scshape ATLAS} collaboration, M.~Aaboud et~al., \emph{{Search for new
  phenomena in high-mass diphoton final states using 37 fb$^{-1}$ of
  proton--proton collisions collected at $\sqrt{s}=13$ TeV with the ATLAS
  detector}},
  \href{http://dx.doi.org/10.1016/j.physletb.2017.10.039}{\emph{Phys. Lett.}
  {\bfseries B775} (2017) 105--125},
  [\href{https://arxiv.org/abs/1707.04147}{{\ttfamily 1707.04147}}].

\bibitem{Aaboud:2017sjh}
{\scshape ATLAS} collaboration, M.~Aaboud et~al., \emph{{Search for additional
  heavy neutral Higgs and gauge bosons in the ditau final state produced in 36
  fb$^{-1}$ of pp collisions at $ \sqrt{s}=13 $ TeV with the ATLAS detector}},
  \href{http://dx.doi.org/10.1007/JHEP01(2018)055}{\emph{JHEP} {\bfseries 01}
  (2018) 055}, [\href{https://arxiv.org/abs/1709.07242}{{\ttfamily
  1709.07242}}].

\bibitem{Khachatryan:2016yec}
{\scshape CMS} collaboration, V.~Khachatryan et~al., \emph{{Search for
  high-mass diphoton resonances in protonÐproton collisions at 13 TeV and
  combination with 8 TeV search}},
  \href{http://dx.doi.org/10.1016/j.physletb.2017.01.027}{\emph{Phys. Lett.}
  {\bfseries B767} (2017) 147--170},
  [\href{https://arxiv.org/abs/1609.02507}{{\ttfamily 1609.02507}}].

\bibitem{Aaboud:2018mjh}
{\scshape ATLAS} collaboration, M.~Aaboud et~al., \emph{{Search for heavy
  particles decaying into top-quark pairs using lepton-plus-jets events in
  proton--proton collisions at $\sqrt{s} = 13$ TeV with the ATLAS detector}},
  {\emph{Submitted to: Eur. Phys. J.} (2018) },
  [\href{https://arxiv.org/abs/1804.10823}{{\ttfamily 1804.10823}}].

\bibitem{Golling:2016gvc}
T.~Golling et~al., \emph{{Physics at a 100 TeV pp collider: beyond the Standard
  Model phenomena}},
  \href{http://dx.doi.org/10.23731/CYRM-2017-003.441}{\emph{CERN Yellow Report}
  (2017) 441--634}, [\href{https://arxiv.org/abs/1606.00947}{{\ttfamily
  1606.00947}}].

\bibitem{Abbott:2017xzg}
{\scshape Virgo, LIGO Scientific} collaboration, B.~P. Abbott et~al.,
  \emph{{GW170817: Implications for the Stochastic Gravitational-Wave
  Background from Compact Binary Coalescences}},
  \href{http://dx.doi.org/10.1103/PhysRevLett.120.091101}{\emph{Phys. Rev.
  Lett.} {\bfseries 120} (2018) 091101},
  [\href{https://arxiv.org/abs/1710.05837}{{\ttfamily 1710.05837}}].

\bibitem{Axen:2018zvb}
M.~F. Axen, S.~Banagiri, A.~Matas, C.~Caprini and V.~Mandic,
  \emph{{Multi-wavelength observations of cosmological phase transitions using
  LISA and Cosmic Explorer}},
  \href{https://arxiv.org/abs/1806.02500}{{\ttfamily 1806.02500}}.

\bibitem{TalkPedro}
T.~Opferkuch, M.~Breitbach, J.~Kopp, E.~Madge and P.~Schwaller, ``Probing light
  hidden sectors with pulsar timing arrays.''
  \url{https://indico.desy.de/indico/event/18498/session/12/contribution/43/material/slides/0.pdf}.

\end{thebibliography}\endgroup

\end{document}